\documentclass[11pt,a4paper]{article}

\usepackage{graphicx} 
\usepackage{xcolor}
\usepackage{amsmath} 
\usepackage{amssymb} 
\usepackage{cite}
\usepackage{float}
\usepackage{braket}
\usepackage[T1]{fontenc}
\usepackage{titlesec}

\titleformat{\paragraph}
{\normalfont\normalsize\bfseries}{\theparagraph}{1em}{}
\titlespacing*{\paragraph}
{0pt}{3.25ex plus 1ex minus .2ex}{1.5ex plus .2ex}
\newcommand{\be}{\begin{equation}}
\newcommand{\ee}{\end{equation}}

\newcommand{\bpm}{\begin{pmatrix}}
\newcommand{\epm}{\end{pmatrix}}

\newcommand{\ii}{\mathrm{i}}
\newcommand{\dd}{\mathrm{d}}
\newcommand{\e}{\mathrm{e}} 

\usepackage{tabularx} 
\usepackage{psfrag} 

\usepackage{eurosym} 
\def\?{\euro{}}
\usepackage{footnote}
\usepackage{tablefootnote}

\usepackage{color} 
\usepackage{mathtools}

\definecolor{linkcolor}{rgb}{0,0,0.6} 
\usepackage[pdftex,colorlinks=true,
pdfstartview=FitV,
linkcolor= linkcolor,
citecolor= linkcolor,
urlcolor= linkcolor,
hyperindex=true,
hyperfigures=false]
{hyperref} 

\usepackage{fancyhdr} 

\numberwithin{equation}{section}
\begin{document}

\begin{titlepage}

	\begin{center}

	\vskip .5in 
	\noindent

	{\Large \bf{Stability of non-supersymmetric vacua from calibrations}}

	\bigskip\medskip

	 Vincent Menet and Alessandro Tomasiello \\

	\bigskip\medskip
	{\small 
Dipartimento di Matematica, Universit\`a di Milano--Bicocca, \\ Via Cozzi 55, 20126 Milano, Italy \\ and \\ INFN, sezione di Milano--Bicocca
		}

	\vskip .5cm 
	{\small \tt vincent.menet, alessandro.tomasiello@unimib.it}
	\vskip .9cm 
	     	{\bf Abstract }
	\vskip .1in
	\end{center}

Supersymmetric vacua are protected from vacuum decay by energy positivity. No such argument is known for any non-supersymmetric vacua. In this paper, we try to extend to the latter a simpler argument based on calibrations, to at least protect them from decays mediated by D-brane bubbles, including their abelian bound states. We examine several classes of AdS$_4$ and AdS$_5$ solutions in type II string theory, including some new ones, involving coset spaces, sphere fibrations, K\"ahler--Einstein manifolds. Many of these vacua have resisted against all the decay channels we were able to assess. We also show how to use calibrations for the stability of D-branes already present in a non-supersymmetric solution.

	\vfill
	\eject

	\end{titlepage}

\tableofcontents

\newpage

\section{Introduction}

Many years of work have yielded a lot of progress in understanding supersymmetric compactifications of string theory. The fermionic supersymmetry parameters define natural $G$-structures, which can be used to classify solutions and sometimes to find them explicitly. 

The lack of such structures makes supersymmetry-breaking vacua much harder to tackle. We have fewer techniques to find and classify them. Many of those we know appear to be unstable or metastable. A vacuum can be unstable because of a tachyonic field, or because of quantum tunneling, via the appearance of a bubble of an alternative vacuum \cite{Coleman:1980aw}. The latter is sometimes also known as non-perturbative instability, to contrast it to tachyons. It is both more common and more subtle, as it is difficult to list all decay channels and check if they actually occur. This phenomenon was suggested as a possible mechanism to reduce the cosmological constant \cite{Brown:1988kg,Bousso:2000xa}. In the context of AdS vacua, it is likely to signal that the dual CFT cannot be unitary; many examples of such decays have been found, starting with \cite{Dowker:1995sg,Maldacena:1998uz,Gaiotto:2009mv}. The membrane version of the weak gravity conjecture even suggests that all non-supersymmetric AdS vacua decay in such a way \cite{Ooguri:2016pdq}. Learning to control such effects would be important for building realistic vacua, and (in the AdS case) for holography.

In this paper, we take a step in that direction: we will extend to supersymmetry-breaking one of the mechanisms that make supersymmetric AdS vacua stable. Powerful positive energy arguments \cite{Witten:1981mf,Gibbons:1982fy,Hull:1983ap,Kowalski-Glikman:1985pgb,Giri:2021eob} protect the latter from \emph{any} decay; replicating this success for supersymmetry-breaking is possible in four dimensions \cite{Boucher:1984yx}, but a lot more challenging for compactifications \cite{Giri:2021eob}. A much simpler argument was found in \cite[Sec.~2]{Giri:2021eob} to show protection from decays where the bubble of new vacuum is a D-brane localized at a value of the spacetime radial coordinate. This easier argument is the one we will extend to supersymmetry-breaking vacua.

Supersymmetric vacua obey a system of \emph{pure spinor equations} \cite{Grana:2005sn} (Sec.~\ref{sec:pure}). Pure spinors are bilinears defined naturally by the supersymmetry parameters. Importantly for us, they can be interpreted \cite{Martucci:2005ht} as \emph{calibrations}, forms whose integrals measure the minimal energy of a brane in a given homology class. One of these calibrations was shown in \cite{Giri:2021eob} to be responsible for the stability of supersymmetric vacua under D-brane bubbles (Sec.~\ref{sec:stab}). 

Here we will look for vacua where this type of calibration exists even if supersymmetry is broken. In a sense, it becomes an auxiliary variable, in the spirit of \emph{fake supersymmetry} techniques \cite{Boucher:1984yx,Freedman:2003ax}. This object provides a quick lower bound to a D-brane's DBI action, which in turn can establish that a given D-brane bubble does not lead to an instability. On the other hand, sometimes we also find instabilities by exhibiting branes that are calibrated, i.e.~that saturate the lower bound. 

This method can also handle abelian bound states, namely branes with an abelian world-sheet flux. It is natural to conjecture that the method would also apply to non-abelian bound states, but this would require more work on the non-abelian DBI action. Presumably these more general objects would be needed in order to cover all possible topological classes for D-branes (which are classified by K-theory \cite{Minasian:1997mm,Witten:1998cd}.) Another possible extension of the formalism would be to cover oblique branes that are not localized in the radial direction, but where the latter is a function of the internal coordinates; these more general objects would look like thick branes from the four-dimensional point of view. 
We hope to return to these points in the future.

As another important limitation, we stress again that our technique in this paper only covers brane bubbles. They are the most commonly found decay modes, but several more types exist, like bubbles of nothing \cite{Witten:1981gj}, their variants \cite{Horowitz:2007pr}, BIonic instabilities \cite{Marchesano:2021ycx}, NS5 instabilities \cite{Apruzzi:2019ecr} among others. Finally, we are working in the supergravity approximation; as usual, string corrections could change some of our conclusions.

We have illustrated the idea on several classes of type II vacua. Some of them were already known, and our contribution is their stability analysis; some of them are new. Here is a list of the classes that we consider:
\begin{itemize}
    \item AdS$_4\times$ a twistor fibration (Sec.~\ref{sub:tw}). Most prominently, this includes $\mathbb{CP}^3$, in which case these vacua were already found in \cite{Koerber:2010rn}. About half of these resisted all the decay channels we were able to check.
    \item AdS$_4\times \mathbb{F}(1,2;3)$ (Sec.~\ref{sub:flag}). These are new, but have similar properties to the previous class, which in fact they include (since some metrics on $\mathbb{F}(1,2;3)$ are twistor fibrations). 
    \item AdS$_4\times$ a K\"ahler--Einstein manifold \cite{Gaiotto:2009mv,Lust:2009zb,Romans:1985tz} (Sec.~\ref{sub:ke6}). A large subset of these is destabilized by D2-branes, but in a certain limit, the only instability channels we find are peculiar bound states with very large world-sheet flux. These are likely to be beyond the domain of validity of our methods.
    \item  AdS$_5\times$ a stretched, regular Sasaki--Einstein manifold \cite{Romans:1984an} (Sec.~\ref{sub:se}). We find no instabilities here.
    \item AdS$_5\times$ an $S^1$-fibration over $S^2\times S^2$ (Sec.~\ref{sub:prodRiem}). These are the so-called $T_{p,q}$-spaces \cite{Romans:1984an}, for which we find again no instabilities.
\end{itemize}
For several of these vacua, the internal space is a round sphere fibration; this might make them vulnerable to bubbles of nothing and their variants. One was indeed found for the AdS$_5\times \mathbb{CP}^3$ vacua of M-theory \cite{Ooguri:2017njy}; it would be interesting to check if this also happens for the twistor vacua of Sec.~\ref{sub:tw}. In that context, the surviving K\"ahler--Einstein vacua of Sec.~\ref{sub:ke6} hold particularly nice features within the set of vacua we investigated: beyond our brane bubble analysis, the possibility of bubbles of nothing and their variants would seem more remote, given the wide variety of available topologies. Moreover, these vacua are protected against BIonic and NS5 instabilities, as they have no NS flux. (Our other AdS$_4$ solutions have no three-cycles, so they are also protected from these.) The cobordism conjecture does predict that there is no \emph{topological} protection against more general bubbles, but that still does not mean that they do nucleate and destabilize the vacuum. Other vacua have so far resisted non-perturbative instabilities \cite{Guarino:2020jwv,Giambrone:2021wsm,Suh:2021icf}.

In a related development, we show in Sec.~\ref{sec:mink} how calibrations can also be used to establish the stability of D$p$-branes already present in a supersymmetry-breaking vacuum. We illustrate this on a non-supersymmetric solution found recently in \cite{Macpherson:2024frt} as a modification of the IIA reduction of the Gaiotto--Maldacena solutions \cite{Gaiotto:2009gz}.

Our work was partially inspired by the idea of using pure spinors and calibrations to \emph{find} supersymmetry-breaking vacua \cite{Lust:2008zd,Legramandi:2019ulq,Menet:2023rnt}. We opted not to use this formalism here, but it has now progressed to the point that it covers almost all Minkowski$_4$ vacua where only one of the three pure spinor equations is satisfied \cite{Menet:2023rnt,Menet:2023rml}. As we will see, this is the same equation that ensures brane bubble stability; so an AdS$_4$ extension of these results would yield $\mathcal{N}=0$ solutions where all D$p$-branes are automatically stable. It would be interesting to pursue this further.

\section{Pure spinors and generalized calibrations}
\label{sec:pure}

We begin with a quick review of some techniques that are useful for supersymmetric vacua.

\subsection{Flux backgrounds and pure spinors}

We consider type II supergravity solutions that are vacua: namely, the ten-dimensional geometry is given by a warped product between a maximally symmetric four-dimensional spacetime $X_4$ and a compact six-dimensional internal manifold $M_6$, with the metric taking the form
\be 
\label{10dmet}
\dd s^2_{10}=\e^{2A}\dd s^2_{4} + \dd s^2_6=\e^{2A(y)}g_{\mu\nu}\dd x^\mu\dd x^\nu+g_{mn}\dd y^m\dd y^n,
\ee
where $x^\mu$, $\mu=0,...,3$ are the external coordinates on $X_4$, and $y^m$, $m=1,...,6$ are the coordinates on $M_6$. 
The NSNS three-form $H$ is purely internal. The ten-dimensional RR field strength takes the form
\be 
F_{10}=F+\e^{4A}\text{vol}_4\wedge\Tilde{F}\label{RR},
\ee
with $\Tilde{F}=\ast\lambda F$, $\ast$ the internal Hodge-star operator, and $\lambda$ the reversal of all form indices: $\lambda \dd x^{m_1} \wedge \ldots \wedge \dd x^{m_k} = \dd x^{m_k} \wedge \ldots \wedge \dd x^{m_1}$.

We specialise the discussion to backgrounds admitting two globally defined ten-dimensional spinors $\epsilon_1$ and $\epsilon_2$, splitting as
  \begin{equation}
\epsilon_1=\zeta\otimes\eta_{1}+c.c.\qquad \epsilon_2=\zeta\otimes\eta_{2} +c.c.\label{10dspinors},
\end{equation}
with $\zeta$ a Weyl spinor of positive chirality on $X_4$, and 
$\eta_{1}$, $\eta_{2}$ Weyl spinors of $M_6$. $\eta_{1}$ always has positive chirality, and in type IIA/IIB, $\eta_2$
has negative/positive chirality.

We introduce the pure spinors $\Phi_1$ and $\Phi_2$, polyforms 
defined via the internal spinors\footnote{One should think of these tensor products in terms of the Fierz identity \be \eta\otimes\chi^\dagger=\sum_{k=0}^6 \frac{1}{k!}\left(\chi^\dagger\gamma_{m_k...m_1}\eta\right)\gamma^{m_1...m_k}.\ee Such tensor products are then isomorphic to polyforms via the Clifford map.}
\begin{subequations}\label{eq:Phi-eta}
\begin{align}
        \Phi_1&=\eta_1\otimes\eta_2^\dagger\\
    \Phi_2&= \eta_1\otimes\eta_2^T,
\end{align}
\end{subequations}
with their norms defined with the Mukai pairing as $\braket{ \Phi_1,\bar\Phi_1} =\braket{ \Phi_2,\bar\Phi_2}= -8i\lVert \eta_1 \rVert\lVert \eta_2 \rVert\text{vol}_6$ and vol$_6$ the string-frame internal volume form. In type IIA/IIB, $\Phi_1$ and $\Phi_2$ are odd/even and even/odd:
\be \Phi_1=\Phi_\mp,\quad \Phi_2=\Phi_\pm \, . \ee
The pure spinors satisfy a compatibility condition:
\be \braket{\Phi_1,V\cdot\Phi_2}=\braket{\Phi_1,V\cdot\bar{\Phi}_2}=0\qquad\forall \ V=v+\xi \text{ sections of } TM_6\oplus T^\ast M_6,\ee
with $ V\cdot \Psi=\iota_v\Psi+\xi\wedge\Psi$ for $\Psi$ a form.

For $\mathcal{N}=1$ supersymmetric backgrounds, the Killing spinor equations satisfied by the internal spinors $\eta_1$ and $\eta_2$ can be written in terms of the pure spinors as\footnote{The $B$ field twist should simply be understood here as the operator $\text{e}^{-B}\Psi\equiv\text{e}^{-B}\wedge\Psi$ for $\Psi$ a form.} \cite{Grana:2005sn}
\begin{subequations}\label{susyads0}
\begin{align}
  \dd(\text{e}^{3A-\phi}\text{e}^{-B}\Phi_2)&= \frac2L \e^{2A-\phi}\text{e}^{-B}\text{Re}\Phi_1\label{susy1ads0}\\ 
  \dd(\text{e}^{2A-\phi}\text{e}^{-B}\text{Re}\Phi_1)&= 0\label{susy2ads0}\\
   \dd(\text{e}^{4A-\phi}\text{e}^{-B}\text{Im}\Phi_1)&=\frac3L \e^{3A-\phi}\text{e}^{-B}\text{Im}\Phi_2- \e^{4A}\text{e}^{-B}\ast\lambda F\label{susy3ads0}. 
\end{align}
\end{subequations}
The second follows from the first upon acting with $\dd$. In the case where the internal space is Minkowski: $X_4=\text{Mink}_4$, the cosmological constant $\Lambda = - 3/L^2$ vanishes and these equations have an interpretation in terms of generalized calibration conditions for different probe branes.

\subsection{Generalized calibrations}

We will use generalized calibrations throughout this paper. These are a natural extension of ordinary calibrations.

A calibration form $\omega$ is a $p$-form of $M_6$ which satisfies two conditions, a differential one and an algebraic one. 
The differential calibration condition simply imposes the closure of the calibration form $\dd \omega=0$. The algebraic condition states that at every point $q\in M_6$, for every $p$-dimensional oriented subspace $\tau$ of the tangent space $T_q M_6$, the following must be respected:
\begin{equation}
\label{calbound}
 \omega|_{\tau} \leq \sqrt{\det g|_{\tau}} \, \dd \tau\equiv \text{vol}_\tau \,  ,
\end{equation}
where $\dd \tau = t^1 \wedge \ldots \wedge t^p$, with $t^\alpha$ a basis for the dual of $\tau$, and $\det g|_{\tau}$ the determinant of the pulled-back metric on $\tau$.
Additionally, at every point there must exist at least one subspace $\tau$ such that the above bound is saturated. 

A submanifold is said to be calibrated by $\omega$ if its corresponding tangent space is such that the above bound is saturated. Through both calibration conditions, we conclude that a calibrated submanifold minimizes the volume of its homology class.

For supersymmetric compactifications without fluxes, there is a direct 
relation between the $M_6$ cycles wrapped by D-branes and calibrations.
Indeed, one can construct calibration forms from the internal spinors of a given background, as spinor bilinears. The closure of such calibration forms then follows from supersymmetry. Given that the energy of these D-branes is merely given by their volume, a D-brane wrapping a calibrated cycle is energy minimizing. Moreover, the saturation of the calibration bound \eqref{calbound} is equivalent to a kappa symmetry condition, such that a D-brane wrapping a calibrated cycle preserves the bulk supersymmetry.

In flux compactification, the energy of a D-brane isn't simply given by its volume, but receives additional background and world-volume fluxes contributions. The generalized calibration theory provides a natural generalization of the standard calibration theory to incorporate these additional contributions, within the context of generalized Complex Geometry \cite{Martucci:2005ht}.
\medskip

More concretely, we consider D-branes in the warped geometry \eqref{10dmet} with a four-dimensional Minkowski spacetime.
These D-branes can wrap an internal cycle $\Sigma$ and they are said to be string, domain-wall or space-filling, when they are codimension two, one and zero respectively in the external space. As discussed in  \cite{Martucci:2005ht}, one can show that a static brane wrapping a cycle $\Sigma$ in the internal manifold of an
$\mathcal{N}=1$ flux backgrounds is supersymmetric if it wraps a calibrated generalized submanifold.
 
To make this statement precise, we recall here the notions of generalized submanifolds and generalized calibration forms.

\medskip

A generalized submanifold is a pair $(\Sigma,\ \mathcal{F})$ with $\Sigma\subset M_6$ a submanifold and $\mathcal{F}$ a two-form. For a D-brane, it is a world-volume two-form satisfying 
 \be 
 \dd \mathcal{F}=H|_\Sigma,\label{Hint}
 \ee
 with $H|_\Sigma$ the pull-back of $H$ on $\Sigma$. The generalized submanifold is said to be a generalized cycle if $\partial\Sigma=\emptyset$. 

\medskip

One can construct  polyforms from the pure spinors characterising the $\mathcal{N}=1$ background:
\begin{subequations}\label{caldef}
 \begin{align}
  \label{caldef-string}
  \omega^{\text{string}}&= \text{e}^{2A-\phi}\text{e}^{-B}\text{Re}\Phi_1\\
  \label{caldef-DW}
 \omega^\text{DW}&= \text{e}^{3A-\phi}\text{e}^{-B}\Phi_2\\
 \label{caldef-sf}
     \omega^{\text{sf}}&=\text{e}^{4A-\phi}\text{e}^{-B}\text{Im}\Phi_1-\text{e}^{-B}\Tilde{C},
 \end{align}
 \end{subequations}
with $\Tilde{C}$ the RR potentials such that 
 $\dd(\text{e}^{-B}\Tilde{C})=\e^{4A}\text{e}^{-B}\Tilde{F}$. As shown in \cite{Martucci:2005ht}, these satisfy the properties of a \emph{generalized} calibration: again this consists of an algebraic and of a differential condition. The algebraic condition is the minimization of the D-brane energy density
\begin{align}
    (\omega\wedge \e^\mathcal{R})|_\Pi\leq\mathcal{E}(\Pi,\mathcal{R})\dd^{p-q+1}\xi
    \ \label{bound}
\end{align}
at every point $p\subset M_6$ and for any generalized submanifold $(\Pi, \mathcal{R})$, with
 \be\mathcal{E}(\Pi,\mathcal{R})=\e^{qA-\phi}\sqrt{\text{det}(g|_\Pi+\mathcal{R})}-\delta_{q-1,3}\e^{4A}\Tilde{C}|_\Pi,
 \ee
$q$ the number of external dimensions, and $\xi$ the world-volume coordinates. 
Additionally, at any point $p\subset M_6$ there must exist at least one generalized submanifold $(\Pi, \mathcal{R})$ such that the above bound is saturated. The differential calibration condition is closure once again. 

The polyforms \eqref{caldef-string}, \eqref{caldef-sf}, and the real and imaginary parts of \eqref{caldef-DW} can be shown to satisfy \eqref{bound} using their definition \eqref{eq:Phi-eta} in terms of spinorial parameters.
Moreover, their closure
\begin{subequations}\label{closecalib}
\begin{align}
  \dd\omega^\text{DW}= 0&\qquad\text{domain-wall BPSness}\label{closecalib1}\\ 
  \dd\omega^\text{string}= 0&\qquad\text{D-string BPSness}\\
   \dd\omega^\text{sf}=0&\qquad\text{gauge BPSness},\label{closecalib3}
\end{align}
\end{subequations}
corresponds to the supersymmetry conditions \eqref{susyads0} with $\Lambda=0$, $L\to\infty$.

As in the standard calibration case, a generalized cycle calibrated by $\omega$ is saturating the calibration bound \eqref{bound}. A D-brane in a $\mathcal{N}=1$ backgrounds is supersymmetric, or BPS, if it wraps a calibrated generalized cycle.

The above generalized calibration forms are associated to space-filling, domain-wall, and string-like D-branes, respectively extended in four, three, and two spacetime dimensions.
The $\mathcal{N}=1$ supersymmetry conditions can thus be thought of as the BPSness for domain-wall, D-string and gauge (space-filling) probe branes. 
\medskip

This clear-cut interpretation of the supersymmetry conditions as calibration conditions for internal calibration forms is lost in the case of $\mathcal{N}=1$ AdS backgrounds. A non-vanishing cosmological constant indeed switches on terms in the supersymmetry conditions \eqref{susyads0} that violate the closure conditions \eqref{closecalib1} and \eqref{closecalib3} of the  would-be domain-wall and space-filling calibration forms. Instead, one can define ten-dimensional calibration forms for networks of branes,\footnote{The network is a brane wrapping an internal cycle and extended in $q$ external dimensions, together with another brane wrapping the same internal cycle and extended in $q-1$ external dimensions. The latter brane can be thought of as the boundary of the former in AdS.} whose calibration conditions are equivalent to the $\mathcal{N}=1$ supersymmetry conditions, see \cite{Koerber:2007jb,Martucci:2011dn}.

However, given the supersymmetry conditions \eqref{susyads0}, we can now define the following domain-wall calibration form\footnote{There can be no string BPS branes in AdS, as the would-be calibration form is exact, which signals the vanishing of the four-dimensional effective tension.}
\begin{align}
 \omega^\text{DW}&= \text{e}^{3A-\phi}\text{e}^{-B}\text{Im}\Phi_2,
 \end{align}
since supersymmetry still ensures $\dd\omega^\text{DW}=0$. In what follows, we will make use of the corresponding calibration theory for AdS domain-wall branes in the case where supersymmetry is broken, but the closure of this calibration form is still preserved. Saturating the bound \eqref{bound} won't be associated with the brane preserving $\mathcal{N}=1$ supersymmetry any more, but with its stability.

\section{Stability criteria for (anti) \texorpdfstring{D$p$-brane bubbles}{Dp-brane bubbles}}
\label{sec:stab}

We now focus on backgrounds with a four-dimensional external space $X_4=\text{AdS}_4$.

\subsection{Non-perturbative stability}
\label{sub:stab}

\begin{figure}[!ht]
    \centering
    \includegraphics[width=0.4\linewidth]{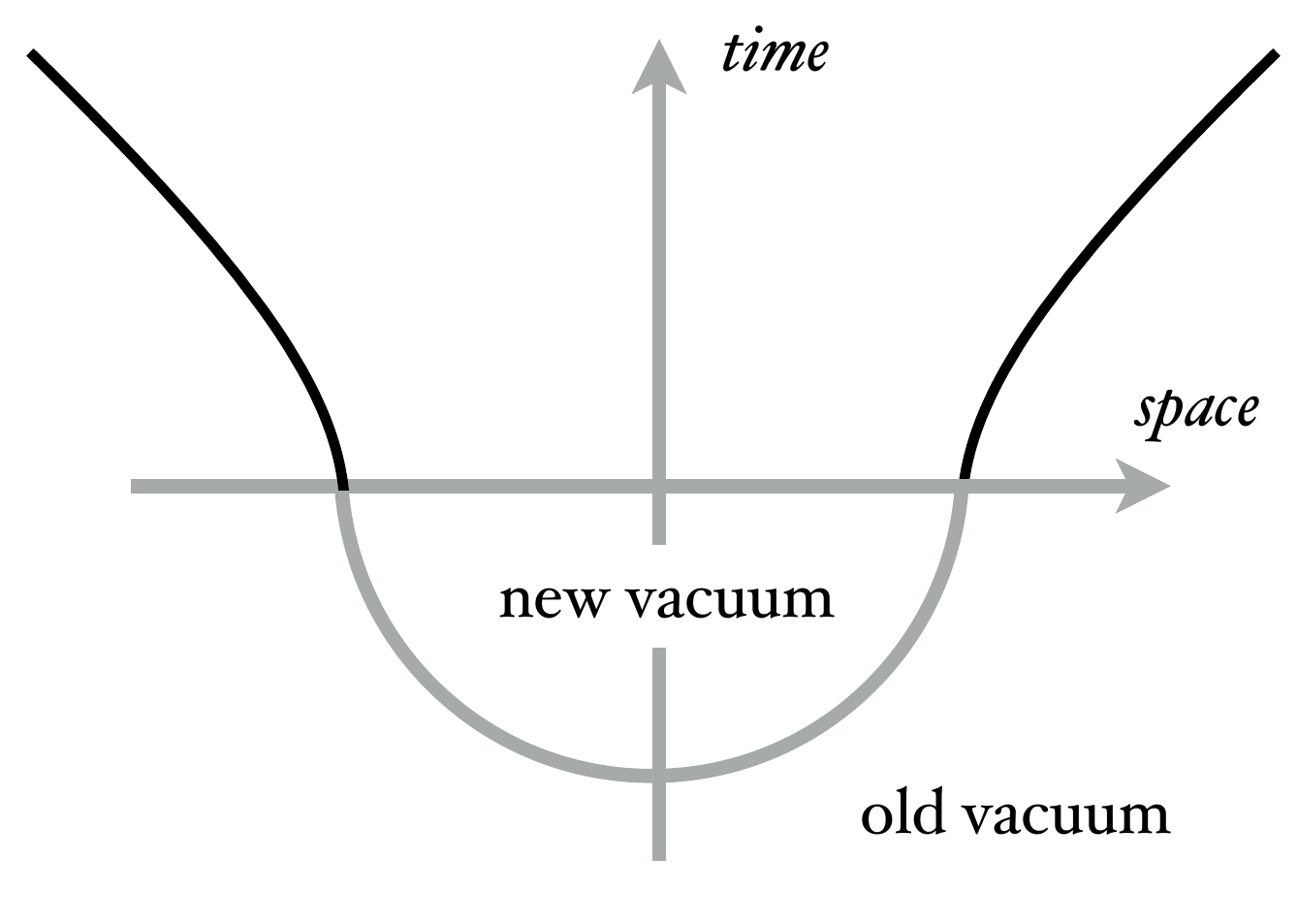}
    \caption{\small A bubble of new vacuum. The bottom gray part represents the instanton in Euclidean time; the upper black part, its subsequent Lorentzian time evolution.}
    \label{fig:bubble}
\end{figure}

We begin with a very quick review of vacuum tunneling. This is a quantum effect, and it can be investigated using the path integral approach. After a Wick rotation, the question becomes whether there can exist an instantonic bubble in Euclidean AdS (localized in Euclidean time $\tau$ as well as space) that encloses the new vacuum inside the old. In gravity this was pioneered in \cite{Coleman:1980aw} for theories with scalars, and extended in \cite{Brown:1988kg} for form fields. A simplified setting is when the bubble is a probe brane, whose back-reaction can be neglected. This case was considered in \cite[Sec.~4]{Maldacena:1998uz} by solving the equations of motion for the bubble's shape as a function of $\tau$. In \cite[Sec.~4.1]{Apruzzi:2019ecr} it was shown that this bubble is in fact a half-sphere (bottom part of Fig.~\ref{fig:bubble}): in line with earlier expectations from \cite{Coleman:1980aw}, it respects an ${\rm SO}(d)$ subgroup of the ${\rm SO}(d,1)$ isometry group of Euclidean AdS. One can make this symmetry manifest by writing the metric as
\begin{equation}\label{eq:eads}
    \dd s^2_{\rm EAdS}= L^2 (\dd r^2 +\sinh^2 r \dd s^2_{S^{d-1}})\,.
\end{equation}
$r$ is an overall radial coordinate, which mixes $\tau$ and the more customary spatial radial coordinates. We can now just \emph{assume} ${\rm SO}(d)$ invariance for the instanton: we take it to be a half-sphere at a fixed $r$. The boundary of this half-sphere (the slice at time $=0$ in Fig.~\ref{fig:bubble}) represents the physical bubble being nucleated by a quantum effect.

Consider a brane action consisting of the usual volume term and a coupling to a top-dimensional gauge field: $S=-\tau \int \dd^{d-1} \sigma \sqrt{-g}- q \int A_{d-1}$. The field strength will be proportional to the volume form: $F_d = h {\rm vol}_d$. Specializing the Wick-rotated action to \eqref{eq:eads} one finds \cite{Apruzzi:2019ecr}
\begin{equation}\label{eq:s-inst}
    S= -L^{d-1} {\rm Vol}(S^{d-1}) (\tau \sinh^{d-1} r  + q h 
    \, c(r)) 
\end{equation}
where $c'(r)= \sinh^{d-1}r$. Thanks to the assumption of ${\rm SO}(d)$ symmetry, finding the instanton has been reduced to extremizing \eqref{eq:s-inst} in $r$; this gives $\tanh r_0 =- \frac{(d-1) \tau}{q h}$ for the radius $r_0$ of the instanton. In particular, the instability does \emph{not} exist if
\begin{equation}\label{eq:dq-stab}
    (d-1)\tau - |q| h\ge 0 \,,\quad \text{or} \quad \frac{|q| h}{(d-1) \tau}\le1 \,.
\end{equation}
We will use both formulations in what follows. Intuitively, the $\tau$ term in the action tends to shrink the instanton, and the $q$ term tends to expand it; the vacuum is unstable when the latter term wins, forcing the bubble into existence.

We will now specialize the above to domain-wall (anti) D$p$-brane wrapping an internal cycle $\Sigma$. The (anti) D$p$-brane carries a closed world-volume gauge field $f_{\rm ws}$, and we introduce the gauge invariant world-volume two-form field $\mathcal{F}=B|_\Sigma+2\pi f_{\rm ws}$, such that $\dd\mathcal{F}=H|_\Sigma$.\footnote{In what follows, if the world-volume flux $f_{\rm ws}$ is non-vanishing, we will sometimes refer to such a D-brane as a bound state: this is because it makes a D$p$ couple to RR potentials $C_k$ with $k<p+1$.  For $f_{\rm ws}=0$, we will sometimes refer to it as a simple D-brane.} Now the parameters above are evaluated as $\tau =T_p\int_\Sigma \dd^{p-d+2}\xi \,\e^{(d-1)A-\phi} \sqrt{|g|_\Sigma+{\cal F}|}$, $qh=T_p L \int_\Sigma \e^{d A} * \lambda F \wedge \e^{-{\cal F}}$, with $T_p$ the tension of the D$p$-brane and $\xi^m$ the internal world-volume coordinates. \eqref{eq:dq-stab} becomes
\begin{equation}\label{boundads}
\Upsilon(\Sigma,\mathcal{F})\equiv T_p \int_\Sigma (d-1)\e^{(d-1)A-\phi}\sqrt{\text{det}(g|_\Sigma+\mathcal{F})}\dd^{p-d+2}\xi\mp L\e^{d A}\ast\lambda F|_\Sigma\wedge \e^{-\mathcal{F}} \ge 0\,,
\end{equation}
with the upper/lower sign for D-branes/anti D-branes respectively. 
The pair $(\Sigma,\mathcal{F})$ is often called a generalized cycle \cite{Martucci:2005ht}, but from now on we mostly omit the world-volume gauge field $\mathcal{F}$ when referring to generalized cycles.

A notable point about \eqref{boundads} is that the coefficients $\tau$ and $q$ in \eqref{eq:dq-stab} have now become dependent on the choice of $\Sigma$. Even when $\Sigma$ is a point, $p=d-2$, $\Upsilon$ is often a function on the internal manifold, because of $A$ and $\phi$; this was important in some applications \cite{Apruzzi:2019ecr,Bena:2020xxb}. For our vacua below, $A$ and $\phi$ will be constant, but for higher-dimensional branes, $p>d-2$, $\Upsilon$ will depend in a complicated way on the choice of the internal cycle $\Sigma$.

So far we have focused on the quantum process that nucleates a bubble of new vacuum. The subsequent evolution can be obtained by an analytic continuation of the above ${\rm SO}(d)$-invariant spherical Euclidean solution, to an ${\rm SO}(d-1,1)$-invariant hyperboloid, representing the evolution of a sphere in AdS \cite{Coleman:1980aw,Brown:1988kg} (upper part of Fig.~\ref{fig:bubble}). As an alternative, one can consider the expansion of branes that are localized along the Poincar\'e radial coordinate \cite{Gaiotto:2009mv,Bena:2020xxb}. This leads again to \eqref{eq:dq-stab} \cite[Sec.~5.1]{Apruzzi:2019ecr}.

Let us stress that the criteria in this subsection regards stability of the background against decays mediated via these bubbles; these are not statements about the stability of the branes themselves.

\subsection{Supersymmetric solutions}

We now specialize to $\mathcal{N}=1$ vacua and $d=4$. 

The supersymmetry condition \eqref{susy3ads0} allows to rewrite the second term in \eqref{boundads}; discarding the total derivative term, we obtain \cite{Giri:2021eob}
\begin{align} 
\Upsilon(\Sigma,\mathcal{F})&=3T_p\int_\Sigma \e^{3A-\phi}(\sqrt{\text{det}(g|_\Sigma+\mathcal{F})}\dd^{p-2}\xi\mp\text{Im}\Phi_2|_\Sigma\wedge \e^{-\mathcal{F}})\label{Eadssusy1}.
\end{align}
On the other hand, for every generalized cycle, we have
\be \left|\text{Im}\Phi_2|_\Sigma\wedge \e^{-\mathcal{F}}\right|_\text{top}\leq\sqrt{\text{det}(g|_\Sigma+\mathcal{F})}\dd^{p-2}\xi.\label{ineqdw} \ee
Taken together, these ensure that \eqref{boundads} are satisfied for both D-branes and anti D-branes, provided that \eqref{susy3ads0} is satisfied. We recover here the fact that $\mathcal{N}=1$ supersymmetric solutions are protected from (anti) D$p$-brane bubbles as a non-perturbative decay channel.

    Let us define the corresponding calibrated and anti-calibrated cycles, respectively respecting
\begin{align}
    \sqrt{\text{det}(g|_{\Sigma_\text{c}}+\mathcal{F}_\text{c})}\dd^{p-2}\xi&=\left.\text{Im}\Phi_2|_{\Sigma_\text{c}}\wedge \e^{-\mathcal{F}_\text{c}}\right|_{\text{top}}\label{kappa1dw}\\
    \sqrt{\text{det}(g|_{\Sigma_\text{ac}}+\mathcal{F}_\text{ac})}\dd^{p-2}\xi&=-\left.\text{Im}\Phi_2|_{\Sigma_\text{ac}}\wedge \e^{-\mathcal{F}_\text{ac}}\right|_{\text{top}}\label{kappa2dw},
\end{align}
 and now possibly wrapped by domain-wall (anti) D$p$-branes.
Using \eqref{Eadssusy1} and these definitions, we therefore have
\begin{align}
    \Upsilon(\Sigma_\text{c},\mathcal{F}_\text{c})= 0 \quad\text{for D-branes,}\qquad\Upsilon(\Sigma_\text{ac},\mathcal{F}_\text{ac})=0\quad\text{for anti D-branes}.
\end{align}
Intuitively this signals that when a (anti) D$p$-brane bubble wraps a (anti-)calibrated generalized cycle, its DBI and WZ contributions even out, and said (anti) D$p$-brane bubble neither expand nor contracts.

\subsection{Non-supersymmetric solutions}
\label{sub:stab-nonsusy}

Let us now turn to the non-supersymmetric case. We keep on using the stability condition \eqref{boundads}, as it is a statement independent from supersymmetry. However, supersymmetry-breaking will prevent the reformulation \eqref{Eadssusy1} of $\Upsilon$.

We consider solutions respecting the following modified pure spinor equations
\begin{subequations}\label{susyads0mod}
\begin{align}
  \dd(\text{e}^{3A-\phi}\text{e}^{-B}\Phi_2)&= \frac2L \e^{2A-\phi}\text{e}^{-B}\text{Re}\Phi_1+\frac2L\text{e}^{-B}\Psi\label{susy1ads0mod}\\ 
  \dd(\e^{2A-\phi}\text{e}^{-B}\text{Re}\Phi_1)&=\dd(\text{e}^{-B}\text{Re}\Psi)\label{susy2adsmod}\\
   \dd(\e^{4A-\phi}\text{e}^{-B}\text{Im}\Phi_1)&=\frac3L\e^{3A-\phi}\text{e}^{-B}\text{Im}\Phi_2- \e^{4A}\text{e}^{-B}\ast\lambda F+\frac3L\text{e}^{-B}\Theta,\label{susy3adsmod}
\end{align}
\end{subequations}
with $\Psi$ and $\Theta$ supersymmetry breaking forms.

Following the procedure of \cite{Giri:2021eob} and using \eqref{susy3adsmod}, the quantities $\Upsilon$ in \eqref{boundads} for a domain-wall D$p$-brane and anti D$p$-brane wrapping a generalized cycle $\Sigma$ can now be reformulated as
\begin{align}
     \Upsilon(\Sigma,\mathcal{F})&=3T_p\int_\Sigma \e^{3A-\phi}(\sqrt{\text{det}(g|_\Sigma+\mathcal{F})}\dd^{p-2}\xi\mp\text{Im}\Phi_2|_\Sigma\wedge \e^{-\mathcal{F}})\pm\Theta|_\Sigma\wedge \e^{-\mathcal{F}}.
\end{align}

Combining these with the bounds \eqref{boundads} and \eqref{ineqdw}, we can write down some inequalities ensuring the stability of the background under both the brane and anti brane decay channels. There are two distinct cases:
\be 
\text{if $\Sigma$ is such that} \int_{\Sigma}\e^{3A-\phi}\text{Im}\Phi_2|_\Sigma\wedge \e^{-\mathcal{F}}>0\nonumber \ee
\be-2\int_{\Sigma}\e^{3A-\phi}\text{Im}\Phi_2|_\Sigma\wedge \e^{-\mathcal{F}}\leq \int_{\Sigma}\Theta|_\Sigma\wedge \e^{-\mathcal{F}}\leq 0\Rightarrow\text{stability against bubbles wrapping}\,  \Sigma\label{boundads1}\ee

and
\be 
\text{if $\Sigma$ is such that}\int_{\Sigma}\e^{3A-\phi}\text{Im}\Phi_2|_\Sigma\wedge \e^{-\mathcal{F}}<0\nonumber \ee
\be0\leq \int_{\Sigma}\Theta|_\Sigma\wedge \e^{-\mathcal{F}}\leq -2\int_{\Sigma}\e^{3A-\phi}\text{Im}\Phi_2|_\Sigma\wedge \e^{-\mathcal{F}}\Rightarrow\text{stability against bubbles wrapping}\,  \Sigma.\label{boundads2}
\ee
For a generic generalized cycle $\Sigma$, these conditions are stronger than stability: in principle a given background could violate them and still be stable under these decay channels.

Using equation \eqref{susy3adsmod}, these two bounds can be reformulated and unified as
\be 
\left|r_\Sigma \equiv
\frac
{L\int_{\Sigma}\e^{4A}\ast\lambda F\wedge \e^{-\mathcal{F}}}
{3\int_{\Sigma}\e^{3A-\phi}\text{Im}\Phi_2|_\Sigma\wedge \e^{-\mathcal{F}}}
\right|\leq 1
\quad\Rightarrow\quad
\text{stability against bubbles wrapping}\,  \Sigma.\label{eq:stabgen}
\ee
We will call $r$ the {\it stability ratio} of the bubble.

If $\text{e}^{-B}\Theta$ is closed, we have
\be \dd(\e^{3A-\phi}\text{e}^{-B}\text{Im}\Phi_2)=0,\label{calibadsN0}\ee
and $\e^{3A-\phi}\text{e}^{-B}\text{Im}\Phi_2$ is a proper calibration form.\footnote{If there exists at least one generalized submanifold such that the inequality \eqref{ineqdw} is saturated.} This eases significantly the stability analysis of $\mathcal{N}=0$ backgrounds satisfying \eqref{calibadsN0}, since the stability ratio in \eqref{eq:stabgen}
now depends only on the homology class of $\Sigma$. By the RR equations of motion, the integrand in the numerator of \eqref{eq:stabgen} is closed, so it also only depends on homology classes.\footnote{Suppose we know  for a basis $\Sigma_a$ in homology that $\frac{q(\Sigma_a) h}{3\tau(\Sigma_a)}\le \frac{L\int_{\Sigma_a}f}{3\int_{\Sigma_a}\psi}\le 1$, where $f$, $\psi$ are the integrands in \eqref{eq:stabgen}. Then for a general element $\Sigma = s_a \Sigma_a$, $s_a>0$, we have $\frac{q(\Sigma) h}{3 \tau(\Sigma)}\le \frac{Ls_a\int_{\Sigma_a}f}{3s_a\int_{\Sigma_a}\psi} \le 1$.} In particular, it vanishes for cycles that are homologically trivial.

Let us now specialise these inequalities to the cases of generalized cycles saturating the bound \eqref{ineqdw} in two distinct ways:
\begin{subequations}\label{kappanonsusyads}
\begin{align}
    \sqrt{\text{det}(g|_{\Sigma_{\Tilde{\text{c}}}}+\mathcal{F}_{\Tilde{\text{c}}})}\dd^{p-2}\xi&=\left.\text{Im}\Phi_2|_{\Sigma_{\Tilde{\text{c}}}}\wedge \e^{-\mathcal{F}_{\Tilde{\text{c}}}}\right|_{\text{top}}\label{kappa1nonsusyads}\\
    \sqrt{\text{det}(g|_{\Sigma_{\Tilde{\text{ac}}}}+\mathcal{F}_{\Tilde{\text{ac}}})}\dd^{p-2}\xi&=-\left.\text{Im}\Phi_2|_{\Sigma_{\Tilde{\text{ac}}}}\wedge \e^{-\mathcal{F}_{\Tilde{\text{ac}}}}\right|_{\text{top}}.\label{kappa2nonsusyads}
\end{align}
\end{subequations}
We call the generalized cycles satisfying \eqref{kappa1nonsusyads} {\it almost calibrated}, and the ones satisfying \eqref{kappa2nonsusyads} {\it almost anti-calibrated} ---introducing the corresponding subscripts $_{\Tilde{\text{c}}}$ and $_{\Tilde{\text{ac}}}$.
These calibration conditions can be formulated as \cite{Martucci:2005ht}
\begin{subequations}\label{eq:alt-cal}
\begin{align}
\label{eq:alt-cal-P2}
\left.\text{Re}\Phi_2|_{\Sigma}\wedge \e^{-\mathcal{F}}\right|_{\text{top}}&=0\\
\label{eq:alt-cal-P1}
\left.(\dd y^m\wedge+g^{mn}\iota_n)\Phi_1|_{\Sigma}\wedge \e^{-\mathcal{F}}\right|_{\text{top}}&=0,
\end{align}
\end{subequations}
 while the orientation of $\Sigma$ determines whether the cycle is almost calibrated or almost anti-calibrated.

A particular case that will be useful later is that of an ${\rm SU}(3)$-structure. Recall that this is defined as a pair $(J,\Omega)$ of a real two-form and complex three-form such that
\begin{equation}
	J \wedge \Omega =0 \, ,\qquad \mathrm{vol}_6 = - \frac16 J^3 = -\frac{\ii}8 \Omega \wedge \bar \Omega\,. 
\end{equation}
The corresponding pure spinors are $\Phi_1=\Omega$, $\Phi_2=\e^{-\ii J}$. \eqref{eq:alt-cal} imply then that the cycle should be holomorphic, and that
\begin{subequations}\label{eq:su3-cal}
\begin{align}
\label{eq:mmms}
    \mathrm{Re}(\e^{- {\mathcal F}} \wedge \e^{\ii \theta}\e^{-\ii J})&=0 \,,\\
\label{eq:F20=0}   
    {\mathcal F}_{2,0}&=0 \,.
\end{align}
\end{subequations}
This system was first identified in \cite{Marino:1999af}. For the K\"ahler case, it was studied mathematically in e.g.~\cite{collins-jacob-yau,collins-shi,Collins:2023gtu}. For small ${\mathcal F}$, one can ignore higher powers in ${\mathcal F}$ term, and one recovers the Hermitian-Yang--Mills (or Donaldson--Uhlenbeck--Yau) equation \cite{donaldson-hym,uhlenbeck-yau}. \eqref{eq:mmms} is now reduced to a stability condition that holds for holomorphic Abelian bundles, and that is more non-trivial in the non-Abelian case.

Coming back to the general case, the above inequalities for generalized almost (anti-)calibrated cycles are now equivalent to the stability of the background against the nucleation and expansion of brane and anti brane bubbles:
\begin{samepage}
\be
-2\int_{\Sigma_{\Tilde{\text{c}}}}\e^{3A-\phi}\text{Im}\Phi_2|_{\Sigma_{\Tilde{\text{c}}}}\wedge \e^{-\mathcal{F}_{\Tilde{\text{c}}}}\leq \int_{\Sigma_{\Tilde{\text{c}}}}\Theta|_{\Sigma_{\Tilde{\text{c}}}}\wedge \e^{-\mathcal{F}_{\Tilde{\text{c}}}}\leq 0\nonumber\ee
\be\Leftrightarrow\label{stabn0}\ee
    \be\text{stability against bubbles wrapping}\,  \Sigma_{\Tilde{\text{c}}}\nonumber\ee
    \end{samepage}
    and
\be
    0\leq \int_{\Sigma_{\Tilde{\text{ac}}}}\Theta|_{\Sigma_{\Tilde{\text{ac}}}}\wedge \e^{-\mathcal{F}_{\Tilde{\text{ac}}}}\leq -2\int_{\Sigma_{\Tilde{\text{ac}}}}\e^{3A-\phi}\text{Im}\Phi_2|_{\Sigma_{\Tilde{\text{ac}}}}\wedge \e^{-\mathcal{F}_{\Tilde{\text{ac}}}}\nonumber\ee
    \be\Leftrightarrow\ee
    \be\text{stability against bubbles wrapping}\,  \Sigma_{\Tilde{\text{ac}}}.\nonumber\ee
These can be once again reformulated as
\be\label{eq:stab}
  \left|r_{\Sigma_{\Tilde{\text{c}}/\Tilde{\text{ac}}}}\right|\leq 1\qquad
   \Leftrightarrow\qquad
    \text{stability against bubbles wrapping}\quad\Sigma_{\Tilde{\text{c}}/\Tilde{\text{ac}}}.\ee
    
In particular, violating these bounds now means that it is energetically favourable for a brane or an anti brane bubble to nucleate and expand, triggering a decay of the background.

Almost (anti-)calibrated generalized cycles therefore appear as particularly simple candidates for the destabilisation of non-supersymmetric background through these decay channels.

\medskip

If $\dd(\e^{3A-\phi}\text{e}^{-B}\text{Im}\Phi_2)=0$ is satisfied on top of the almost (anti-)calibration condition \eqref{kappanonsusyads}, the generalized cycle is now (anti-)calibrated and the stability ratio depends again only on its homology class.

\medskip

\begin{figure}[!ht]
    \centering
\includegraphics[width=.8\linewidth]{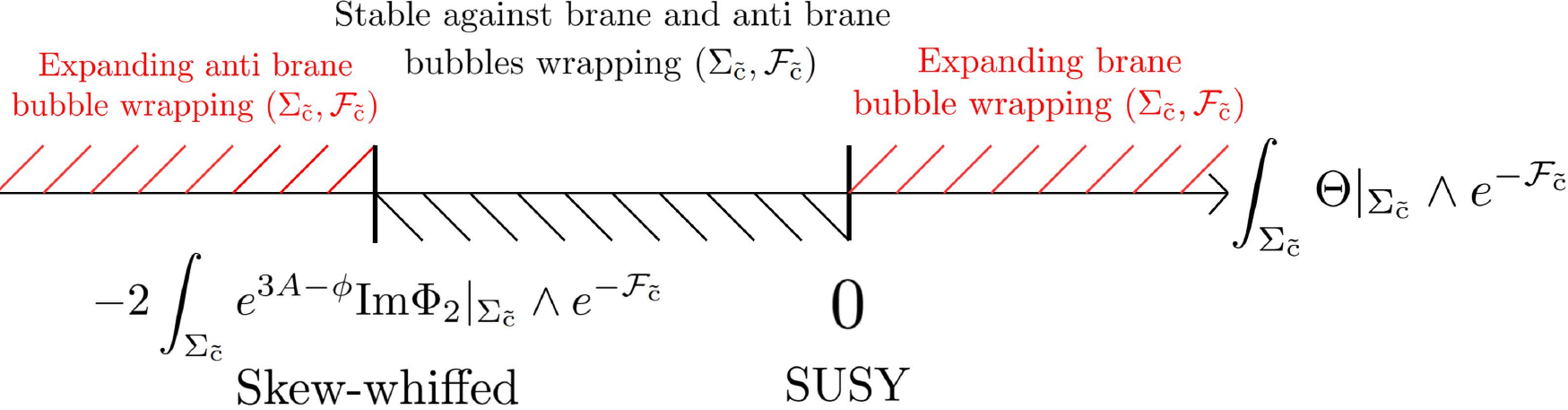}
    \caption{\small Behaviour of anti brane and brane bubbles wrapping an almost calibrated generalized cycle $(\Sigma_{\Tilde{\text{c}}},\mathcal{F}_{\Tilde{\text{c}}})$ in terms of the value of the integrated supersymmetry breaking term.}
   \label{fig:branebubble}
\end{figure}

Interestingly, the saturation of the upper bound corresponds to the case of supersymmetric backgrounds, or backgrounds at least preserving the supersymmetry equation \eqref{susy3ads0}, and the saturation of the lower bound of \eqref{stabn0} corresponds to the non-supersymmetric solutions which are usually called \emph{skew-whiffed}, where all the RR fluxes take a minus sign with respect to the supersymmetric case. This is illustrated in Fig.~\ref{fig:branebubble}, for branes wrapping an almost calibrated generalized cycle.
From the stability criteria written in terms of the stability ratio \eqref{eq:stabgen}, we see that the numerator can be interpreted as a WZ like term, contributing to the expansion of the bubble, while one can think of the denominator as an estimate for the DBI term, contributing to its contraction. The criterion then states that the expanding force shouldn't win against the contracting one.

This highlights one strength of this formalism: the bounds directly depend on a pure spinor of the background. In the supersymmetric case, the pure spinors are tightly constrained by the pure spinor equations \eqref{susyads0}. However, departing from supersymmetry, one has quite a lot of freedom in picking $\Phi_2$, the only requirements being its compatibility with the other pure spinor and the compatibility of the pair with the generalized metric \cite{Grana:2005sn}. This freedom can be used to choose convenient DBI estimates, or even combine them to span the parameter space of some solutions in the most efficient way. We will make use of this consideration throughout our study of the non-perturbative stability of various non-supersymmetric AdS flux vacua in the next section.

\section{\texorpdfstring{AdS$_4$ solutions}{AdS4 solutions}}

We start by considering several AdS$_4$ vacua in IIA supergravity. We will consider some known solutions, whose non-perturbative stability we will analyze using our method, and some new ones. The known solutions are AdS$_4\times$ a twistor space, which includes the $\mathbb{CP}^3$ case considered in \cite{Koerber:2010rn} (Sec.~\ref{sub:tw}), and AdS$_4\times {\rm KE}_6$ \cite{Gaiotto:2009mv,Lust:2009zb,Romans:1985tz} in Sec.~\ref{sub:ke6}. The new ones are AdS$_4\times $ the flag manifold $\mathbb{F}(1,2;3)$ (Sec.~\ref{sub:flag})

The setup was discussed in Sec.~\ref{sec:pure}; in particular, the metric is as in \eqref{10dmet}, and the three-form $H$ is purely internal. The IIA equations of motion for the NSNS fields are\footnote{The inner product of two $k$-forms $\alpha^a_k\equiv \frac1{k!} \alpha^a_{m_1\ldots m_k} \dd x^{m_1}\wedge\ldots \wedge \dd x^{m_k} $ is defined as $\alpha^1\cdot \alpha^2 \equiv \frac1{k!} \alpha^1_{m_1\ldots m_k} \alpha^{2\, m_1\ldots m_k}$. The pointwise norm of a $k$-form is then $|\alpha|^2 \equiv \alpha \cdot \alpha$. Finally, the total pointwise norm of the polyform $F=\sum_k F_k$ is just $|F|^2=\sum_k |F_k|^2$.}  
\begin{subequations}\label{eq:4d-eom}
\begin{align}
	\label{eq:4d-eom-intgr}&R_{mn}-4 (\nabla_m \partial_n A + \partial_m A \partial_n A)+ 2 \nabla_m \nabla_n \phi  - \frac12 \iota_m H \cdot \iota_n H \\
	&\nonumber\hspace{4cm} =\frac14 \e^{2 \phi}(2\iota_m F \cdot \iota_n F - g_{mn} |F|^2)\,;\\
    	\label{eq:4d-eom-extgr}&\e^{-2 A} \Lambda - \nabla^2 A - 4 |\dd A|^2 + 2 \dd A \cdot \dd \phi = -\frac14 \e^{2 \phi} |F|^2 \,;\\
	\label{eq:4d-eom-dil}&2 \nabla^2 \phi - 4|\dd \phi|^2 + 8 \dd A \cdot \dd \phi  + |H|^2 = \frac12 \e^{2 \phi} \sum_k (5-k) |F_k|^2 \,;\\
	\label{eq:4d-eom-H}
	& \dd\left(\e^{4A - 2\phi} \ast H \right) = \e^{4A} \left( -F_0 \ast F_2 - F_2 \wedge \ast F_4 - F_4 \wedge \ast F_6 \right)\,.
\end{align}
The Bianchi identities and equations of motion for the RR fields are:
\begin{equation}
(\dd - H \wedge) F = 0 = (\dd-H \wedge) (\e^{4A} * \lambda F)\,.
\end{equation}
\end{subequations}
(The operators $\dd$ and $\ast$ are on $M_6$.)

\subsection{\texorpdfstring{AdS$_4\times$ twistor spaces}{AdS4 x twistor spaces} }
\label{sub:tw}

We will now consider twistor spaces $M_6=\mathrm{Tw}(\mathrm{QK}_4)$; that is, the total space of the twistor bundle over a quaternionic K\"ahler four-manifold $\mathrm{QK}_4$. In practice, we will focus almost exclusively on the case $M_6 = \mathbb{CP}^3$.

The reason is that there are only two \emph{smooth} $\mathrm{QK}_4$ manifolds in four dimensions: $\mathbb{CP}^3=\mathrm{Tw}(S^4)$, and the flag manifold $\mathbb{F}(1,2;3)=\mathrm{Tw}(\mathbb{CP}^2)$ \cite{hitchin-qK}. Both are also homogeneous; for the latter, this property gives rise to additional solutions, so it deserves its own separate treatment in the next subsection. There are additional quaternionic K\"ahler four-manifolds with orbifold singularities \cite[12.5]{boyer-galicki-book}. 
 
For $\mathbb{CP}^3$, the solutions in this subsection were discussed in \cite{Koerber:2010rn}. Our new contribution is the study of its non-perturbative stability, as an application of the method outlined in Sec.~\ref{sec:stab}. There are also some supersymmetric vacua in this set, which were first found in \cite{Tomasiello:2007eq} and later reformulated in \cite{Koerber:2008rx} for $\mathrm{Tw}(S^4)=\mathbb{CP}^3$, again by viewing it as a homogeneous space. 

\subsubsection{Twistor geometry}

The twistor bundle $\mathrm{Tw}(\mathrm{QK}_4)$ is the sphere bundle inside the bundle $\Lambda^2_- T^* (\mathrm{QK}_4)$ of anti-self-dual two-forms. Since the latter has rank three, the fiber of $\mathrm{Tw}(\mathrm{QK}_4)$ is $S^2$. We define coordinates $y_a$, $a=1,\,2,\,3$ such that $y_a y_a=1$, and their covariant derivatives $D y_a= \dd y_a + \epsilon_{abc}A_b y_c$, with $A$ a connection for the bundle. The latter can be taken to satisfy $\dd A_a + \epsilon_{abc}A_b \wedge \omega_c=0$, where $\omega_a$ make up a basis of anti-self-dual two-forms. Since $\mathrm{QK}_4$ is quaternionic K\"ahler, the curvature $F_a = \dd A_a + \epsilon_{abc}A_b \wedge A_c$ satisfies $F_a= 4\omega_a$, in a normalization where the Ricci scalar is $R_4=6$. One can now define the forms
\begin{subequations}\label{eq:j-psi-twistor}
\begin{align}
	&j_\mathrm{F}\equiv \frac12 \epsilon_{abc} y_a D y_b \wedge D y_c \, ,\qquad j_\mathrm{B}\equiv y^a j_a\,;\\
	&\psi = D y_a \wedge j_a  \, ,\qquad \tilde \psi = \epsilon_{abc} y_b D y_c \wedge j_a\,.
\end{align}
\end{subequations}
$j_\mathrm{F}$ and $j_\mathrm{B}$ can be thought of as being along the fiber and base, respectively. These forms satisfy the algebraic conditions 
\begin{equation}
	j_\mathrm{F}^2=0= j_\mathrm{B}^3 \, ,\qquad
	j_\mathrm{F} \wedge \psi = j_\mathrm{B} \wedge \psi =
	j_\mathrm{F} \wedge \tilde\psi = j_\mathrm{B} \wedge \tilde\psi = 0\,.   
\end{equation}
One can also show that
\begin{equation}\label{eq:d-j-psi-twistor}
	\dd j_\mathrm{F}= 4 \psi \, ,\qquad \dd j_\mathrm{B}= \psi \, ,\qquad \dd \tilde \psi = 2 j_\mathrm{F} \wedge j_\mathrm{B} +  8 j_\mathrm{B}^2\,.
\end{equation}
For more information, see \cite[Sec.~7.5.2]{book}. 

The metric we will consider is
\begin{equation}\label{eq:tw-met}
	\dd s^2_6 = R^2 \left(\frac 2 \sigma \dd s^2_{{\rm QK}_4}+ \frac14 D y^a D y^a \right)\,,
\end{equation}
The peculiar parameterization for the metric coefficients is chosen so as to match with that in \cite{Tomasiello:2007eq}. 

For later use we also introduce the $\mathrm{SU}(3)$-structure
\begin{equation}\label{eq:SU3-tw}
\begin{split}
	J &= R^2 \left( \frac{2}{\sigma} j_\mathrm{B} + \frac{1}{4} j_\mathrm{F} \right) \equiv J_B + J_F\\
	\Omega &= \frac{R^3}{\sigma} \left( -\psi + i \tilde{\psi} \right)\,.
\end{split}	
\end{equation}

\subsubsection{The solutions}
\label{ssub:ads-tw-sol}

The Ansatz now consists in taking the fluxes to be proportional to the forms in (\ref{eq:j-psi-twistor}), (\ref{eq:SU3-tw}) and their products:
\begin{equation}\label{eq:cp3-flux}
	F_2 = f_{2\mathrm{B}} J_\mathrm{B} + f_{2 \mathrm{F}} J_\mathrm{F} \, ,\quad
	F_4 = f_{4 \mathrm{B}} \frac{J_B^2}{2} + f_{4 \mathrm{F}} J_\mathrm{B} \wedge J_\mathrm{F} \, ,\quad
	F_6 = f_6 \frac{J^3}{6} = -f_6 \, \text{vol}_6 \, ,\quad
	H = h \, \text{Re} \, \Omega\,.
\end{equation}
We can take 
\begin{equation}
	B= -\frac{h R^3}{4 \sigma} j_\mathrm{F}\,.
\end{equation}
The total flux is
\begin{equation}
	* \lambda F = f_6 - (f_{4\mathrm{F}} J_\mathrm{B} + f_{4\mathrm{B}}J_\mathrm{F})+ (f_{2\mathrm{B}} J_\mathrm{B} \wedge J_\mathrm{F} + f_{2\mathrm{F}} J_\mathrm{B}^2/2) + f_0 (-J_\mathrm{B}^2 \wedge J_\mathrm{F}/2)\,.
\end{equation}

The Ricci tensor for the metric \eqref{eq:tw-met} can be found in this language in \cite[(3.16)]{Tomasiello:2007eq}; for $\mathbb{CP}^3$, it can also be obtained by considering it as a coset space ${\rm Sp}(2)/{\rm Sp}(1)\times {\rm U}(1)$, using a formula in the next section.

We consider no warping, $A=0$, and the equations of motion (\ref{eq:4d-eom}) now become purely algebraic:
\begin{subequations}\label{eq:tw-eom}
\begin{align}
	\frac{\sigma(6-\sigma)}{g_s^2 R^2} - \frac{h^2}{g_s^2} &= \frac14 (- f_0^2 +f_{2\mathrm{F}}^2 + f_{4 \mathrm{B}}^2 + f_6^2)\,,\\
	\frac{4+\sigma^2}{g_s^2 R^2} - \frac{h^2}{g_s^2} &= \frac14 (- f_0^2 -2 f_{2\mathrm{B}}^2+f_{2\mathrm{F}}^2- f_{4\mathrm{B}}^2+ 2f_{4 \mathrm{F}}^2 + f_6^2)\,;\\
	-\frac4{g_s^2} \Lambda &= f_0^2 +2 f_{2\mathrm{B}}^2+f_{2\mathrm{F}}^2 + f_{4\mathrm{B}}^2+ 2f_{4 \mathrm{F}}^2 + f_6^2 \,;\\
	\frac{8}{g_s^2} h^2 &= 5 f_0^2 + 6 f_{2\mathrm{B}}^2+3f_{2\mathrm{F}}^2 +f_{4\mathrm{B}}^2+ f_{4 \mathrm{F}}^2 - f_6^2 \,;\\
	\frac{4h}{R g_s^2} &= f_0 f_{2\mathrm{B}} + f_{2\mathrm{B}} f_{4\mathrm{B}} + f_{2\mathrm{F}} f_{4\mathrm{F}}+ f_{4\mathrm{F}} f_6 \,;\\
	\frac{4 \sigma h}{R g_s^2} &= f_0 f_{2\mathrm{F}} + 2 f_{2\mathrm{B}} f_{4\mathrm{F}} + f_{2\mathrm{F}} f_{4\mathrm{F}}+ f_{4\mathrm{B}} f_6 \,;\\
    -f_0 h R&= 2 f_{2\mathrm{B}}+ \sigma f_{2\mathrm{F}}\,,\\
    f_6 h R  &= f_{4\mathrm{B}} + 2 f_{4\mathrm{F}}\,.
\end{align}
\end{subequations}

We studied these algebraic equations numerically; our results appear to be in excellent agreement with those in \cite[Fig.~2]{Koerber:2010rn}. 

The supersymmetric solution \cite{Tomasiello:2007eq} reads 
\begin{align}\label{susyCP3}
	&f_0 = 5 \frac m{g_s} \, ,\qquad f_{2\mathrm{B}} = \frac{3 \sigma -2}{2 g_s R} \, ,\qquad f_{2\mathrm{F}} = \frac{-5 \sigma +6}{2 g_s R} 
	\, , \qquad
	 f_{4\mathrm{B}} = f_{4\mathrm{F}}= 3 \frac m{g_s} \, ,\qquad f_6= -\frac{3(2+\sigma)}{2g_s R}\,\nonumber\\
     &h=-2m \, ,\qquad m=\frac1{2R}\sqrt{(2-\sigma)(5 \sigma-2)}\,,\qquad \Lambda = -\frac{12}{5R^2}(2 \sigma +1) \,,\quad \tan\theta=\frac{2+\sigma}{hR }\,.
\end{align}
We see that this only exists for $\sigma\in [2/5,2]$. The Romans mass $F_0$ vanishes at the endpoints of this interval. For these two cases, the solution can be uplifted to eleven dimensions. For $\mathbb{CP}^3$, we obtain in this way AdS$_4\times S^7$, with the $S^7$ round ($\sigma=2$) and squashed ($\sigma=2/5$). The $\sigma=2$ case is thus the famous ${\mathcal N}=6$ solution \cite{Nilsson:1984bj,Watamura:1983hj,Sorokin:1985ap,Aharony:2008ug}. As found in \cite{Koerber:2010rn}, non-supersymmetric solutions exist for a slightly larger interval in $\sigma$, as we will see in more detail later. 

Flux quantization demands
\begin{equation}\label{eq:fq}
    (2\pi)^{1-k}\int_{B_k} (\e^{-B} F)_k \in \mathbb{Z}\,.
\end{equation}
As in \cite[Sec.~3.2]{Aharony:2010af}, we take $B= B_0 + b(j_\mathrm{F}-4 j_\mathrm{B})$; recalling $\dd (j_\mathrm{F}-4 j_\mathrm{B})=0$, $b$ is a free parameter. The other free parameters are $\sigma$, $g_s$, $L\equiv \sqrt{-3/\Lambda}$. These four are fixed by the four conditions in \eqref{eq:fq}. We first compute the periods $n^0_k\equiv (2\pi)^{1-k}\int_{B_k}(\e^{-B_0} F)_k$; they are of the form $n_k^0= L^{k-1}/g_s \psi_k$, where $\psi_k$ are multi-valued functions of $\sigma$. The actual flux quanta are then given by a $b$-transform: $n_0=n_0^0$, $n_2= n_2^0- b n_0$, and so on. $b$ can be eliminated from the system thanks to two invariants under $b$-transform:  $I_4=n_2^2-2n_0 n_4= (n_2^0)^2-2n_0^0 n_4^0$, $I_6= n_2^3+ 3n_0 n_6- 3 n_0 n_2 n_4= (n_2^0)^3+ 3n_0^0 n_6^0- 3 n_0^0 n_2^0 n_4^0$ \cite{Aharony:2010af}. 

The quotient $I_4^3/I_6^2$ then only depends on $\sigma$, and determines it in terms of the flux quanta. We obtained this function numerically; it is multi-valued, and contains the supersymmetric result in \cite{Aharony:2010af} as one of its branches. The limits studied there, $n_6\ll n_2^3/n_0^2$ and $n_6\gg n_2^3/n_0^2$ (for $n_4=0$) were motivated by holography; $I_4^3/I_6^2\to 1$ and $0$ respectively. In the first limit we find that $\sigma$ can have the same values of the supersymmetric case, namely $2$, $2/5$ and one more value near $2/5$. In the second limit, besides the supersymmetric case $\sigma=1$, several additional values appear. Finally, $L= I_6^{1/6} n_0^{-1/2}(\psi_6^3+ 3 \psi_0^2 \psi_6 -3 \psi_0 \psi_2 \psi_4)^{-1/6} \psi_0^{1/2}$, and $g_s = I_6^{-1/6} n_0^{-1/2}(\psi_6^3+ 3 \psi_0^2 \psi_6 -3 \psi_0 \psi_2 \psi_4)^{1/6} \psi_0^{1/2}$. The functions of $\sigma$ that appear in these expressions are of order one on most of the interval.

\subsubsection{Stability: bubbles of simple branes}
Our stability analysis will take advantage of the fact that there are two possible choices for $J$:
\begin{equation}
	 J_\pm = R^2\left(\pm\frac2 \sigma j_\mathrm{B} +\frac14 j_\mathrm{F} \right)\equiv \pm J_\mathrm{B}+ J_\mathrm{F}\,,
\end{equation}
so two possible pure spinors: $\Phi_+ = \e^{\ii \theta}\e^{-\ii J_+}$ or $\Phi_+ = \e^{\ii \theta}\e^{-\ii J_-}$. We can then simply pick the most advantageous one to evaluate our stability criteria.\footnote{The odd pure spinor compatible with $\Phi_+ = \e^{\ii \theta}\e^{-\ii J_-}$ is only locally defined, but we won't be needing it for our construction.}

We first look at simple domain-wall D$p$-branes, before switching on a world-sheet flux.

\begin{figure}[ht!]
    \centering
\includegraphics[width=0.6\linewidth]{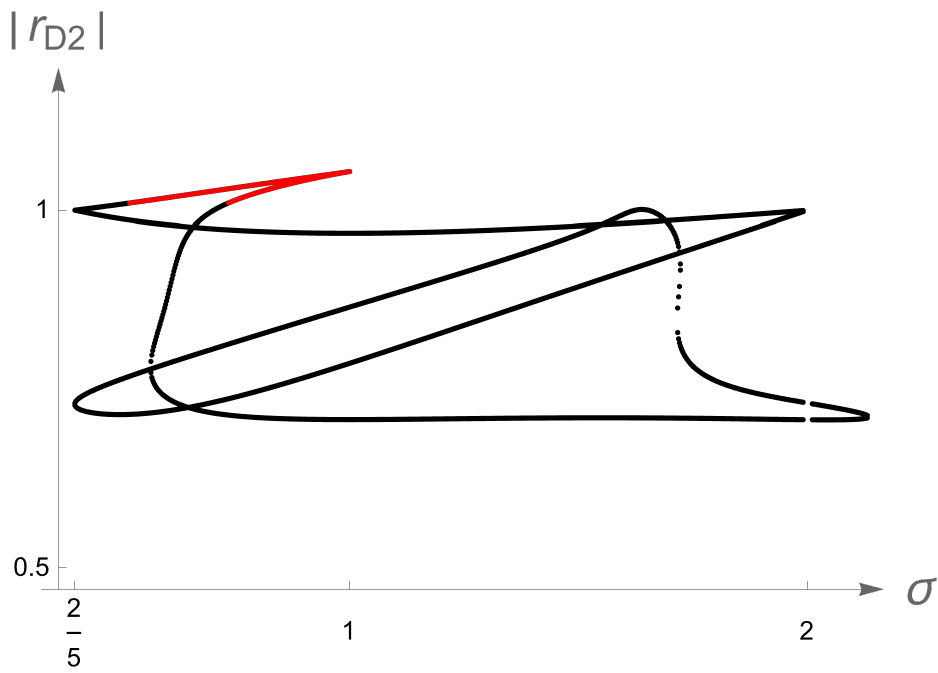}
    \caption{\small Stability ratio of D2 bubbles for the supersymmetric solutions and various non-supersymmetric solutions. The red line indicates instability of the given solution under the nucleation and expansion of such bubbles, and the black lines signify stability. Note that the $\sigma$ interval for non-supersymmetric solutions is wider than the supersymmetric one, as discussed in \cite{Koerber:2010rn}.}
    \label{fig:D2CP3}
\end{figure}

\subparagraph{D2.} Here we will evaluate the stability bound derived in the previous section for D2 bubbles. To make our criterion most efficient, we minimize our stability ratio: we maximize $(\mathrm{Im}\Phi_+)_0=\sin \theta$ in $\theta$ by taking $\theta=\pi/2$. In doing so, we trivially satisfy the almost calibration condition \eqref{kappa1nonsusyads} for points of $M_6$. In this case, we can therefore use the criterion (\ref{eq:stabgen}), which is both necessary and sufficient for stability. The corresponding stability ratio is
\begin{equation}
	r_{\mathrm{D}2} = \frac13 
    g_s L f_6\,.
\end{equation}
Note that here, the almost calibration condition \eqref{kappa1nonsusyads} is satisfied for $\theta=\pi/2$, which is different from the $\theta$ set by supersymmetry in \eqref{susyCP3}. This shows that the domain-wall D2 branes are not BPS in the $\mathcal{N}=1$ case.
The criterion for the stability of D2 brane bubbles is illustrated in Fig.~\ref{fig:D2CP3} for the supersymmetric solution and different non-supersymmetric solutions.

We conclude from Fig.~\ref{fig:D2CP3} that most non-supersymmetric solutions are stable against the nucleation and expansion of D2 bubbles, with one solution being unstable in the region $\sigma\in[1/2,3/4]$ and two in the region $\sigma\in[3/4,1]$.

\subparagraph{D4.} We consider here D4 bubbles wrapping a two-cycle $B_{2}$. $\mathrm{Im}(\e^{-B}\wedge\Phi_+)_2=- \sin \theta B + \cos \theta J$ is in general not closed. As discussed earlier, this would cause the problem that the integral $\int_{B_2} \mathrm{Im}(\e^{-B}\wedge\Phi_+)_2$ would depend on $B_2$, and not on its homology class alone; in other words, we would have to perform the integral over infinitely many possibilities. Fortunately, $\dd (\mathrm{Im}\e^{-B}\wedge \Phi_+)_2 =0$ can be arranged by choosing $\theta$ appropriately: this is, in fact, what happens in the supersymmetric case, where the pure spinor is  $\e^{\ii \theta}\e^{-\ii J_+}$ and the angle is given by $\tan\theta = (\sigma+2)/hR$. This provides one estimate for the DBI term entering the stability ratio. Another comes from the pure spinor $\e^{\ii \theta} \e^{-\ii J_-}$, with $\theta$ now such that $\tan\theta = (\sigma-2)/hR$. 
In both cases separately, both $(\mathrm{Im}\e^{-B}\wedge \Phi_+)_2$ and $(\e^{-B}\wedge * \lambda F)_2$ are proportional to $-4 j_\mathrm{B}+ j_\mathrm{F}$, which is indeed closed by (\ref{eq:j-psi-twistor}). In the stability ratio, the integral of this two-form on 
$B_2$ is present in both the numerator and denominator, and cancels out.

However, in choosing $\theta$ these ways, no
two-cycle can be almost calibrated with respect to $(\mathrm{Im}\e^{-B}\wedge \Phi_+)_2$. We therefore evaluate our criterion \eqref{eq:stabgen}, sufficient but not necessary for stability. We do so by picking the pure spinor built out of $J_-$, since it yields a smaller stability ratio than its $J_+$ counterpart. All in all this gives
\begin{equation}
	r_{\mathrm{D}4} = \frac{g_s L f_{4\mathrm{F}} }{3 h R} \sqrt{h^2 R^2+ (\sigma-2)^2}\,.\label{r4CP3}
\end{equation}

\subparagraph{D6.} We consider now D6 bubbles wrapping a four-cycle $B_{4}$. $(\mathrm{Im}\e^{-B}\wedge \Phi_+)_4$ is automatically closed for any $\theta$. To compute its integral over $B_4$, we can use the fact that $4 j_\mathrm{B}^2 + j_\mathrm{F}\wedge j_\mathrm{B}$ is exact. Alternatively, $-4 j_\mathrm{B}+ j_\mathrm{F}$ is a closed two-form, so it can be taken to be the generator of the only cohomology class up to rescaling; so it is the Poincar\'e dual of $B_4$. Either way, we conclude $\int_{B_4}(p j_\mathrm{B}^2 + q j_\mathrm{F}\wedge j_\mathrm{B})= \int_{B_4} (p-4q) j_\mathrm{B}^2$. Finally, we are allowed to choose $\theta$ so as to maximize $\int_{B_4}(\mathrm{Im}\e^{-B}\wedge \Phi_+)_4$. By doing so, we minimize our stability ratio, but here we remain agnostic as to whether or not $B_4$ is an almost calibrated cycle. Indeed, maximizing the pure spinor four-form in $\theta$ is equivalent to satisfying \eqref{eq:alt-cal-P2}, 
but we don't know whether or not \eqref{eq:alt-cal-P1} holds.
Our criterion is therefore once again the one stronger than stability, \eqref{eq:stabgen}. Simplifying $\int_{B_4} j_\mathrm{B}^2$, we obtain
\begin{equation}
	r_\mathrm{D6} = \frac{g_s L (f_{2\mathrm{F}} + f_{4\mathrm{F}}hR-f_{2\mathrm{B}})}{3 \sqrt{h^2 R^2 + (\sigma-1)^2}}\,.
\end{equation}

\subparagraph{D8.} Finally we consider D8 bubbles wrapping the whole internal space. In this case, the integrals are straightforward. maximizing in $\theta$, we end up almost-calibrating the internal space with respect to $(\mathrm{Im}\e^{-B}\wedge\Phi_+)_6$, since both \eqref{eq:su3-cal} are satisfied. We obtain the following stability ratio entering \eqref{eq:stabgen}
\begin{equation}
	r_\mathrm{D8} = \frac{g_s L(f_{2\mathrm{F}} h R - f_0 \sigma)}{3 \sqrt{h^2 R^2 + \sigma^2}}\,.
\end{equation}

\begin{figure}[ht!]
    \centering
    \includegraphics[width=0.6\linewidth]{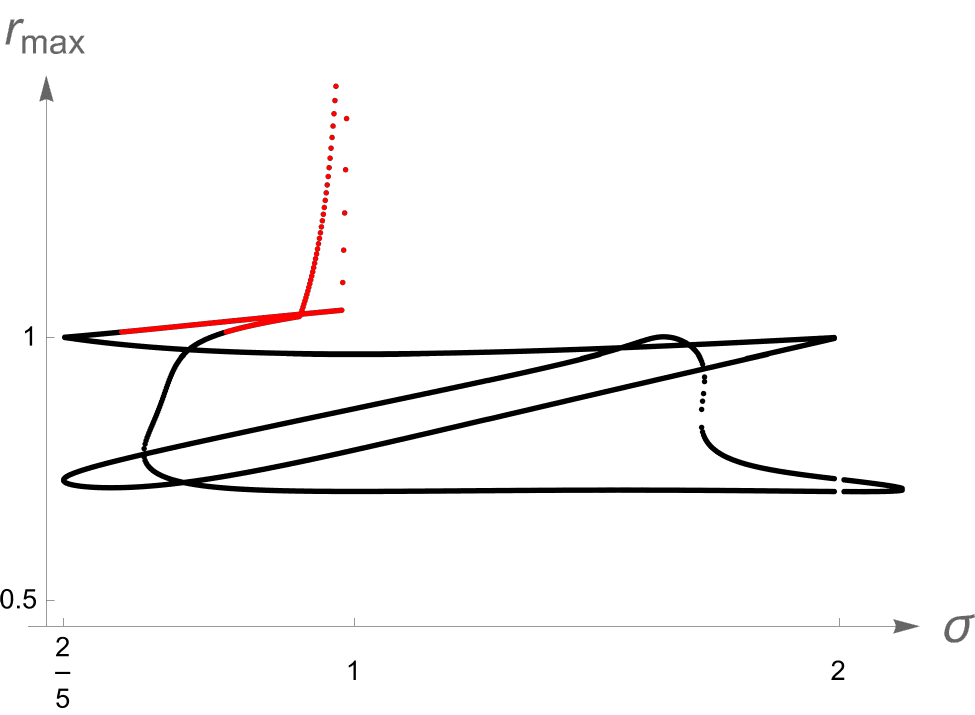}
    \caption{\small Combined stability ratio for D2, D4, D6, and D8 bubbles, for the supersymmetric solution and various non-supersymmetric ones. The red line
indicates the inconclusiveness of our criteria to determine whether or not the given solution is stable under the nucleation and expansion of bubbles,
and the black lines signify stability against all bubbles.}
    \label{fig:combCP3bubble}
\end{figure}

\subparagraph{Combining all bubbles}  Let us bring these results together and define a combined stability ratio $r_{\text{max}}=\text{max}\{|r_{\text{D}2}|,|r_{\text{D}4}|,|r_{\text{D}6}|,|r_{\text{D}8}|\}$
for each solution at a given value of $\sigma$. Violating the stability bound with $r_{\text{max}}$ does not necessarily mean that the given solution is unstable, since it combines ratios entering bounds either strictly equivalent to stability or more restrictive. Fig.~\ref{fig:combCP3bubble} illustrates the combined stability analysis. 

Fig.~\ref{fig:combCP3bubble} shows that most non-supersymmetric solutions are stable against the nucleation and expansion of any pure brane bubbles. 

At $\sigma=1$, the geometry becomes nearly-K\"ahler, and some solutions have a vanishing NSNS flux. For D4 bubbles in the $H\rightarrow 0$ branch approaching such solutions, our DBI estimate becomes particularly inaccurate, resulting in the divergence of the stability ratio \eqref{r4CP3}, as seen in the $\sigma=1$ vicinity of Fig.~\ref{fig:combCP3bubble}. Our D4 stability analysis therefore remains inconclusive for these solutions, but these are precisely the ones that are destabilised by D2 bubbles anyway.

\subsubsection{Stability: bubbles of bound states}

\begin{figure}[!ht]
    \centering
    \includegraphics[width=0.5\linewidth]{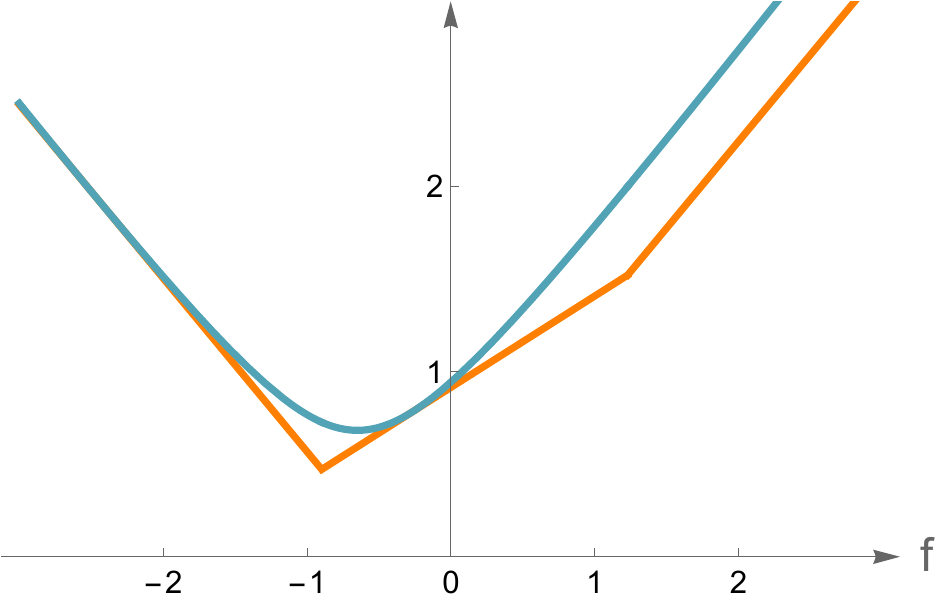}
    \caption{\small Lower and upper bounds on the minimal DBI action for a cycle on the fiber, shown here for one of the solutions at $\sigma = 1.3$.}
    \label{fig:dbi-2cycle}
\end{figure}

We switch on a world-sheet flux $2\pi f_\mathrm{ws}= f(-4 j_\mathrm{B}+ j_\mathrm{F})$, and from now on, we denote the D8/D6/D4/D2, D6/D4/D2, and D4/D2 bound states simply as D8, D6, D4 bound states respectively.
\subparagraph{D4.} As for the pure D4, we need to fix $\theta$ so as to make $(\mathrm{Im}\e^{-{\mathcal F}}\Phi_+)_2$ closed. By doing so we again prevent any two-cycle to be almost calibrated by $(\mathrm{Im}\e^{-{\mathcal F}}\Phi_+)_2$. But we face a new issue: the integral of $(\mathrm{Im}\e^{-{\mathcal F}}\Phi_+)_2$ would always vanish for at least one choice of $f$.\footnote{In fact $f$ is restricted by world-sheet flux quantization and such a value could not be chosen exactly; but one could get arbitrarily close to it at large $R$.} In fact:
\begin{equation}\label{eq:dbi-2cycle}
	(\mathrm{Im}\e^{-{\mathcal F}}\Phi_+)_2 = \frac{g_s L (\mp h R^3 + 2 f \sigma (\sigma \pm 2))}{2 \sigma \sqrt{h^2 R^2 + (\sigma+2)^2}}
\end{equation}
where the signs are relative to the choice $J=J_\pm$. Fortunately, we are not forced to choose one of the signs once and for all: both choices give a lower bound on the minimum value of the DBI action in the homology class, and for each solution and choice of $f$ we may take the largest of the two. In Fig.~\ref{fig:dbi-2cycle} we show the maximum over the two signs of the absolute value of $\int_{B_2} (\mathrm{Im}\e^{-{\mathcal F}}\Phi_+)_2 $, taking $B_2$ the fiber over a point; this gives the lower piecewise-linear plot. For reference, we have also directly evaluated the DBI action on the fiber, which gives $\pi \sqrt{R^4+(4f-R^3 h/\sigma)^2}$. We don't know for sure if the DBI is in fact minimized over this cycle representative, or perhaps over another that is homologous to it. The real value of the DBI minimum might lie anywhere between the curve and the piecewise-linear function.

Overall we get 
\begin{equation}\label{eq:CP3-rD4D2}
	r_\mathrm{D4}= \max_{s\in \{\pm 1\}}\frac{g_s L (-f_{4\mathrm{F}}R^2 + 2 f f_6 \sigma)}{-3shR^3 +6 f \sigma (\sigma+2s)}\sqrt{h^2 R^2 + (\sigma+2s)^2}\,.
\end{equation}

To check which solutions are stable according to the criterion in (\ref{eq:stabgen}), we now need to check that $|r_\mathrm{D4}|<1$ for every $f$. It is enough to maximize in $f$, so as to check the worst-case scenario.\footnote{An efficient way to do so is to notice that it is monotonic in $f$ for both $s=\pm1$. So we really only need to maximize among the values of $f$ where the two branches intersect, and the limits $f\to \pm \infty$.}

\begin{figure}[!ht]
    \centering
    \includegraphics[width=0.6\linewidth]{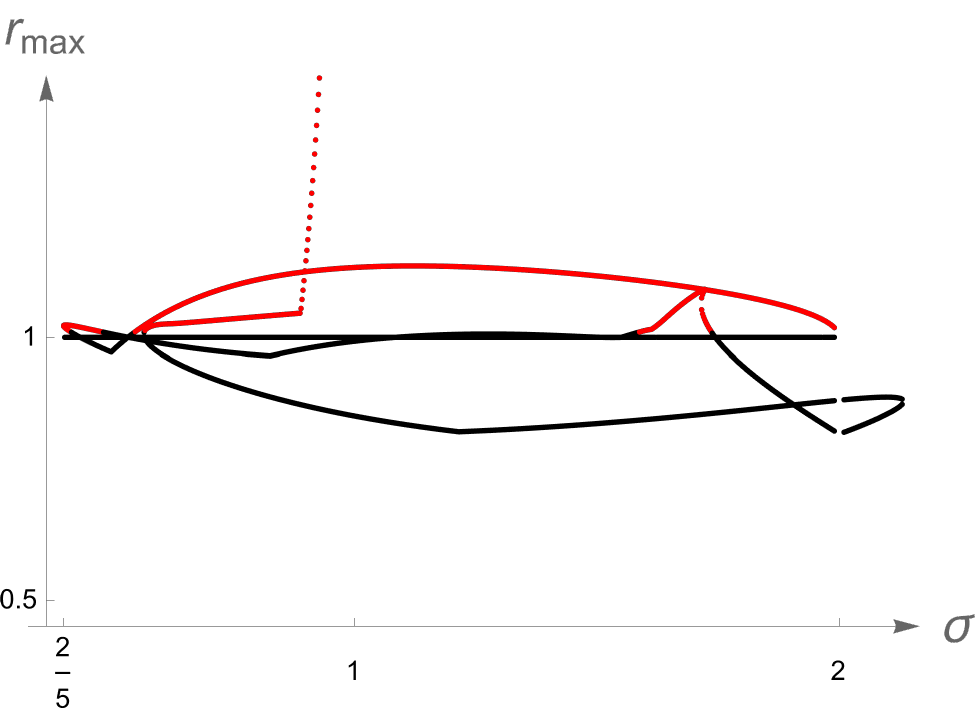}
    \caption{\small Combined stability ratio for bound states, for the supersymmetric solution and various non-supersymmetric ones. The red line
indicates the inconclusiveness of our criteria to determine whether or not the given solution is stable under the nucleation and expansion of bound states,
and the black lines signify stability against all the $2\pi f_{\rm ws}=f (4j_{\rm B}-  j_{\rm F})$ bound states.}
    \label{fig:BSCP3}
\end{figure}

\subparagraph{Other bound states.} For D6 bound states, when we choose $2\pi f_{\rm ws}=f (4j_{\rm B}-  j_{\rm F})$, it turns out that the choice $J=J_-$ always gives the largest value of $(\mathrm{Im}\e^{-{\mathcal F}}\Phi_+)_4$. For the vast majority of solutions, these bound states do not yield an instability. These results contribute to Fig.~\ref{fig:BSCP3}.

A general four-cycle $B_4$ has a larger cohomology $H^2(B_4)$ than that of its ambient $\mathbb{CP}^3$, and in principle one might want to consider $f_{\rm ws}$ to be a general element of it. We were not able to find a protection mechanism for these more general objects.  This is an incompleteness in our approach, beyond our omission of non-Abelian bound states, mentioned in the Introduction. On the other hand, we also have no indication that either of the calibration conditions \eqref{eq:su3-cal} can be solved on such configurations. We will be able to say much more about these objects for the K\"ahler--Einstein solutions of Sec.~\ref{sub:ke6}.

For D8 bound states, both choices $J=J_+$ and $J=J_-$ give the same result for $2\pi f_{\rm ws}=f (4j_{\rm B}-  j_{\rm F})$. In both cases the resulting functions are quite complicated and we do not write them down explicitly. For each solution, we maximize in $\theta$ just like in the bubble case, and we maximize in $f$ numerically. Again these seldom cause instabilities, and are presented as part of Fig.~\ref{fig:BSCP3} below.

\subparagraph{Combining bound states.} Let us now bring together all the bound states we analyzed. We define once again a combined stability ratio $r_{\text{max}}=\text{max}\{|r_{\text{D}2}|,|r_{\text{D}4}|,|r_{\text{D}6}|,|r_{\text{D}8}|\}$ for each solution at a given value of $\sigma$. Fig.~\ref{fig:BSCP3} illustrates the combined stability analysis.

Fig.~\ref{fig:BSCP3} shows that some non-supersymmetric solutions remain stable under the nucleation and expansion of any $2\pi f_{\rm ws}=f (4j_{\rm B}-  j_{\rm F})$ bound states. We witness the same divergence behaviour for the $r_\mathrm{D4}$ stability ratio of solutions approaching the nearly-K\"ahler $H=0$ solutions. The $r_\text{max}=r_\mathrm{D4}=1$ black line corresponds to the supersymmetric solutions, given that having $r_\mathrm{D4}=1$ is equivalent to the supersymmetry equation \eqref{susy3ads0}.\footnote{This employs the supersymmetric pure spinor $\e^{\ii \theta} \e^{-\ii J_+}$. Given that the stability ratio lies at the intersection of the pure spinor values for the D4 bound states, as mentioned in the previous footnote, this is effectively what we are using.}

One could refine our results by studying further the stability properties of the solutions having a stability ratio larger than one, for which we didn't reach a conclusion. This could be done by direct evaluation of the DBI contribution for D4 bound states wrapping the fiber for instance. We don't pursue this further as here we are using our stability criteria to show that some solutions are actually stable against the $2\pi f_{\rm ws}=f (4j_{\rm B}-  j_{\rm F})$ bound-state bubbles.

Finally, let us mention that a partial perturbative stability analysis of these vacua has been conducted in \cite{Koerber:2010rn}; loosely speaking, they found tachyonic instabilities for some solutions in the vicinity of $\sigma=\frac12$ and $\sigma=2$, we thus conclude from Fig.~\ref{fig:BSCP3} that there are solutions resisting both these tachyonic instabilities and the  $2\pi f_{\rm ws}=f (4j_{\rm B}-  j_{\rm F})$ bound-state bubbles.

\subsection{\texorpdfstring{AdS$_4\times$ flag manifold}{AdS4 x flag manifold}}
\label{sub:flag}

We now consider the so-called flag manifold $\mathbb{F}(1,2;3)$. A (complete) flag in $\mathbb{C}^3$ is a complex plane, together with a line belonging to it. $\mathbb{F}(1,2;3)$ is the space of all such flags. It has a rich geometry, which can be treated from multiple points of view; see \cite{Altavilla_2022} for a recent mathematical review. It is topologically $\mathrm{Tw}(\mathbb{CP}^2)$, and as such it was covered in Sec.~\ref{sub:tw}. However, it is also a homogeneous space $\mathrm{SU}(3)/\mathrm{U}(1)\times \mathrm{U}(1)$; this gives access to more forms, and thus to a richer space of solutions. This possibility was actually mentioned in \cite{Koerber:2010rn}, but not explicitly realized; we are now going to do so. This parallels the developments for supersymmetric solutions, which were first found using the twistor metric \cite{Tomasiello:2007eq} and then extended to a more general metric using the homogeneous structure \cite{Koerber:2008rx}.

\subsubsection{Coset space geometry}
\label{ssub:coset}

Given a fiber bundle $E\buildrel\pi\over\to M$, if a form on the total space $E$ is invariant and horizontal (annihilated by both $\iota_v$ and $L_v = \{ \dd, \iota_v\}$, for $v$ any vertical vector field), then it is the pullback under $\pi$ of a form on $M$. In other words, such a form is basically a form on the base. 

Consider now the particular case $G\to G/H$. On a group manifold $G$, there are left-invariant one-forms $\lambda^i$ obtained by expanding $g^{-1}\dd g$ over the generators. In general few (if any) of these are invariant and horizontal under $G\to G/H$. However, it is more common for some higher forms $\frac1{k!}\omega_{i_1\ldots i_k} \lambda^{i_1}\wedge \lambda^{i_k}$ to be invariant and horizontal. In the context of homogeneous spaces these are often just called invariant. For more details see for example \cite[Sec.~3]{Koerber:2008rx} or \cite[Sec.~4.4.2]{book}.

For our case $G=\mathrm{SU}(3)$, $H=\mathrm{U}(1)\times \mathrm{U}(1)$, there are three invariant two-forms $j_a$ and two invariant three-forms $\psi$, $\tilde \psi$. They satisfy 
\begin{equation}\label{dflag}
	j_1 \wedge j_2 \wedge j_3 = -\frac14 \psi \wedge \tilde \psi \, ,\qquad
	\dd j_a = \psi \, ,\qquad \dd \tilde \psi = 4(j_1 \wedge j_2 + j_2 \wedge j_3 + j_3 \wedge j_1)\,.
\end{equation}
The two-forms can be written locally as $j_1=\lambda^1 \wedge \lambda^2$, $j_2 = \lambda^3 \wedge \lambda^4$, $j^3= \lambda^5 \wedge \lambda^6$. In terms of these, we write the metric as
\begin{equation}
	\dd s^2 = \alpha_1^2 (\lambda_1^2 + \lambda_2^2) + \alpha_2^2 (\lambda_3^2 + \lambda_4^2) + \alpha_3^2 (\lambda_5^2 + \lambda_6^2)\,.
\end{equation}
We can also introduce the $\mathrm{SU}(3)$-structure
\begin{equation}
	J= J_1 + J_2 + J_3 \, ,\qquad J_a\equiv \alpha_a^2 j_a \,;\qquad \Omega= \alpha_1 \alpha_2 \alpha_3 (- \psi+\ii \tilde \psi)\,.
\end{equation}

This geometry reduces to that in \ref{sub:tw} on the three loci
\begin{equation}\label{eq:flag-tw-loci}
	\left\{\alpha_1^2= \alpha_2^2 = \frac{R^2}\sigma, \ \alpha_3^2=R^2\right\} \, ,\quad
	\left\{\alpha_2^2= \alpha_3^2 = \frac{R^2}\sigma, \ \alpha_1^2=R^2\right\} \, ,\quad
	\left\{\alpha_1^2= \alpha_3^2 = \frac{R^2}\sigma, \ \alpha_2^2=R^2\right\}\,.
\end{equation}
Indeed, there are three ways to fiber $\mathbb{F}(1,2;3)$ over  $\mathbb{CP}^2$ \cite[Sec.~5]{Altavilla_2022}.

\subsubsection{The solutions}

Our Ansatz for the form fields consists in expanding them over the invariant forms:
\begin{equation}\label{eq:flag-flux}
	F_2= f_{2a} J_a \, ,\qquad
	F_4 =  f_{41} J_2 \wedge J_3+\text{cycl.} \, ,\qquad
	F_6 = f_6 J_1 \wedge J_2 \wedge J_3\,;\qquad H=h \mathrm{Re} \Omega\,.
\end{equation}
We can take
\be B=-h\alpha_1\alpha_2\alpha_3\,  j_1.\ee
The total flux is
\be * \lambda F =f_6 - f_{4a} J_a +(f_{21} J_2\wedge J_3+\text{cycl.}) - f_0\, J_1\wedge J_2\wedge J_3.\ee
To compute the Ricci tensor, we used the following formula \cite[(7.33)]{besse}: 
\begin{equation}
	R_{ab}= e_a^j e_b^l \left[ -\frac12 f^k{}_{ji} f_{kl}{}^i-\frac12 f^i{}_{jk} f^k{}_{li} - f^i{}_{(j|\alpha} f^\alpha{}_{l)i}
    +\frac14 f_{jik}f_l{}^{ik}
+ f^i{}_{ki} f_{(jl)k}    \right]\,.
\end{equation}
$f$ are the structure constants, $i$ are indices along $M=G/H$, $\alpha$ along $H$. 

Once again the equations of motion become algebraic:
\begin{subequations}\label{eq:flag-eom}
\begin{align}
	\frac4{g_s^2}\left(\frac{\alpha_1^4- \alpha_2^4 - \alpha_3^4 + 6 \alpha_2^2 \alpha_3^2}{ \alpha_1^2 \alpha_2^2 \alpha_3^2} -\frac{h^2}{g_s^2}\right) &= 
	-f_0^2 + f_{21}^2 - f_{22}^2 - f_{23}^2 - f_{41}^2 + f_{42}^2 + f_{43}^2 +f_6^2\quad \text{and cycl.}\,;\\
	-\frac 4{g_s^2} \Lambda &= f_0^2 + f_{2a}f_{2a} + f_{4a}f_{4a} + f_6^2
\,;\\
	\frac8{g_s^2}h^2 & = 5 f_0^2 + 3 f_{2a}f_{2a} + f_{4a}f_{4a} - f_6^2\,;
	\\ 
	4\frac{\alpha_3}{\alpha_1 \alpha_2}\frac{h^2}{g_s^2} & = 
	f_0 f_{23} + f_{21} f_{42} + f_{22} f_{41} + f_{43} f_6 \qquad \text{and cycl.}\,;\\
	\sum_a f_{2a} \alpha_a^2 &=- h \alpha_1 \alpha_2 \alpha_3 f_0 \,,\\
	\sum_a f_{4a} \alpha_a^2 &= h \alpha_1 \alpha_2 \alpha_3 f_6 \,.
\end{align}
\end{subequations}
Notice that these admit more solutions than (\ref{eq:tw-eom}) even on the twistor loci (\ref{eq:flag-tw-loci}), because the flux Ansatz (\ref{eq:flag-flux}) is more general than (\ref{eq:cp3-flux}). With additional restrictions on the flux parameters, (\ref{eq:flag-eom}) do reduce to (\ref{eq:tw-eom}), so the twistor solutions are indeed part of the solutions to the new system.

While the $\alpha_a$ parameterization is nice and symmetric, to study solutions we also find it practical to switch to an overall radius $R$ and two shape parameters, similar to the $R$ and $\sigma$ in Sec.~\ref{sub:tw}:
\begin{equation}
	\alpha_1^2 = \frac{R^2}{\sigma_1} \, ,\qquad \alpha_2^2 = \frac{R^2}{\sigma_2}\, ,\qquad \alpha_3^2 = R^2\,.
\end{equation}
The twistor loci (\ref{eq:flag-tw-loci}) are now $\{\sigma_1= \sigma_2\}$, $\{\sigma_1=1\}$, $\{\sigma_2=1\}$.

The symmetry among the $\alpha_a$ is of course not lost; it becomes 
\begin{equation}\label{symflag}
	(\sigma_1, \sigma_2, R) \mapsto \begin{array}{c}
		 (\sigma_2, \sigma_1, R)\\
		 \left(\frac1{\sigma_1}, \frac{\sigma_2}{\sigma_1}, \frac R{\sqrt{\sigma_1}}\right)\\
		 \left(\frac{\sigma_1}{\sigma_2}, \frac1{\sigma_2}, \frac R{\sqrt{\sigma_2}}\right)
	\end{array} \,.
\end{equation}
A fundamental region in the $(\sigma_1,\sigma_2)$ plane is $\{ \sigma_1\geq \sigma_2, \sigma_2\geq 1 \}$. The boundary of this region consists of the half-lines $\{\sigma_1 \geq 1, \sigma_2=1\}$ and $\{\sigma_1= \sigma_2, \sigma_2\ge1\}$, which are part of the twistor locus. They include the solutions of (\ref{eq:tw-eom}) with $\{ \sigma\ge 1\}$ and $\{\sigma \le 1\}$, respectively.

The supersymmetric solution \cite{Koerber:2008rx} is 
\begin{equation}
\begin{split}
	&f_0 = 5 \frac m{g_s} \, ,\qquad 
	f_{21} = \frac{-5 \alpha_1^2 + 3 \alpha_2^2 +3 \alpha_3^2}{2 g_s \alpha_1 \alpha_2 \alpha_3} \quad \text{and cycl.}\,,\qquad
	 f_{4i}=3\frac m{g_s} \, ,\qquad f_6 = -3\frac{\tilde m}{g_s}\\
	&m=\frac{\sqrt{6(\alpha_1^2 \alpha_2^2 + \alpha_1^2 \alpha_3^2 + \alpha_2^2 \alpha_3^2) - 5 (\alpha_1^4 + \alpha_2^4 + \alpha_3^4)}}{\sqrt{20} \alpha_1 \alpha_2 \alpha_3}
	\, ,\qquad \tilde m = -\frac{\alpha_1^2 + \alpha_2^2 + \alpha_3^2}{2 \alpha_1 \alpha_2 \alpha_3} \,,\\
	& h=-2m \, ,\qquad \tan\theta = \frac{\tilde m}m\,.
\end{split}	
\end{equation}

\subsubsection{Stability: bubbles of  bound states}
As for the $\mathbb{CP}^3$ case, our stability analysis will take advantage of the fact that there are different possible choices for $J$: the first is
\be J= J_1+J_2+J_3,\label{J}\ee
and three others correspond to flipping the sign of $J_1$, $J_2$, or $J_3$. We thus have four pure spinors $\Phi_+ = \e^{\ii \theta}\e^{-\ii J}$ at our disposal, with the four different versions of $J$.\footnote{The odd pure spinors compatible with the different even pure spinors constructed out of $J$'s with sign flips are only locally defined, but we won't be needing them for our construction.}

Here we directly consider bubbles of bound states, since bubbles of simple branes are merely a particular case. By \eqref{dflag}, the most general closed world-sheet flux can be written as
\be 2\pi f_\mathrm{ws}= f_1(j_1 -j_2)+f_2(j_1-j_3)\, .\label{wsfFlag}\ee 

\begin{figure}[!ht]
    \centering
    \includegraphics[width=0.6\linewidth]{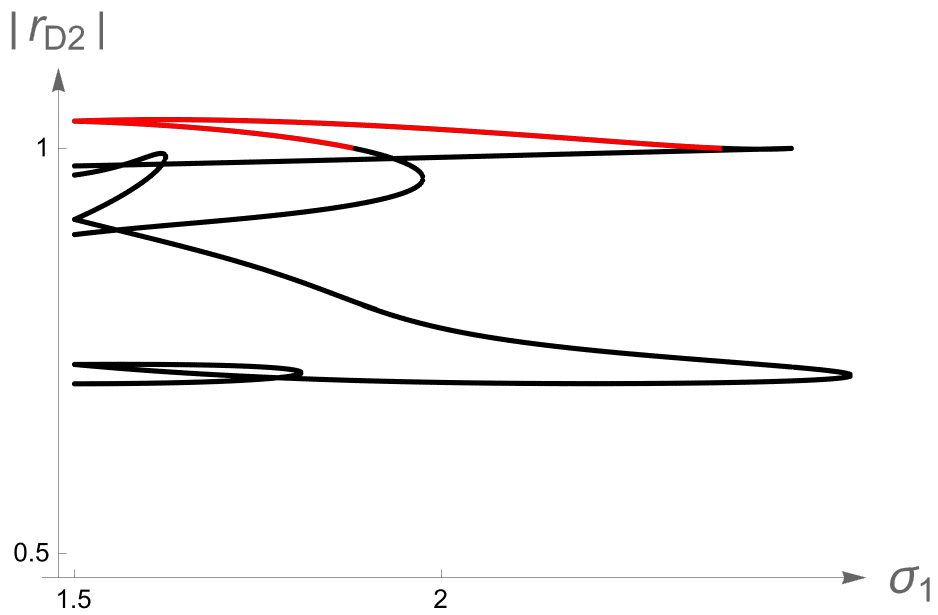}
    \caption{\small Stability ratio of D2 bubbles for the supersymmetric solutions and various non-supersymmetric solutions, with $\sigma_2=1.5$. The red line indicates instability of the given solution under the nucleation and expansion of such bubbles, and the black lines signify stability.}
    \label{fig:D2flag}
\end{figure}

\subparagraph{D2.} We start by considering D2 bubbles. All available versions of the pure spinor have $(\mathrm{Im}\Phi_+)_0=\sin \theta$. We again maximize in $\theta$ so the points of $M_6$ are again almost calibrated with respect to $(\mathrm{Im}\Phi_+)_0$, yielding a stability ratio similar to the $\mathbb{CP}^3$ case:
\begin{equation}
	r_{\mathrm{D}2} = \frac13 g_s L f_6\,.
\end{equation}
The criterion for the D2 bubbles is therefore necessary and sufficient for stability. It is illustrated in Fig.~\ref{fig:D2flag} for the supersymmetric solution and different non-supersymmetric solutions.

\subparagraph{D4.} There are two classes of homologically distinct two-cycles, with Poincar\'e duals proportional to $j_1-j_2$ and $j_1-j_3$. We evaluate the stability ratio for both classes separately.

As for the $\mathbb{CP}^3$ case, we need to fix $\theta$ so as to make $(\mathrm{Im}\e^{-{\mathcal F}}\Phi_+)_2$ closed, so that the stability ratio doesn't depend on the specific two-cycle considered in each class. We do so for each of the pure spinors constructed out of the four versions of $J$. For the pure spinor constructed out of $J$ \eqref{J}, this yields the stability ratios
\begin{subequations}\label{rD4flag}
\begin{align}
r_{\mathrm{D4}}|_{j_1-j_2}=\frac{
  h R \sqrt{\sigma_1} \left( f_{42} R^2 - f_1 f_6 \sigma_2 \right) 
  \sqrt{1 + \frac{ \left( \sigma_1 + \sigma_2 + \sigma_1 \sigma_2 \right)^2 }{ h^2 R^2 \sigma_1 \sigma_2 }}
}{
  3(h R^3 \sqrt{\sigma_1} - f_1 \sqrt{\sigma_2} \left( \sigma_1 + \sigma_2 + \sigma_1 \sigma_2 \right))
}\,,\\
r_{\mathrm{D4}}|_{j_1-j_3}=\frac{
  h R \sqrt{\sigma_1 \sigma_2} \left( f_{43} R^2 - f_2 f_6\right) 
  \sqrt{1 + \frac{ \left( \sigma_1 + \sigma_2 + \sigma_1 \sigma_2 \right)^2 }{ h^2 R^2 \sigma_1 \sigma_2 }}
}{
  3(h R^3 \sqrt{\sigma_1 \sigma_2} - f_2  \left( \sigma_1 + \sigma_2 + \sigma_1 \sigma_2 \right))
}\,.
\end{align}
\end{subequations}
 Given that neither class of two-cycles is almost calibrated with respect to $(\mathrm{Im}\e^{-{\mathcal F}}\Phi_+)_2$, we now need to evaluate our criterion \eqref{eq:stabgen}, and to do so for every world-volume fluxes. 

The stability ratios \eqref{rD4flag} have the same structure than in the $\mathbb{CP}^3$ case: they each depend on a single world-volume flux, and they have a pole for one of its value. Fortunately we are not forced to choose among our four different pure spinors once and for all: all four choices give a lower bound on the minimum value of the DBI action in the homology class, and for each solution and value of the world-volume flux we may select whichever is the largest.\footnote{In practice we do so by comparing pairs of pure spinors for every world-volume flux value, and selecting the best ratios among the pairs. This might get optimised by comparing more DBI estimates simultaneously but as we will see later on the resulting stability ratios will be small enough to reach stability conclusions.} To ensure that our stability criterion is respected for all possible world-volume flux, it suffices to consider the worst case scenario where a given value maximizes the stability ratio. We do so for each solutions and values of $\sigma_1$ and $\sigma_2$.
\subparagraph{D6.} For bubbles of bound states wrapping a four cycle $B_4$, we have $\int_{B_4}(a j_1\wedge j_2 + b j_1\wedge j_3+ c j_2\wedge j_3)=\int_{B_4}((a-c) j_1\wedge j_2 + (b-c) j_1\wedge j_3)$, which comes from the fact that $j_1 \wedge j_2 + j_2 \wedge j_3 + j_3 \wedge j_1$ is exact. We therefore calculate the stability ratios for the $j_1\wedge j_2$ and $j_1\wedge j_3$ components separately, with their integral over $B_4$ again dropping out of the stability ratios. We maximize the DBI estimates in $\theta$ as before, and numerically maximize in both world-sheet fluxes $f_1$ and $f_2$ to encompass all cases of world-volume flux values. Just like in the $\mathbb{CP}^3$ case, the $B_4$ we consider aren't necessarily almost calibrated so these stability ratios are the ones of the criterion \eqref{eq:stabgen}. It turns out that they are the smallest when evaluated with the pure spinor having the sign of $J_1$ flipped.

When considering world-volume fluxes more general than \eqref{wsfFlag}, the situation is closely analogous to the $\mathbb{CP}^3$ case: a general four-cycle $B_4$ can have a larger cohomology $H^2(B_4)$ than that of its ambient $\mathbb{F}(1,2;3)$, which would give access to more general possibilities for $f_{\rm ws}$. As for the $\mathbb{CP}^3$ case, we were not able to find a protection mechanism for these objects. However, we also have no indication that either of the calibration conditions \eqref{eq:su3-cal} can be solved on such configurations. We refer the reader to the K\"ahler--Einstein solutions of Sec.~\ref{sub:ke6}, for which we are able to say much more about the analogous objects.

\subparagraph{D8.} Finally we consider bubbles of bound states wrapping the whole internal space. maximizing in $\theta$, we end up almost calibrating the internal space with respect to $(\mathrm{Im}\e^{-B}\wedge\Phi_+)_6$, just like in the $\mathbb{CP}^3$ case. The resulting stability ratio thus enters the criterion \eqref{eq:stab}. We again maximize in both world-volume fluxes. This time the pure spinor constructed out of the standard $J$ suffices to reach small enough stability ratios for most solutions.

\begin{figure}[!ht]
    \centering
    \includegraphics[width=0.6\linewidth]{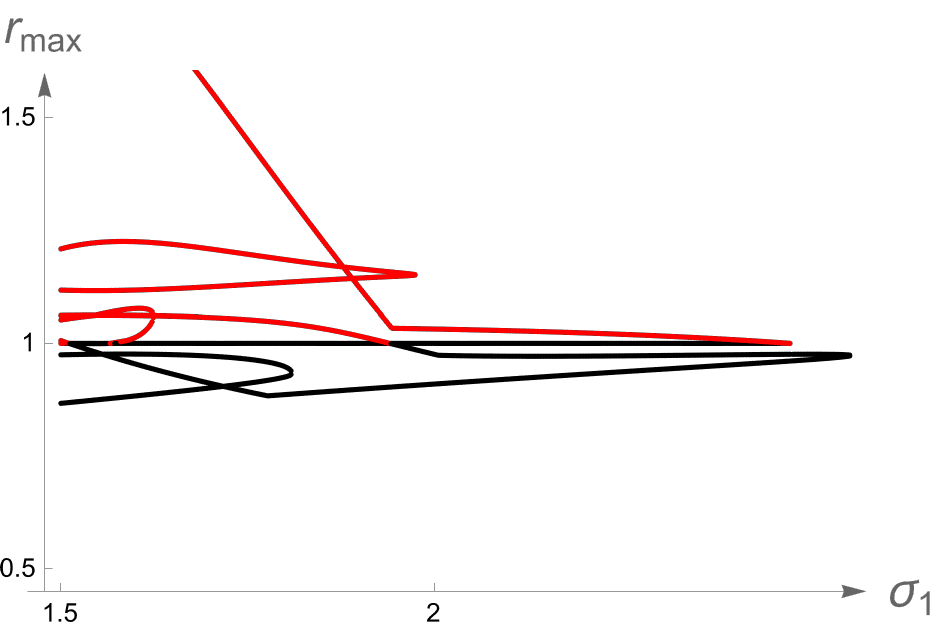}
    \caption{\small Combined stability ratio for bound states, for the supersymmetric solution and various non-supersymmetric ones, with $\sigma_2=\frac32$. The red line
indicates the inconclusiveness of our criteria to determine whether or not the given solution is stable under the nucleation and expansion of bound states,
and the black lines signify stability against all $2\pi f_\mathrm{ws}= f_1(j_1 -j_2)+f_2(j_1-j_3)$ bound states.}
    \label{fig:BSflag}
\end{figure}

\subparagraph{Combining bound states} Let us once again bring the bound states results together and define a combined stability ratio $r_{\text{max}}=\text{max}\{|r_{\text{D}2}|,|r_{\text{D}4}|,|r_{\text{D}6}|,|r_{\text{D}8}|\}$
for each solution at a given value of $\sigma_1$ and $\sigma_2$. Fig.~\ref{fig:BSflag} illustrates the combined stability analysis.

\begin{figure}[!ht]
    \centering
    \includegraphics[width=0.5\linewidth]{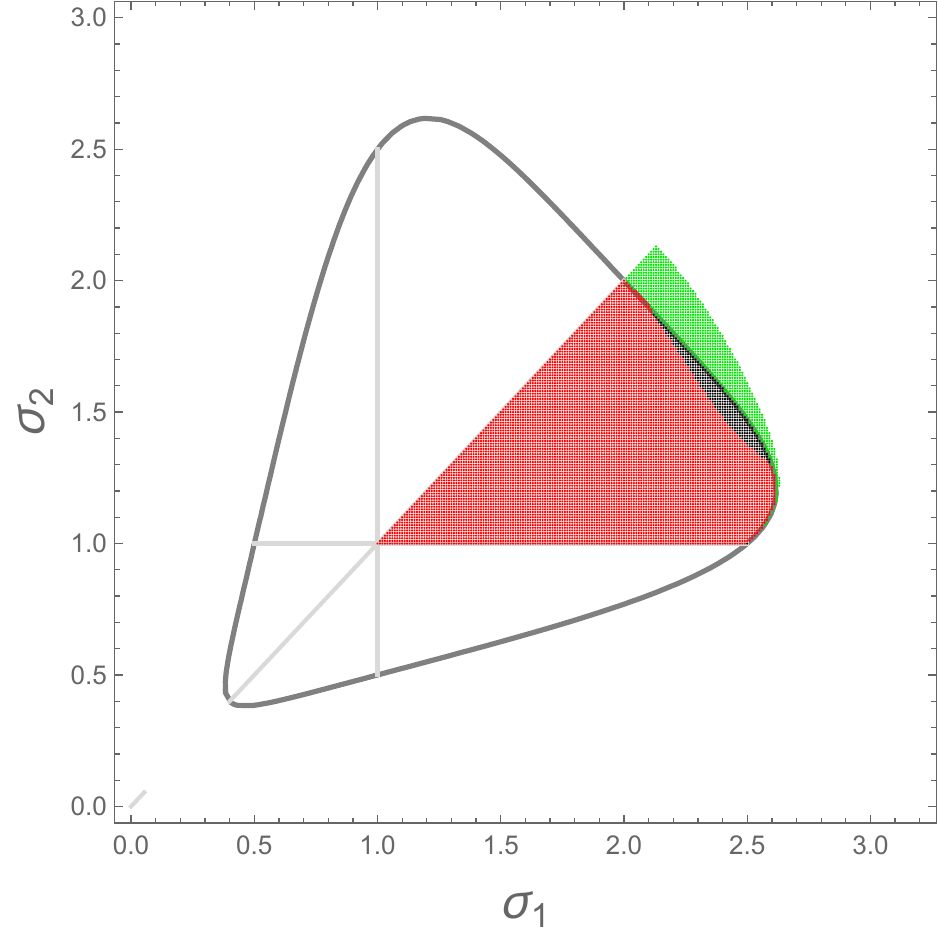}
    \caption{\small Different regions of the $\sigma_1\sigma_2$ plane: the inside of the grey triangle-like contour corresponds to the region admitting supersymmetric solutions. The sub-regions delimited by light grey lines are equivalent up to the symmetries \eqref{symflag}. We therefore analyse only the fundamental region $\{ \sigma_1\geq \sigma_2, \sigma_2\geq 1 \}$. Along the light grey lines lie the twistor solutions.
    The red region admits at least one unstable non-supersymmetric solution; the black one admits at least one solution with inconclusive stability results; in the green one, all solutions are stable against the bound states with world-sheet fluxes $2\pi f_\mathrm{ws}= f_1(j_1 -j_2)+f_2(j_1-j_3)$.}
    \label{fig:combplane}
\end{figure}

Fig.~\ref{fig:BSflag} shows that some non-supersymmetric solutions remain stable under the nucleation and expansion of any $2\pi f_\mathrm{ws}= f_1(j_1 -j_2)+f_2(j_1-j_3)$ bound states.

One could refine our results by studying further the stability properties of the solutions having a stability ratio larger than one, for which we didn't reach a conclusion. We don't pursue this further as here we were mainly interested in using our stability criteria to show that some solutions are actually stable against the $2\pi f_\mathrm{ws}= f_1(j_1 -j_2)+f_2(j_1-j_3)$ bound-state bubbles.

The $r_\text{max}=1$ line corresponds to the supersymmetric solutions. From Fig.~\ref{fig:BSflag} we see that the region of the $\sigma_1$ interval admitting supersymmetric solutions coincides precisely with the one where there is at least one unstable/inconclusive solution, since one of these $\mathcal{N}=0$ solution merges with the supersymmetric branch at its extremity. Going beyond the $\sigma_2=\frac32$ slice, this is true in the whole $\sigma_1$, $\sigma_2$ plane, as we can see in Fig.~\ref{fig:combplane}.

\subsection{\texorpdfstring{AdS$_4 \times {\rm KE}_6$ solutions}{AdS4 x KE6 solutions}}\label{sub:ke6}

Non-supersymmetric solutions on K\"ahler--Einstein manifolds $\mathrm{KE}_6$ were already considered in \cite{Gaiotto:2009mv,Lust:2009zb}, and even partially in \cite{Romans:1985tz}. This is one of the first examples where bubble stability was addressed \cite[Sec.~4.1.2]{Gaiotto:2009mv}, but only for D2-branes and near the supersymmetric limit in parameter space, for reasons related to the AdS/CFT correspondence. In that limit, a conclusion of instability was reached.
Here we will greatly expand that analysis, finding possible islands of stability near the skew-whiffed solutions (already mentioned in Sec.~\ref{sub:stab-nonsusy}). 

There exist many examples of K\"ahler--Einstein manifolds. Some famous ones are $\mathbb{CP}^3$ and the $d<5$ Fermat equations $\{ \sum_{I=1}^5 Z_I^d=0\}\subset \mathbb{CP}^4$, and the flag manifold ${\mathbb F}(1,2;3)$. The problem has famously been reduced \cite{yau-Kstabconj,tian-Kstabconj,chen-donaldson-sun,tian-Kstab} to the notion of K-stability. This has been used to great effect in recent years; see e.g.~\cite[Sec.~6]{calabi-fano} for a large set of examples.

\subsubsection{The solutions}

We normalize the internal curvature such as
\begin{equation}\label{eq:ke-ricci}
	R_{mn}= \frac 8{R^2} g_{mn}\,.
\end{equation}
The fluxes are taken to be proportional to the K\"ahler form: 
\begin{equation}
	F_2 = f_2 J \, ,\qquad F_4 = f_4 \frac{J^2}2 \, ,\qquad F_6 = f_6 \frac{J^3}6 \,.
\end{equation} 
We take $H=0$, as there is no natural three-form in this general class of manifolds. Two cases where such a three-form exists are $\mathbb{CP}^3$ and $\mathbb{F}(1,2;3)$, which were considered in Sec.~\ref{sub:tw}. As we saw, those spaces admit a richer space of solutions, generically not K\"ahler. $\mathbb{CP}^3$ admits a K\"ahler structure for $\sigma=2$, while $\mathbb{F}(1,2;3)$ does when two of the $\alpha_a^2$ are half of the third.

Considering no warping, $A=0$, the equations of motion (\ref{eq:4d-eom}) reduce to
\begin{subequations}\label{eq:ke-eom}
\begin{align}
	 -\frac4{g_s^2 R^2} &= -f_0^2 -f_2^2 + f_4^2 + f_6^2 \,,\\
	\frac{32}{g_s^2 R^2} &= - f_0^2 - f_2^2 + f_4^2 + f_6^2\,,\\
	\label{eq:ke-dil}
	 0&= 5f_0^2 +9 f_2^2 + 3 f_4^2 - f_6^2\,,  \\
	\label{eq:ke-d*h}
	0 &= f_0 f_2 + 2 f_2 f_4 + f_4 f_6\,. 
\end{align}
\end{subequations}
Following \cite{Gaiotto:2009mv}, we can solve (\ref{eq:ke-d*h}) for $f_4=-f_0 f_2 /(f_6+2 f_2)$, as (\ref{eq:ke-dil}) ensures that $f_6\neq -2f_2$. (\ref{eq:ke-dil}) then determines $f_0$. The remaining two give $R$ and $\Lambda$. We can keep $f_6$, $R$, and $z\equiv f_2/f_6$\footnote{This is the inverse of the quantity by the same name introduced in \cite{Gaiotto:2009mv}, and turns out to be more convenient.} as free parameters, and give everything else in terms of them:
\begin{equation}
\begin{split}
	f_0 = &\frac{(1-9z^2)(1+2z)^2}{5+20z+23z^2} \, ,\qquad
	f_2= z f_6 \, ,\qquad 
	f_4 = -\frac{z \sqrt{1-9z^2}}{5+20z+23z^2} \, ;\\
	& g_s^2 = 8 \frac{5+20z+23z^2}{R^2 (1+z)^4} \, ,\qquad L\equiv \sqrt{-\Lambda/3} = \frac R2\,.
\end{split}
\end{equation}
$L$ is the usual AdS radius. We see that $z\in[-1/3,1/3]$. Given a solution, there are three more obtained by reversing the sign of $f_0$ and that of $f_6$. We can thus assume $f_0$, $f_6>0$ without loss of generality.

For $z=1/3$, $f_0=F_0=0$, and the solution can be uplifted to eleven-dimensional supergravity; $f_4=0$ and $f_2$ becomes part of the internal metric, so the only flux left is $G_7$, signaling a Freund--Rubin solution AdS$_4\times M_7$. It turns out that $M_7$ in this case is a Sasaki--Einstein and the solution is supersymmetric. For $z=-1/3$, once again $f_0=f_4=0$, but the other fluxes have the opposite sign; so this time in eleven dimensions we obtain the skew-whiffed solution, which is non-supersymmetric. 

Flux quantization proceeds similarly to the procedure outlined for the twistor solutions in Sec.~\ref{ssub:ads-tw-sol}. The main difference is the possible presence of many two- and four-cycles. Given our normalization \eqref{eq:ke-ricci}, $J= \frac\pi4 c_a {\rm PD}(B_{2a})/R^2$, where $c_a\in \mathbb{Z}$ are the components of the first Chern class and ${\rm PD}(B_{2a})$ are the Poincar\'e duals of the two-cycles. $B$ is closed. Flux quantization is satisfied if $n_2\equiv g_2 f_2^b R^2/8$, $n_4\equiv g_4 f_4^b \frac{R^4}{2^{16}\pi}$, $n_6\equiv g_6 f_6^b \frac{R^6}{3\cdot 2^{12}\pi^2}$ are integers, where $g_2 ={\rm gcd}(c_a)$, $g_4={\rm gcd}(D^{abc}c_b c_c)$, $g_6=D^{abc} c_a c_b c_c/6$, $D^{abc}$ is the triple intersection form, and $(f^b)_k =(\e^b f)_k$. In terms of the invariants introduced in Sec.~\ref{ssub:ads-tw-sol}, $I_4= f_6^2 \tilde I_4(z) \frac{R^4}{64}$, $I_6= f_6^3 \tilde I_6 \frac{R^6}{2^9}$.  $I_4^3/I_6^2$ is now a (complicated) function of $z$ and determines it; the remaining parameters are fixed by $f_6^2= \frac{5+20z+23z^2}{(1-9z)^2(1+2z)^2} \frac{n_0^2}{4\pi^2}$, $R^2 = 4\pi \frac{f_0 \sqrt{I_4}}{n_0 f_6\sqrt{\tilde I_4}}$, $g_s^2= \frac{32\pi^2}{n_0^2 R^2}\frac{(1-9z^2)(1+2z)^2}{(1+z)^4}$, $b=\frac{f_2}{f_0} - \frac{16\pi n_2}{g_2 R^2 n_0}$. We will soon see that the region $z\sim -1/3$ is important for stability; in that limit, and taking for simplicity $n_4=0$, $n_0>0$, $n_2 n_6 <0$,
\begin{equation}
    \delta z\equiv z+\frac13 \sim  -\frac{8 g_2^3}{81 g_6}\frac{n_0^2 n_6}{n_2^3} \,,\qquad
    R^4\sim \frac{-512\pi^2 g_2 n_6}{3g_6 n_2} \,,\qquad
    g_s^2 \sim  \pi\frac{g_2^3}{g_6}\sqrt{-\frac{2n_6}{3n_2^5}} \,,\qquad
    f_6 \sim \frac{n_0}{\pi\sqrt{3} \delta z} \,.
\end{equation}

\subsubsection{Stability: bubbles of simple branes}

We now study non-perturbative stability using the methods in Sec.~\ref{sec:stab}. We take 
\begin{equation}
	\Phi_+= \e^{\ii \theta}\e^{-\ii J}\,.
\end{equation}
On a $2k$-cycle $B_{2k}$, $\mathrm{Im} \Phi_+$ is proportional to $J^k$ times a trigonometric function. We can maximize in $\theta$ independently for each $k$. This leads to $\max_\theta (\mathrm{Im} \Phi_+)_{2k}= J^k/k!$. The form $J$ and its powers are closed and therefore proper calibrations.

Locally we can complete $\Phi_+$ to a compatible pair of pure spinors by taking $\Phi_-=\Omega$. In general this is not globally defined as a form, but only as a section of the trivial bundle $\Lambda^{3,0}\otimes K^{-1}$, with $K$ the canonical bundle. Applying \eqref{kappanonsusyads} now reduces to \emph{Wirtinger's theorem} in K\"ahler geometry: for a holomorphic cycle $B_{2k}$, 
\begin{equation}\label{eq:wirtinger}
    {\rm vol}(B_{2k}) = \left|\int_{B_{2k}}\frac{J^k}{k!}\right|\,.   
\end{equation}
For homology classes that have a holomorphic representative, our criterion (\ref{eq:stab}) is then both necessary and sufficient for stability. Not all homology classes have this property.

\begin{figure}
    \centering
    \includegraphics[width=0.7\linewidth]{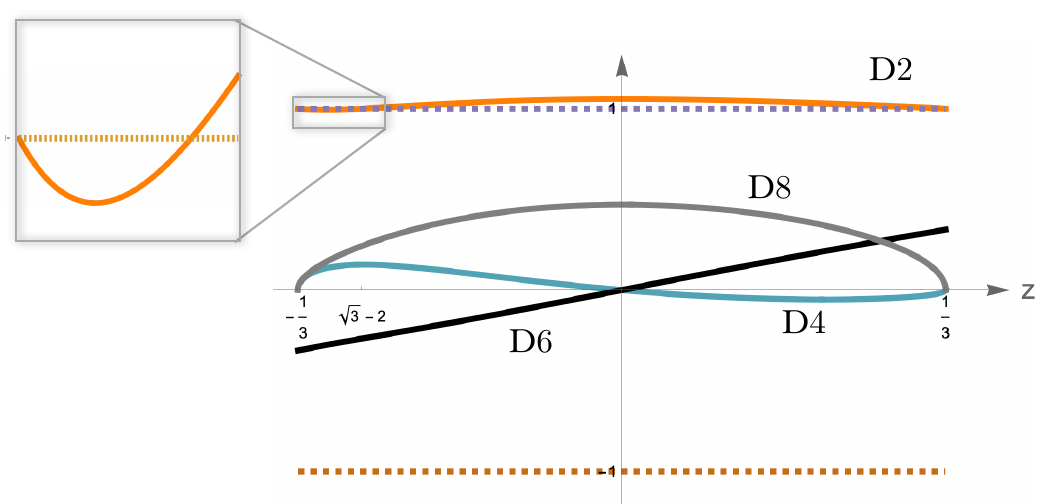}
    \caption{\small The stability ratio \eqref{eq:ke-stab} for various D$p$-branes.\protect\footnotemark}
    \label{fig:ke-stab}
\end{figure}

When we integrate $*\lambda f$ and Im$\Phi_+$ over the cycle, a factor $\int_{B_{2k}} J^k$ appears both in the numerator and denominator of (\ref{eq:stab}), and cancels out. This leads to evaluating the quantities
\begin{equation}\label{eq:ke-stab}
	\frac{g_s R}6 f_{8-p}
\end{equation}
for the stability of a D$p$-brane bubble. These functions are plotted in Fig.~\ref{fig:ke-stab}. Their absolute values are $\le 1$, except for D2-branes, which have that property only for\footnote{As mentioned earlier, the D2 case was analyzed in \cite{Gaiotto:2009mv}, but only for $z\ge 0$. This was presumably motivated by that paper's focus on the holographic interpretation of these solutions for $\mathrm{KE}_6= \mathbb{CP}^3$, which was only justified near the supersymmetric case, $z\sim 1/3$.}
\begin{equation}
	z\in \left[-\frac13,-2+\sqrt{3}\right]\,.
\end{equation}
(Recall that $z=-1/3$ corresponds to the skew-whiffed solution.)

\footnotetext{Here we didn't take the absolute value of the stability ratio like we usually do, for aesthetic reasons.}

\subsubsection{Stability: bubbles of bound states}

We now repeat the previous analysis with a non-zero ${\mathcal F}= B+2\pi f_\mathrm{ws}$. When $f_\mathrm{ws}\neq 0$, these represent bound states.

\subparagraph{D4.}

On a two-cycle $B_2$, $h^2(B_2,\mathbb{R})=1$: all closed two-forms are proportional up to exact forms, so we can just take ${\mathcal F}=f J$. $\mathrm{Im}(\e^{- {\mathcal F}} \wedge \Phi_+)_2 = - (\sin \theta f+ \cos \theta) J$; maximizing this in $\theta$ we obtain $\sqrt{1+f^2}J$. Once again $\int_{B_2} J$ simplifies in the stability ratio, which becomes
\begin{equation}\label{eq:ke-rd4d2}
	r_{\mathrm{D4}}= \frac{g_s R}6 \frac{f_4 + f f_6}{\sqrt{1+f^2}}\,.
\end{equation} 
Both $f$ and the $f_k$ are restricted by flux quantization, but at large $R$ their values cover almost all possible real values. In any case, let us look now for the worst case scenario for stability, and maximize (\ref{eq:ke-rd4d2}) in $f$. This yields 
\begin{equation}\label{eq:ke-d4inst}
	\max_{f\in \mathbb{R}} r_{\mathrm{D4}}= \frac{g_s R}6 \sqrt{f_4^2+f_6^2}\,.
\end{equation}
This function turns out to be $\ge 1$. In particular, near $z=-1/3$, the function in (\ref{eq:ke-d4inst}) is $\sim 1+\frac{27}4 \delta z^2$, with $\delta z \equiv z+1/3$. So stability would appear to be violated for any $z>-1/3$. 

\begin{figure}
    \centering \includegraphics[width=0.5\linewidth]{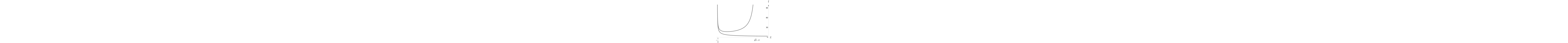}
    \caption{\small The window of instability ratio for D4 bound states diverges as $z\to -1/3$, the skew-whiffed limit.}
    \label{fig:D4D2-KE}
\end{figure}

However, it is instructive to investigate the values of $f$ for which this happens. This is the region between the two curves in Fig.~\ref{fig:D4D2-KE}. We see that it diverges near $z=-1/3$; asymptotically, 
\begin{equation}\label{eq:f-window-D4D2-KE}
	f_- < f < f_+ \, ,\qquad f_\pm \sim \frac2{\sqrt{3 \delta z}}\pm 1\,.
\end{equation}
In other words, for $z\sim -1/3$, the window of instability is located at very large values of $f$. In this regime, it is not clear whether we should trust the leading order brane action we have been using. (See \cite{Tseytlin:1999dj} for a review of corrections to brane actions.) So it is not clear that this window represents an actual instability, when $z$ is very close to the skew-whiffed limit. 

\subparagraph{D6.}

A holomorphic four-cycle $B_4$ is simply-connected, by the Lefschetz hyperplane theorem (see e.g.~\cite[Sec.~1.2]{griffiths-harris}). However, $H^2(B_4, \mathbb{R})$ can be larger than that of the $\mathrm{KE}_6$ itself.\footnote{For a $B_4 \subset \mathbb{CP}^3$ of degree $d$, the Chern classes are $c_1=(4-d)\omega$, $c_2=(6-4d+d^2) \omega^2$, with $\omega$ the hyperplane class in $H^2(\mathbb{CP}^3)$. Now the dimension $h_2(B_4)=-2+\chi(B_4)= -2+\int_{B_4} c_2  = -2+d(6-4d +d^2)$. On $H^2$, the intersection form has signature $(\frac{h_2-\sigma}2,\frac{h_2+\sigma}2)$, $\sigma= \frac13\int_{B_4}(c_1^2-2 c_2)=\frac13 d (4 -d^2)$, but on $H^{1,1}$ it has signature $(1,h^{1,1}-1)$. As a famous example, for $d=4$ we have the quartic K3; it has $h_2=22$, signature $(3,19)$ on $H^2$, and $(1,19)$ on $H^{1,1}$.} Given ${\mathcal F} \in H^2(B_4, \mathbb{R})$, let us introduce $k_1$, $k_2$ such that $\int_{B_4} {\mathcal F} \wedge {\mathcal F} \equiv k_2 f^2\int_{B_4}J^2$ and $\int_{B_4} {\mathcal F} \wedge J= k_1 f \int_{B_4} J^2$. Due to \eqref{eq:F20=0}, $f_{\rm ws}$ is $(1,1)$ (and in particular belongs to the Picard lattice $H^{1,1}(B_4)\cap H^2(B_4,\mathbb{Z})$). By the \emph{signature theorem} \cite[(IV.2.13)]{barth-peters-vandeven}, the intersection pairing has signature $(1,h^{1,1}-1)$ on $H^{1,1}$. Applying this to our definitions, one obtains 
\begin{equation}\label{eq:k2<k1}
    k_2\le k_1^2\,.
\end{equation}

Now $\int\mathrm{Im}(\e^{- {\mathcal F}} \wedge \Phi_+)_4=(\sin\theta (k_2 f^2-1)+ 2\cos \theta f)\int_{B_4} J^2$. Maximizing the coefficient in $\theta$:
\begin{equation}\label{eq:ke-rd6d4}
	r_{\mathrm{D6}}= \frac{g_s R}6 \frac{f_2 + 2 k_1 f f_4 + k_2 f^2 f_6}{((k_2 f^2-1)^2 + 4 k_1^2 f^2)^{1/2}}\,.
\end{equation}
Similarly to (\ref{eq:f-window-D4D2-KE}), we find that instabilities would be present in a window $f_- < f < f_+$, where
\begin{equation}
	\frac19(2+3f_\pm^2(3k_1^2-k_2))-\frac{k_1}{2\sqrt3}(-1+3 k_2f_\pm^2) f\sqrt{\delta z} \sim 0 \,.
\end{equation}
By \eqref{eq:k2<k1}, again there are no finite solutions in the $\delta z \to 0$ limit: rather, $f_\pm$ diverge in that limit. For the particular case $k_1=k_2=1$, $f_\pm \sim \frac{4(\pm 1+\sqrt3)}{3 \sqrt{\delta z}}- \frac98(\pm 23 + 8\sqrt 3) \sqrt{\delta z}$. So it is again not clear whether these instabilities should be trusted.

\subparagraph{D8.}
Now we introduce $k_1$, $k_2$, $k_3$ such that $\int_{\mathrm{KE}_6} {\mathcal F}^3 = k_3 f^3 \int_{\mathrm{KE}_6} J^3$, $\int_{\mathrm{KE}_6} {\mathcal F}^2 \wedge J = k_2 f^2 \int_{\mathrm{KE}_6} J^3$, $\int_{\mathrm{KE}_6} {\mathcal F} \wedge J^2 =k_1 f \int_{\mathrm{KE}_6} J^3$. Again the signature theorem implies \eqref{eq:k2<k1}. The stability ratio is
\begin{equation}\label{eq:ke-rd8d6}
	r_{\mathrm{D8}}= \frac{g_s R}6 \frac{f_0 + 3 k_1 f f_4 + 3 k_2 f^2 f_4 + k_3 f^3 f_6}{((1-3k_2 f^2)^2 + (3 k_1 f - k_3 f^3))^{1/2}}\,.
\end{equation}
In the limit $z\to -1/3$, the instability window is $f_-<f<f_+$, with 
\begin{equation}\label{eq:D8-D6-inst}
    1+ (8 k_1^2-6 k_2)f_\pm^2 + (9 k_2^2 -4 k_1 k_3) f_\pm^4 + \sqrt{3}f_\pm(1+3 k_2f_\pm^2)(k_3 f_\pm^2-k_1) \sqrt{\delta z} \sim 0 \,.    
\end{equation}
For ${\mathcal F}=f J$, $k_1=k_2=k_3=1$, the instability window is divergent as $z\to -1/3$, as for the previous bound states. For general $k_i$, taking into account \eqref{eq:k2<k1} we find that the instability window could become finite only if
\begin{equation}\label{eq:k1k3>k2}
    k_1 k_3 > 9k_2^2\,.
\end{equation}
It might be that this never happens, perhaps along the lines of the Khovanskii--Teissier inequalities \cite[Sec.~1.6]{lazarsfeld1}. In any case, \eqref{eq:k1k3>k2} is not an issue when $h^2(\mathrm{KE}_6)=1$, such as for $\mathbb{CP}^3$, since in that case ${\cal F}=f J$ is the only possibility. More generally, \eqref{eq:k1k3>k2} is less likely to be problematic when $H^2(\mathrm{KE}_6)$ is small. When \eqref{eq:k1k3>k2} is satisfied, we cannot tell whether there is an actual instability; this would involve solving the six-dimensional case of (\ref{eq:mmms}), namely
\begin{equation}
    \cos \theta (- {\mathcal F}^3 + 3 {\mathcal F}\wedge J^2) + \sin \theta (3 {\mathcal F}^2 \wedge J - J^3)=0\,.
\end{equation}

This ends our treatment of non-perturbative stability for these solutions. Regarding tachyons, the $\rm{KE}_6=\mathbb{CP}^3$, $z=\pm1/3$ solutions are perturbatively stable; moreover, all masses lie strictly above the BF bound \cite{Nilsson:1984bj}. By continuity, a neighborhood of those two points is then expected to be also perturbatively stable \cite{Gaiotto:2009mv}. It would be interesting to check how generally this property holds for other K\"ahler--Einstein manifolds.

\subsubsection{\texorpdfstring{A generalization: AdS$_4\times (S^2)^3$}{A generalization: AdS4xS2xS2xS2}}

We now briefly discuss a natural generalization of the K\"ahler--Einstein case: we consider $S^2\times S^2 \times S^2$ with a product metric $\dd s^2_6= \sum_{a=1}^3 R_a^2 \dd s^2_{S^2_a}$. Defining $J_a$ to be the volume forms of the three $S^2$, we can take the fluxes to be
\begin{equation}
	F_2 =  f_{2a} J_a \, ,\qquad F_4 =  f_{41} J_2 \wedge J_3+\text{cycl.} \, ,\qquad F_6 = f_6 J_1 \wedge J_2 \wedge J_3 \, ,\qquad H=0\,.
\end{equation}
The equations of motion are very similar to (\ref{eq:ke-eom}), and can be solved in a similar fashion. It is convenient to define $z_a \equiv f_{2a}/f_6$, $m_k\equiv \sum_{a=1}^3 z_a^k$, $p\equiv z_1 z_2 z_3$, and $D\equiv m_2 -6 m_4 + 2 m_6 -12 p + 4 m_2^2 -m_2 m_4 -4 p m_2 + 6 p^2 + 5 (1-m_2+2p)^2$. The solution then reads
\begin{equation}
\begin{split}
	 \frac{F_0}{f_6} &=(1-m_2 +2 p) \sqrt{\frac{1-3m_2}D} \, ,\\
    \frac{f_{41}}{f_6}&= (z_1 (1-z_1^2 + z_2^2 + z_3^2)-2 z_2 z_3) \sqrt{\frac{1-3m_2}D} \, ,\\
	\frac{\Lambda}{g_s^2 f_6^2}&= \frac{m_2-1}2 +\frac{1-3m_2}D (1-m_2 +2p)^2,\, \\
    \frac1{g_s R_1} &= f_6 (1+z_1^2-z_2^2-z_3^2)\sqrt{\frac{(1-z_1^2)^2+(z_2^2-z_3^2)^2}{D}} 
	\,,\\
\end{split}	
\end{equation}
plus the cyclically similar equations for the other $f_{4a}$, $R_a$. The free parameters can be taken to be $f_6$, $g_s$, the $z_a$, plus the three parameters in $B=b_a J_a$. These eight parameters are then fixed by flux quantization. The solution is restricted to the two-ball 
$\{m_2 = \sum_a z_a^2 \le 1/3\}$ in $z_a$ space.

Given the limited physical interest of the $(S^2)^3$ geometry, we don't lay out a comprehensive study of bound-state bubble instability here, we merely investigate how simple bubble decay constrains the two-ball parameter space of solutions.

As in the  K\"ahler--Einstein case, we take
\begin{equation}
	\Phi_+= \e^{\ii \theta}\e^{-\ii J}\,,
\end{equation}
with $J=J_1+J_2+J_3$. Maximizing in $\theta$, the stability ratio for D2 bubbles yields
\be 
|r_{\mathrm{D}2}|\equiv\frac{g_s |f_6|}{\sqrt{-3\Lambda}}=\frac{1}{\sqrt{3}}\left[\frac{1-m_2}2 +\frac{3m_2-1}D (1-m_2 +2p)^2\right]^{-1/2}\,.
\ee
We don't display the expressions of the stability ratios for other simple D$p$ bubbles,\footnote{They can be calculated as before, maximizing in $\theta$.} as they are lengthy and yield no regions of instabilities within the the two-ball parameter space of solutions.

\begin{figure}[!ht]
    \centering
    \includegraphics[width=0.5\linewidth]{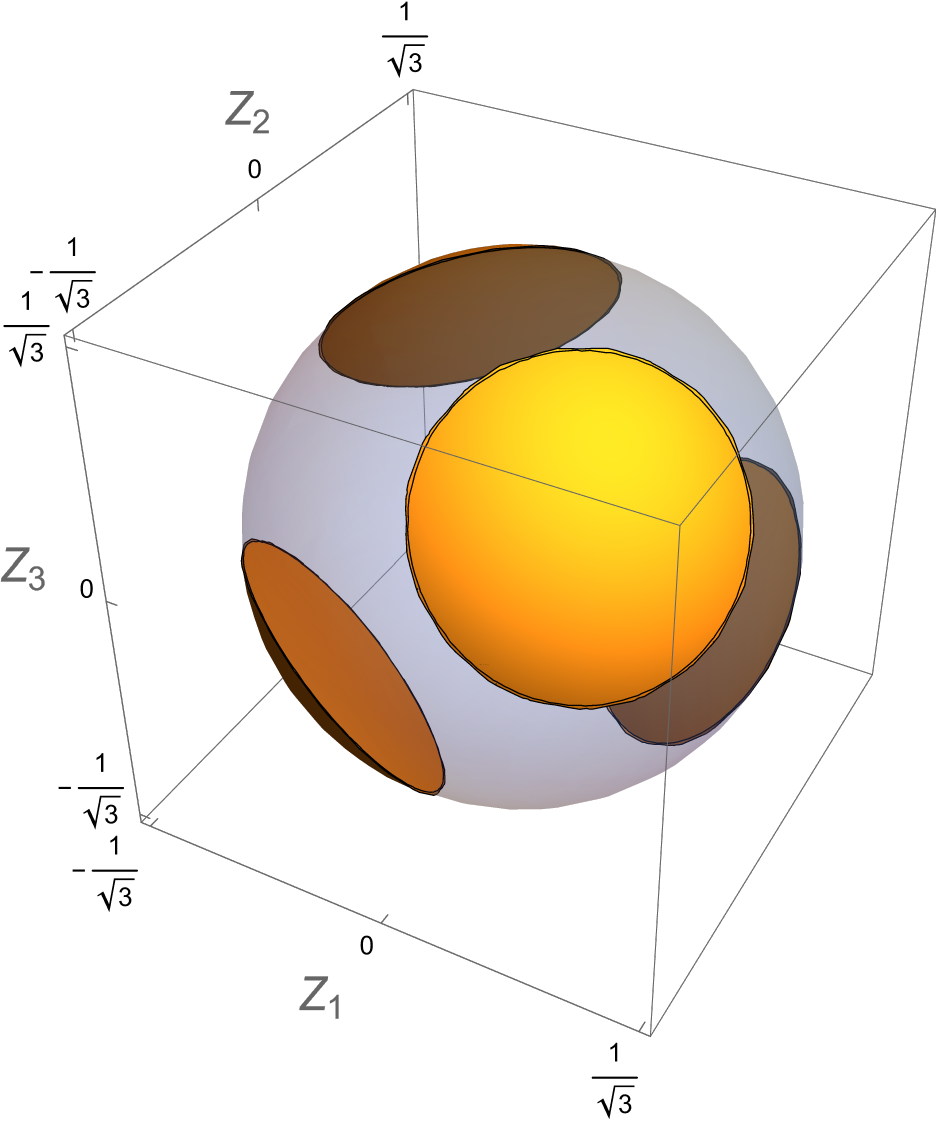}
    \caption{\small Region of stability for all simple D$p$-branes for the AdS$_4\times (S^2)^3$ solution.}
    \label{fig:S2_3}
\end{figure}

Fig.~\ref{fig:S2_3} shows the subregions of stability against simple bubble decay.
This can be thought of as the generalization of the stability result for simple D$p$ branes shown in Fig.~\ref{fig:ke-stab}. Only D2 brane bubbles can decay, and they are stable only near the boundary of the parameter space.

\section{\texorpdfstring{AdS$_5$ vacua}{AdS5 vacua}}

We now consider AdS$_5$ vacua in IIB. The setup is similar to the previous sections: the metric is $\e^{2A}\dd s^2_{\mathrm{AdS}_5} + \dd s^2_{M_5}$, the warping $A$ and the dilaton only depend on the internal coordinates, and the NSNS flux is purely internal. 

\subsection{General formalism}
\label{sub:ads5}

The internal spaces we will consider are $S^1$-fibrations over K\"ahler--Einstein manifolds $\mathrm{KE}_4$ (Sec.~\ref{sub:se}) and over a product of Riemann surfaces (Sec.~\ref{sub:prodRiem}). Some of these solutions were already found in \cite{Romans:1984an}; some are new.

In terms of internal quantities, the IIB equations of motion and Bianchi identities are
\begin{subequations}\label{eq:5d-eom}
\begin{align}
	\label{eq:5d-eom-intgr}&R_{mn}-5 (\nabla_m \partial_n A + \partial_m A \partial_n A)+ 2 \nabla_m \nabla_n \phi  - \frac12 \iota_m H \cdot \iota_n H \\
	&\nonumber \hspace{4cm} =\frac14 \e^{2 \phi}(2\iota_m F \cdot \iota_n F - g_{mn} |F|^2)\,;\\
	\label{eq:5d-eom-extgr}&\e^{-2 A} \Lambda - \nabla^2 A - 5 |\dd A|^2 + 2 \dd A \cdot \dd \phi = -\frac14 \e^{2 \phi} |F|^2 \,;\\
	\label{eq:5d-eom-dil}&2 \nabla^2 \phi - 4|\dd \phi|^2 + 10 \dd A \cdot \dd \phi  + |H|^2 = \frac12 \e^{2 \phi} \sum_k (5-k) |F_k|^2 \,;\\
	\label{eq:5d-eom-H}
	& \dd\left(\e^{5A - 2\phi} \ast H \right) = \e^{5A} \left( F_1 \wedge \ast F_3 + F_3 \ast F_5 \right)\,;\\
	\label{eq:5d-RR}& (\dd - H \wedge) f = 0 = (\dd-H \wedge) (\e^{5A} * \lambda f)\,.	
\end{align}
\end{subequations}

The pure spinor equations can be obtained from (\ref{susy3ads0}) with the usual trick \cite{Gauntlett:2006ux} of rewriting the solution as a Mink$_4\times M_6$ warped product $\e^{2A_4}\dd s^2_{\mathrm{Mink}_4}+ \dd s^2_6$, with $\e^{2A}= \rho^2 \e^{2A_4}$ and ``internal'' metric $\dd s^2_{M_6}= L^2\e^{2A}\frac{\dd r^2}{r^2} + \dd s^2_{M_5}$. Here $\rho$ is the radial coordinate of AdS$_5$, and
\begin{equation}\label{eq:L-AdS5}
	L=\sqrt{-\Lambda/4}
\end{equation}
is its curvature radius. The corresponding pure spinors decomposition reads, in IIB:\footnote{This was done in IIB in a slightly different language \cite{Gabella:2010laf}, and in this language but in IIA \cite{{Apruzzi:2015zna}}.}
\begin{equation}\label{eq:Phi-psi}
	\Phi_1 = L \e^A \frac{\dd \rho}\rho \wedge \psi^1_- + \ii \psi^1_+ \, ,\qquad
	\Phi_2 = L \e^A \frac{\dd \rho}\rho \wedge \psi^2_- + \ii \psi^2_+ \,.
\end{equation}
The forms $\psi^a_\pm$ can also be written as bilinears of the spinors on $M_5$ obtained by reducing those on $M_6$. 
The Minkowski$_4$ pure spinor equations (which we will see in the next section) can be obtained from \eqref{susyads0} by sending $\Lambda\to 0$, $L\to \infty$; together with \eqref{eq:Phi-psi} they now give \cite[(2.15)]{Apruzzi:2015zna}
\begin{subequations}\label{eq:susyads5}
\begin{align}
	\label{eq:susyads5-1}
	\dd_H (\e^{3A- \phi} \mathrm{Re} \psi^2_-) &= -\frac2L \e^{2A-\phi} \mathrm{Im} \psi^2_+\,,\\
		\label{eq:susyads5-2}
	\dd_H (\e^{4A- \phi} \psi^1_+) &= \ii\frac3L \e^{3A-\phi} \psi^1_-\,,\\
		\label{eq:susyads5-3}
	\dd _H(\e^{5A- \phi} \mathrm{Im} \psi^2_-)&=\frac4L \e^{4A-\phi} \mathrm{Re} \psi^2_+ + \e^{5A} * \lambda F \,.
\end{align}
\end{subequations}
Now $F$ is the flux along $M_5$; the total flux is determined as usual via $F_{10}= F + \e^{5A}\mathrm{vol}_{\mathrm{AdS}_5} \wedge *\lambda F$. 

We can now repeat the logic of Sec.~\ref{sec:stab}. The sufficient criterion for stability (\ref{eq:stabgen}) is modified to 
\begin{equation}\label{eq:stab-ratio-AdS5}
	0\leq \left|r_\Sigma \equiv\frac
	{\int_{\Sigma}\e^{5A}\ast\lambda F\wedge \e^{-\mathcal{F}}}
	{\int_{\Sigma}\frac4L\e^{4A-\phi}\text{Re}\psi^2_+|_\Sigma\wedge \e^{-\mathcal{F}}}
	\right|\leq 1
	\quad\Rightarrow\quad\text{stability against bubbles wrapping}\,  (\Sigma,\mathcal{F}).
\end{equation}

Just as for AdS$_4$, all the examples we will consider in this section have constant dilaton and $A=0$, and have no sources. There is a stability result that is common to all such AdS$_5$ solutions; we can point it out already now. (\ref{eq:5d-eom-extgr}), (\ref{eq:L-AdS5}) imply 
\begin{equation}\label{eq:Lf5}
	\frac4{g_s L}= \sqrt{\sum_k |f_k|^2}\ge |f_5|\,.
\end{equation}
The volume of a point is just 1; a point is calibrated if $\mathrm{Re}\psi^2_+=1$ on it. Now for D3-branes the bound (\ref{eq:stab-ratio-AdS5}) reads 
\begin{equation}\label{eq:rD3}
	r_\mathrm{D3}=\frac{f_5}{\sqrt{\sum_k |f_k|^2}} \le 1 \,.
\end{equation}
In other words, D3 bubbles are always stable in this class.

\subsection{Circle bundles over K\"ahler--Einstein}
\label{sub:se}

First we consider $S^1$-fibrations over K\"ahler--Einstein manifolds $\mathrm{KE}_4$. These admit the so-called regular Sasaki--Einstein metrics, which are known to admit supersymmetric Freund--Rubin vacua (see \cite{Sparks:2010sn} and \cite[Sec.~7.4]{book} for reviews), but also non-supersymmetric vacua \cite[Sec.~3]{Romans:1984an}, whose stability we will analyze.

\subsubsection{The fibrations}
\label{ssub:fibr}

A Sasaki--Einstein manifold $\mathrm{SE}_5$ is defined by its cone $C(\mathrm{SE}_5)$ being a Calabi--Yau. One can decompose the $\mathrm{SU}(3)$-structure on the latter in terms of the radial direction $\rho$:
\begin{equation}\label{eq:JO-dec-SE}
	J = \rho \dd \rho \wedge \eta + \rho^2 j \, ,\qquad \Omega = \rho^2 (\dd \rho+ \ii \rho \eta) \wedge \omega\,.
\end{equation}
This gives rise to an $\mathrm{SU}(2)$-structure $(\eta, j, \omega)$ on $\mathrm{SE}_5$. These are respectively a real one-form, a real two-form and a complex two-form such that
\begin{equation}\label{eq:alg-se}
		\eta \cdot j = \eta \cdot \omega = 0 \, ,\qquad \omega^2=0 \, ,\qquad j \wedge \omega = 0 \, ,\qquad \omega \wedge \bar \omega = 2 j^2 \neq 0 \,.
\end{equation}
Closure of $J$ and $\Omega$ on $C(\mathrm{SE}_5)$ then gives the differential conditions
\begin{equation}\label{eq:d-se}
	\dd \eta = 2 j \, ,\qquad \dd \omega = 3\ii\, \eta \wedge \omega\,.
\end{equation}
The metric determined by this $\mathrm{SU}(2)$-structure can be written as $\dd s^2_4 + \eta^2$, where $\dd s^2_4= \sum_{a=1}^4 e^a e^a$ and $j=e^1 \wedge e^3 + e^2 \wedge e^4$, $\omega=(e^1 + \ii e^3)\wedge (e^2 + \ii e^4)$. When the orbits of the vector field dual to $\eta$ are circles of the same size, the $\mathrm{SE}_5$ is said to be \emph{regular}: it is then the total space of a circle bundle over $\dd s^2_4$. If we introduce a coordinate $y\cong y + \Delta y$  on the $S^1$ fiber, we can write $\eta = \dd y + a$. (\ref{eq:d-se}) imply\footnote{The Ricci form $\rho$ is defined as usual by $R_{mn}= \rho_{pn}I^p{}_m$, with $I$ the complex structure.}
\begin{equation}
	\dd a= 2 j \, ,\qquad \rho=6j\,.
\end{equation}
We see that the base metric $\dd s^2_4$ is Einstein. (\ref{eq:d-se}) implies that the form $j$ is horizontal and invariant (Sec.~\ref{ssub:coset}), so it is the pull-back of a form on the base. Moreover we can write $\omega= \e^{-3\ii y}\omega_0$, for $\omega_0$ a $(2,0)$-form on the base. $j$ and $\omega_0$ now show that the base is in fact a K\"ahler--Einstein manifold $\mathrm{KE}_4$. In general $\omega_0$ is not globally well-defined on this base (it is a section of the canonical $K_{\mathrm{KE}_4}$), but $\omega$ is well-defined on $\mathrm{SE}_5$.

The first Chern class $c_1(\mathrm{KE}_4)$ is then represented on a basis of two-cycles $B_{2i}\subset \mathrm{KE}_4$ by  $\frac1{2\pi}\int_{B_{2i} } \rho = n_i\in \mathbb{Z}$. Integrality of $c_1$ of the fibration then gives $\frac1{\Delta y}\int_{B_{2i}}\dd a \in \mathbb{Z}$, which fixes $\Delta y= \frac{2\pi}3 \ell$, where $\ell$ has to divide $\mathrm{gcd}(n_i)$.

K\"ahler--Einstein four-manifolds are classified \cite{Tian:1987if}: they are $\mathbb{CP}^2$, $\mathbb{CP}^1\times \mathbb{CP}^1$, and del Pezzo manifolds $\mathrm{dP}_k$ for $3\le k \le 8$. These are all simply connected. The Gysin sequence
\begin{equation}\label{eq:gysin}
\begin{split}
	0&\to H^1(\mathrm{KE}_4, \mathbb{Z})=0 \to H^1(\mathrm{SE}_5,\mathbb{Z})\to H^0(\mathrm{KE}_4,\mathbb{Z}) = \mathbb{Z} \\
	&\buildrel{c_1\wedge}\over\to H^2(\mathrm{KE}_4,\mathbb{Z})\to H^2(\mathrm{SE}_5,\mathbb{Z})\to H^1(\mathrm{KE}_4,\mathbb{Z})=0\\
	&\to H^3(\mathrm{KE}_4,\mathbb{Z})\to H^3(\mathrm{SE}_5,\mathbb{Z})\to H^2(\mathrm{KE}_4,\mathbb{Z})\to \ldots
\end{split}
\end{equation}
now shows that $H^k(\mathrm{SE}_5,\mathbb{Z})=\{\mathbb{Z},0, \mathbb{Z}^{b-1} \oplus \mathbb{Z}_m , \mathbb{Z}^{b-1} , \mathbb{Z}_m, \mathbb{Z} \}$, for some $m$ and $b=\mathrm{dim}H^2(\mathrm{KE}_4,\mathbb{R})$  \cite[Sec.~4.3]{Morrison:1998cs}. Using Poincar\'e duality and the universal coefficient theorem, we also obtain $H_k(\mathrm{SE}_5,\mathbb{Z})=\{\mathbb{Z}, \mathbb{Z}_m, \mathbb{Z}^{b-1}, \mathbb{Z}^{b-1} \oplus \mathbb{Z}_m  , 0, \mathbb{Z} \}$.

Instead of the ordinary Sasaki--Einstein metric we just described, we are going to consider its variant defined as
\begin{equation}
	\dd s^2_5 = R^2(\dd s^2_\mathrm{KE_4} + \sigma^2 \eta^2)\,,
\end{equation}
where $\sigma$ is a real parameter; one often calls this a \emph{squashed} or \emph{stretched} SE if $\sigma <1 $ or  $>1$ respectively. We have also introduced an overall radius $R$. We will often use $e^5= \sigma\eta$ to complete the vielbein $e^{1,\ldots,4}$ for the $\mathrm{KE}_4$. The Ricci tensor for such a metric reads
\begin{subequations}
\begin{align}
	\label{eq:Ricci-fibration}
	R_{ab}e^a e^b &= \sum_{a=1}^4 \left(R^{\mathrm{KE}_4}_{ab} -\frac{\sigma^2}2 (\dd a)_{ac}(\dd a)_b{}^c\right) e^a e^b +
	\frac{\sigma^2}2 |\dd a|^2 e_5^2\\
	&= (6-2 \sigma^2)\sum_{a=1}^4 e_a^2 - 4\sigma^2 e_5^2\,.
\end{align}
\end{subequations}
The Einstein case is recovered when $\sigma=1$.

\subsubsection{The solutions}

As we mentioned, we are now going to assume constant dilaton, $A=0$, and no sources. Moreover, it is natural to consider an Ansatz where the fluxes are expanded on the $\mathrm{SU}(2)$-structure forms:\footnote{As we saw in (\ref{eq:susyads5}), we only consider here the part of the RR flux; the total $F_{10}$ is self-dual, and in particular it also includes a component $f_5 \mathrm{vol}_{\mathrm{AdS}_5}$ for the five-form.}
\begin{equation}
	F_1=0 \, ,\qquad F_3 = R^3 e^5 \wedge (f_3 j + \tilde f_3 \mathrm{Re} \omega) \, ,\qquad F_5= f_5 \mathrm{vol}_5= -\frac12 f_5 R^5 j^2 \wedge e^5\,.
\end{equation}
(A possible term in $\mathrm{Im}\omega$ in $F_3$ can be reabsorbed by multiplying $\omega$ by a phase.) The RR equation $\dd *_5 F_3 + H \wedge F_5=0$ then implies $f_3=0$ and 
\begin{equation}\label{eq:H-se}
	H= h R^3 e^5 \wedge \mathrm{Im} \omega\,,
\end{equation}
with $h=-\frac{3 \tilde f_3}{R \sigma f_5}$.\footnote{Looking at this equation alone would seem to suggest another branch with $f_5=\tilde f_3=0$ and undetermined $H$, but this is actually eliminated by (\ref{eq:5d-eom-dil}).} The full set of equations of motion (\ref{eq:5d-eom}) gives
\begin{subequations}\label{eq:se-eom}
\begin{align}
	&\frac{6-2 \sigma^2}{g_s^2 R^2}-\frac{h^2}{2 g_s^2} = \frac14 f_5^2\,,&
	&\frac{4\sigma^2}{g_s^2 R^2}-\frac{h^2}{g_s^2} = \frac12 \tilde f_3^2 + \frac14 f_5^2\,;\\
	&\frac{16}{g_s^2 L^2} = 2 \tilde f_3^2 + f_5^2\,;&
	&\frac{h^2}{g_s^2}= \tilde f_3^2 \,;\\
	&\tilde f_3 \left( R^2 f_5 - \frac9{g_s^2 \sigma^2 f_5}\right)=0\,;&
	\label{eq:se-eom-h}
	&h=-\frac{3 \tilde f_3}{R \sigma f_5},\, \qquad\qquad f_3=0\,.
\end{align}
\end{subequations}
The first in (\ref{eq:se-eom-h}) gives rise to two branches; both survive the remaining equations in (\ref{eq:se-eom}), and eventually read
\begin{equation}
	\sigma= \sqrt{\frac32} \, ,\qquad
	f_3=0 \, ,\qquad \tilde f_3 = \frac{\sqrt{3}s_2}{g_s R}  \, ,\qquad
	 f_5 = \frac{3 s_1}{g_s R \sigma} \, ,\qquad
	 h=-s_1 g_s \tilde f_3\,,\qquad
	(\text{Branch } 1)
\end{equation}
where $s_i\in \{+1,-1\}$ are two signs, and
\begin{equation}
	\sigma=1 \, ,\qquad f_3 = \tilde f_3 =0 \, ,\qquad 
	f_5 = \pm\frac4{g_sR} \, ,\qquad h=0 \, .\qquad
	(\text{Branch } 2)
\end{equation}

Branch 2 consists of the supersymmetric and skew-whiffed solutions (with the minus and plus sign, respectively). Branch 1 breaks supersymmetry with a stretched SE; it was found in a slightly different language in \cite[Sec.~3]{Romans:1984an}. Notice that the case $s_1=1$ has $G_3= F_3-\ii \e^{-\phi} H= \tilde f_3 \omega$.

We also briefly discuss flux quantization. We can take
\begin{equation}\label{eq:B-se}
	B= -\frac13 h \sigma R^3 \mathrm{Re} \omega\,.
\end{equation}
The total internal closed flux is then
\begin{equation}
	\e^{-B} F = R^3 \tilde f_3 e^5 \wedge \mathrm{Re} \omega + R^5\left(-\frac12 g_5 -\frac13 h \sigma R \tilde f_3\right) e^5 \wedge j^2\,.
\end{equation}
The three-form part is exact (by (\ref{eq:d-se})), so its periods are automatically zero. The five-form part $(\e^{-B} F)_5$ is zero on Branch 1. On Branch 2 we get
\begin{equation}
	\mp \frac{ R^4 \sigma}{8 \pi^4 g_s} \mathrm{Vol}_0 = n_5\ \in \mathbb{Z} 
\end{equation}
where $\mathrm{Vol}_0$ is the volume of the original $\mathrm{SE}_5$ at $R=\sigma=1$.

\subsubsection{Stability}

Using (\ref{eq:B-se}) and the value of $h$ below (\ref{eq:H-se}), we compute
\begin{equation}\label{eq:se-eF*lF}
	\e^{- {\mathcal F}} \wedge * \lambda F= \e^{-2\pi f_\mathrm{ws}}\wedge \left(f_5-\frac{R^2 \tilde f_3^2}{2 f_5} j^2 \right)\,.
\end{equation}
By (\ref{eq:d-se}), $j^2$ is exact. 

For the denominator of (\ref{eq:stabgen}), in view of (\ref{eq:JO-dec-SE}) it might look natural to consider $\psi^2_+ = \e^{\ii \theta}\e^{-\ii R^2 j}$. The problem with this is that there would be no choice of $\theta$ such that $\e^{-{\mathcal F}}\wedge\mathrm{Re} \psi^2_+$ would be closed. As we have seen with the AdS$_4$ solutions, the integral of a non-closed quantity would then depend on the representative of the homology class, and thus would be  difficult to bound. For an $\mathrm{SU}(2)$-structure in four dimensions, $j$, $\mathrm{Re} \omega$ and $\mathrm{Im} \omega$ have similar properties, and can be rotated into each other. This inspires the following choice:
\begin{equation}
	\psi^2_+ = \e^{\ii \theta}\e^{-\ii R^2 \mathrm{Re} \omega}\,.
\end{equation}

We have already evaluated the stability ratio (\ref{eq:stab-ratio-AdS5}) for a D3 in (\ref{eq:rD3}), finding stability. For D5 bound states, as we mentioned,
\begin{equation}\label{eq:se-eFpsi2}
	(\e^{-{\mathcal F}}\wedge\mathrm{Re} \psi^2_+)_2= - \cos \theta\, 2\pi f_\mathrm{ws} +R^2 \left(-\cos \theta \frac{\tilde f_3}{f_5} + \sin \theta \right) \mathrm{Re} \omega\,
\end{equation}
is not closed in general; but it is if we choose $\cos \theta= \frac{f_5}{\sqrt{\tilde f_3^2 + f_5^2}}$. Now only the term $f_\mathrm{ws}$ remains in (\ref{eq:se-eFpsi2}). It also appears in (\ref{eq:se-eF*lF}), and the integrals $\int_{B_2} f_\mathrm{ws}$ simplify in the stability ratio (\ref{eq:stab-ratio-AdS5}), giving
\begin{equation}\label{eq:se-rD5}
	r_\mathrm{D5}= \frac{f_5}{\cos\theta\sqrt{2 \tilde f_3^2 + f_5^2}}
	= \frac{\sqrt{\tilde f_3^2 + f_5^2}}{\sqrt{2 \tilde f_3^2 + f_5^2}}\,.
\end{equation}
This is $<1$, so D5 bound states are also stable.

We can carry out a similar computation for D7 bound states; now $\theta$ will be fixed (to the same value as above) by requiring that $f_\mathrm{ws} \wedge \mathrm{Re} \omega$ terms drop out from  $(\e^{-{\mathcal F}}\wedge\mathrm{Re} \psi^2_+)_4$, and we end up with the same result as in (\ref{eq:se-rD5}). However, there are in fact no four-cycles, as we saw below (\ref{eq:gysin}).

\subsection{Circle bundles over product of Riemann surfaces}
\label{sub:prodRiem}

We now consider the total space of an $S^1$-bundle over $\Sigma_1 \times \Sigma_2$, two Riemann surfaces. When the latter have genus zero, we obtain the so-called $T^{p_1,p_2}$-spaces. The $p_1=p_2=1$ case admits a famous Sasaki--Einstein structure related \cite{Candelas:1989js} to the conifold; our attempts in this case have produced solutions that are specializations either of the previous subsection, or of the results below. Thus in the following we  focus on $p_1\neq p_2$. 

\subsubsection{The fibrations}

We introduce two-forms $j_a=\mathrm{vol}^0_{\Sigma_a}$, the volume forms with respect to metrics normalized as $R_{mn}^{0\,\Sigma_a}= \kappa_a g_{mn}^{0\,\Sigma_a}$, with $\kappa_a \in \{-1,0,1\}$. 

We take the bundle connection to be $\eta = \dd y + p_1 a_1 + p_2 a_2$, where $\dd a_a = j_a$. 
Notice that for $p_1\neq p_2$ we cannot introduce an analogue of the $\omega$ of the previous subsection.\footnote{In Sec.~\ref{ssub:fibr}, $\mathrm{SE}_5$ was basically the circle bundle inside the canonical $K_{\mathrm{KE}_5}$; $\omega_0$ was a section of this bundle, and $\e^{p\ii y}$ makes it well-defined globally exactly for $p=3$. Writing a $(1,0)$-form $h_a$ on each $\Sigma_a$, the $(2,0)$-form $\omega_0 = h_1 \wedge h_2$ can be similarly repaired on each $\Sigma_a$ with a different choice of exponent, but not on both simultaneously unless $p_1=p_2$.}
Gauss--Bonnet gives $\int_{\Sigma_a} j_a = 2\pi \chi_a \equiv 4\pi (1-g_a)$ for $g_a\neq 1$ (and no constraint for $g_a=1$). Quantization of $c_1$ gives $\frac{p_a}{\Delta y} \int_{\Sigma_a} j_a \in \mathbb{Z}$.

We consider the metric
\begin{equation}
	\dd s^2 = R^2 \left( \dd s^2_{0\,\Sigma_1} + \alpha^2 \dd s^2_{0\,\Sigma_2} + \sigma^2 \eta^2\right)\,.
\end{equation}
Its Ricci tensor can be obtained from (\ref{eq:Ricci-fibration}). 
 The role of the two $\Sigma_a$ can be exchanged by implementing
\begin{equation}\label{eq:prod-exch}
	R\to R \alpha  \, ,\qquad \alpha\to \frac1 \alpha \, ,\qquad \sigma\to \frac\sigma \alpha \, ,\qquad \kappa_1 \leftrightarrow \kappa_2 \, ,\qquad p_1 \leftrightarrow p_2\,.
\end{equation}

\subsubsection{The solutions}

We assume the form fields to be 
\begin{equation}\label{eq:prod-flux}
\begin{split}
	&F_1 =0 \, ,\qquad F_3 = R^3 (f_{31} e^5 \wedge j_1 + f_{32} \alpha^2 e^5 \wedge j_2)\,, \\
	&F_5= f_5 \mathrm{vol}_5 \, ,\qquad
	H= R^3 (h_1 e^5 \wedge j_1 + h_2 \alpha^2 e^5 \wedge j_2)\,.
\end{split}
\end{equation}
The Bianchi identity $\dd H=0$ imposes right away that
\begin{equation}
	p_2 h_1 = - p_1 h_2 \alpha^2\,.
\end{equation}

The equations of motion give
\begin{subequations}
\begin{align}
	&\frac1{g_s R^2}\left(\kappa_1-\frac{p_1^2\sigma^2}2 \right)-\frac{h_1^2}{2g_s^2} = \frac14(f_{31}^2 - f_{32}^2+f_5^2) \,,&
	& \frac{16}{g_s^2 L^2} = f_{31}^2 + f_{32}^2 + f_5^2\,,\\
	 &\frac1{g_s R^2}\left(\kappa_2-\frac{p_2^2\sigma^2}{2 \alpha^4} \right)-\frac{h_2^2}{2g_s^2} 
	 = \frac14(-f_{31}^2 + f_{32}^2+f_5^2)\,,& 
	 &p_2 f_{31}= - p_1 f_{32} \alpha^2 \,, \\
	 &\frac{\sigma^2}{2 g_s^2 R^2}\left(p_1^2 + \frac{p_2^2}{\alpha^4}\right) -\frac{h_1^2+h_2^2}{2 g_s^2} = \frac14 ( f_{31}^2 + f_{32}^2 + f_5^2) \,,&
	 &\frac{h_1^2+h_2^2}{g_s^2} = f_{31}^2 + f_{32}^2\,,\\
	 \label{eq:prod-f5fh} &f_5 f_{31}= f_5 f_{32}= 0\,, & &f_5 h_1 = f_5 h_2 =0 \,.
\end{align}	
\end{subequations}

We obtain the solution
\begin{equation}
\begin{array}{l}
	f_{31}=f_{32}=h_1=h_2=0 \, ,\qquad \kappa_1 = \kappa_2=1 \, ,\\[5pt] \sigma^2=\frac{2 \alpha^2}{p_1^2 \alpha^4+2p_2^2}\, ,\qquad
	f_5^2=\frac{4(p_1^2 \alpha^4+p_2^2)}{g_s^2 R^2 \alpha^2 (p_1^2 \alpha^4 + 2 p_2^2)}\,,\\[10pt]
	p_1^2 \alpha^4 (\alpha^2-2)+p_2^2(2\alpha^2-1)=0\,.
\end{array}
\end{equation}
$\alpha$ is determined implicitly by the last equation. These were found by Romans \cite[Sec.~2]{Romans:1984an}: they are of Freund--Rubin type, with $M_5$  Einstein (but not Sasaki). 

Flux quantization gives 
\begin{equation}
	\frac{\Delta y R^4}{3\pi^2 g_s}\chi_1 \chi_2 \in \mathbb{Z} \,,
\end{equation}
where $\chi_a= \chi(\Sigma_a)=(2-2g_a)$. 

\subsubsection{Stability}

These solutions were found to be perturbatively unstable \cite{Gubser:2001zr}, so our bubble stability analysis will be brief. 

We take $\psi^2_+= \e^{\ii \theta}\e^{-\ii J}$, with $J= R^2(j_1 \pm\alpha^2 j_2)$. D3 stability was considered in (\ref{eq:rD3}). Note that $f_5$ is the only non-zero RR field, so $r_\mathrm{D3}=1$ and the D3 is only marginally stable. This is of course at danger of becoming unstable once string corrections are taken into account. 

For D5 bound states, we take $2\pi f_\mathrm{ws}= f(-q_2 j_1 + q_1 j_2)$. For the generic case $p_a\neq 0$, again by the Gysin sequence (\ref{eq:gysin}) there is only one two-cycle $B_2\subset M_5$; its Poincar\'e dual is proportional to $\eta \wedge (p_1 j_1 - p_2 j_2)$, which is closed but not exact.  Now
\begin{align}
	\int_{B_2} \frac4L \e^{4A - \phi} \e^{- {\mathcal F}}\wedge \mathrm{Re} \psi^2_+ &= |f_5|\int_{B_2} (- \cos \theta {\mathcal F} + \sin \theta J) \\
	\nonumber
	&= |f_5|(-\cos \theta f p_a q_a + \sin \theta R^2 (-p_2\pm p_1 \alpha^2))\int_{M_5}\eta \wedge j_1 \wedge j_2\,.
\end{align}
On the other hand
\begin{equation}
	\int_{B_2} \e^{-\mathcal F} \wedge * \lambda F= - f f_5 \int_{B_2} (-q_2 j_1+q_1 j_2)=-f f_5 p_a q_a \int_{M_5}\eta \wedge j_1 \wedge j_2\,.
\end{equation}
Maximizing in $\theta$, we find
\begin{equation}
	r_\mathrm{D5}= \frac{2 f p_a q_a}{\sqrt{(p_a q_a)^2 f^2 + R^4 (\mp p_1 \alpha^2 +p_2)^2}}\,.
\end{equation}
which is always $<1$. 

Just as in the SE solutions of the previous subsection, there are no four-cycles in this geometry, and hence no D7 instabilities.

\section{The Minkowski case}
\label{sec:mink}

Throughout this paper, we took advantage of the fact that the WZ action of domain-wall D$p$-branes can be reformulated in terms of a corresponding calibration form.
When the cosmological constant vanishes, it is the WZ action of space-filling D$p$-branes that can be reformulated in a similar fashion, yielding stability results for such branes. We will therefore no longer discuss the possibility of non-perturbative decay through branes, but the stability of background space-filling D$p$-branes themselves.

\subsection{Space-filling (anti) \texorpdfstring{D$p$-brane}{Dp-brane} stability}\label{mink}

\subparagraph{Supersymmetric case.} We specialise the discussion to the case of a four-dimensional Minkowski external space: $X_4=\text{Mink}_4$. $\mathcal{N}=1$ supersymmetric solutions satisfy the pure spinor equations
\begin{subequations}\label{susymink}
\begin{align}
  \dd(\text{e}^{3A-\phi}\text{e}^{-B}\Phi_2)&= 0\label{susy1}\\ 
  \dd(\e^{2A-\phi}\text{e}^{-B}\text{Re}\Phi_1)&= 0\label{susy2}\\
   \dd(\e^{4A-\phi}\text{e}^{-B}\text{Im}\Phi_1)&= \e^{4A}\text{e}^{-B}\ast\lambda F\label{susy3}\,.
\end{align}
\end{subequations}
These are the $\Lambda=0$, $L\to\infty$ limit of \eqref{susyads0}; we mentioned earlier in \eqref{closecalib} that they can be recovered as closure of calibrations.  
We locally pick a gauge for the RR potentials such that
\be \e^{4A-\phi}\text{Im}\Phi_1=\Tilde{C}\label{susy3pot},\ee
with $\dd(\text{e}^{-B}\Tilde{C})=\e^{4A}\text{e}^{-B}\ast\lambda F$. The choice \eqref{susy3pot} is therefore consistent with the supersymmetry condition \eqref{susy3}.

Let us now consider a space-filling D$p$-brane wrapping an internal cycle $\Sigma$. The D$p$-brane energy per unit of (Euclidean) external volume is
\be E(\Sigma,\mathcal{F})=T_p\int_\Sigma \e^{4A-\phi}\sqrt{\text{det}(g|_\Sigma+\mathcal{F})}\dd^{p-3}\xi-\Tilde{C}|_\Sigma\wedge \e^{-\mathcal{F}}.\ee
Through \eqref{susy3pot}, it can be written as
\begin{align} E(\Sigma,\mathcal{F})&=T_p\int_\Sigma \e^{4A-\phi}(\sqrt{\text{det}(g|_\Sigma+\mathcal{F})}\dd^{p-3}\xi-\text{Im}\Phi_1|_\Sigma\wedge \e^{-\mathcal{F}})\label{EDpsusy}.
\end{align}
On the other hand, we have
\be\left|\text{Im}\Phi_1|_\Sigma\wedge \e^{-\mathcal{F}}\right|_{\text{top}}\leq \sqrt{\text{det}(g|_\Sigma+\mathcal{F})}\dd^{p-3}\xi,\label{ineq}\ee
for every generalized cycle $\Sigma$, and with $_{\text{top}}$ selecting the top-form component on $\Sigma$.

Through \eqref{EDpsusy} and \eqref{ineq}, we therefore have that
\begin{align} E(\Sigma,\mathcal{F})&\geq 0.
\end{align}

Let us now focus on calibrated generalized cycles, the ones saturating the bound \eqref{ineq}:
\begin{align}
    \sqrt{\text{det}(g|_{\Sigma_\text{c}}+\mathcal{F}_\text{c})}\dd^{p-3}\xi&=\left.\text{Im}\Phi_1|_{\Sigma_\text{c}}\wedge \e^{-\mathcal{F}_\text{c}}\right|_{\text{top}}.
    \label{kappa1}
\end{align}
Using \eqref{EDpsusy}, we therefore have
\begin{align}
    E(\Sigma_\text{c},\mathcal{F}_\text{c})&= 0.
\end{align}
We recover here the fact that a space-filling D$p$-brane wrapping a calibrated generalized cycle minimizes its energy within its generalized homology class \cite{Martucci:2005ht}.

 As mentioned above, a D$p$-brane wrapping a calibrated generalized cycle preserves the background supersymmetry.\footnote{Indeed, spelled in the $\kappa$-symmetry formalism, \eqref{kappa1} is simply the supersymmetry condition for a D$p$-brane $\Gamma_\kappa \epsilon = \epsilon$, with $\epsilon^\text{T}=(\eta_1\, \eta_2)$.}

\subparagraph{Non-supersymmetric case.}
Let us now consider non-supersymmetric type II backgrounds respecting the following modified pure spinor equations:
\begin{subequations}
\begin{align}
  \dd(\text{e}^{3A-\phi}\text{e}^{-B}\Phi_2)&= \text{e}^{-B}\Psi \label{nonsusy1}\\ 
  \dd(\e^{2A-\phi}\text{e}^{-B}\text{Re}\Phi_1)&= \text{e}^{-B}\Upsilon\label{nonsusy2}\\
   \dd(\e^{4A-\phi}\text{e}^{-B}\text{Im}\Phi_1)&= \e^{4A}\text{e}^{-B}\ast\lambda F+\dd(\text{e}^{-B}\Xi)\label{nonsusy3},
\end{align}
\end{subequations}
where $\Psi,\ \Upsilon$ and $\Xi$ are supersymmetry breaking forms. We introduce the following local gauge for the RR potentials
\be 
\label{susybr3}
\e^{4A-\phi}\text{Im}\Phi_1-\Tilde{C}=\Xi,
\ee
now consistent with the supersymmetry breaking equation \eqref{nonsusy3}.

    Let us now consider space-filling D$p$-branes and anti D$p$-branes wrapping a generalized cycle $\Sigma$. Their energies per unit of external volume are 
\begin{align}
E(\Sigma,\mathcal{F})=T_p\int_\Sigma \e^{4A-\phi}\sqrt{\text{det}(g|_\Sigma+\mathcal{F})}\dd^{p-3}\xi\mp\Tilde{C}|_\Sigma\wedge \e^{-\mathcal{F}},
\end{align}
and the upper/lower sign is for D-branes/anti D-branes respectively.
Through \eqref{susybr3}, these can now be written as
\begin{align}
E(\Sigma,\mathcal{F})&=T_p\int_\Sigma \e^{4A-\phi}(\sqrt{\text{det}(g|_\Sigma+\mathcal{F})}\dd^{p-3}\xi\mp\text{Im}\Phi_1|_\Sigma\wedge \e^{-\mathcal{F}})\pm\Xi|_\Sigma\wedge \e^{-\mathcal{F}}\label{Enonsusy}
\end{align}
Let us focus on the generalized cycles saturating the bound \eqref{ineq} in two distinct ways:
\begin{subequations}
\begin{align}
    \sqrt{\text{det}(g|_{\Sigma_{\Tilde{\text{c}}}}+\mathcal{F}_{\Tilde{\text{c}}})}\dd^{p-3}\xi&=\left.\text{Im}\Phi_1|_{\Sigma_{\Tilde{\text{c}}}}\wedge \e^{-\mathcal{F}_{\Tilde{\text{c}}}}\right|_{\text{top}}\label{kappa1nonsusy}\\
    \sqrt{\text{det}(g|_{\Sigma_{\Tilde{\text{ac}}}}+\mathcal{F}_{\Tilde{\text{ac}}})}\dd^{p-3}\xi&=-\left.\text{Im}\Phi_1|_{\Sigma_{\Tilde{\text{ac}}}}\wedge \e^{-\mathcal{F}_{\Tilde{\text{ac}}}}\right|_{\text{top}}\label{kappa2nonsusy}
\end{align}
\end{subequations}
We call the generalized cycles satisfying \eqref{kappa1nonsusy} {\it almost calibrated}, and the ones satisfying \eqref{kappa2nonsusy} {\it almost anti-calibrated} ---hence the corresponding subscripts $_{\Tilde{\text{c}}}$ and $_{\Tilde{\text{ac}}}$.

The energies of a (anti) D$p$-brane wrapping an almost (anti-)calibrated cycle are thus
\begin{subequations}
\begin{alignat}{2}
\text{D-brane:}& \qquad\quad E(\Sigma_{\Tilde{\text{c}}},\mathcal{F}_{\Tilde{\text{c}}})&&=T_p\int_{\Sigma_{\Tilde{\text{c}}}}\Xi|_{\Sigma_{\Tilde{\text{c}}}}\wedge \e^{-\mathcal{F}_{\Tilde{\text{c}}}}\label{Enonsusycalib}\\
\text{anti D-brane:}&\qquad E(\Sigma_{\Tilde{\text{ac}}},\mathcal{F}_{\Tilde{\text{ac}}})&&=-T_p\int_{\Sigma_{\Tilde{\text{ac}}}}\Xi|_{\Sigma_{\Tilde{\text{ac}}}}\wedge \e^{-\mathcal{F}_{\Tilde{\text{ac}}}}.
\end{alignat}
\end{subequations}
Let us focus on D$p$-branes only and consider a D$p$-brane wrapping a generalized cycle $(\Sigma',\mathcal{F}')$ in the generalized homology class of an almost calibrated generalized cycle $(\Sigma_{\Tilde{\text{c}}},\mathcal{F}_{\Tilde{\text{c}}})$. This means that there is a generalized cycle $(\Tilde{\Sigma},\Tilde{\mathcal{F}})$ such that $\partial \Tilde{\Sigma}=\Sigma'-\Sigma_{\Tilde{\text{c}}}$ and $\Tilde{\mathcal{F}}|_{\Sigma_{\Tilde{\text{c}}}}=\mathcal{F}_{\Tilde{\text{c}}}$, $\Tilde{\mathcal{F}}|_{\Sigma'}=\mathcal{F}'$.

The energy of such a D$p$-brane is then
\begin{subequations}
\begin{align} E(\Sigma',\mathcal{F}')\geq T_p\int_{\Sigma'}\Xi|_{\Sigma'}\wedge \e^{-\mathcal{F}'}&=T_p\int_{\Sigma_{\Tilde{\text{c}}}}\Xi|_{\Sigma_{\Tilde{\text{c}}}}\wedge \e^{-\mathcal{F}_{\Tilde{\text{c}}}}+T_p\int_{\partial\Tilde\Sigma}\Xi|_{\partial\Tilde\Sigma}\wedge \e^{-\Tilde{\mathcal{F}}}\\
&= E(\Sigma_{\Tilde{\text{c}}},\mathcal{F}_{\Tilde{\text{c}}})+T_p\int_{\Tilde\Sigma}\dd(\Xi\wedge \e^{-\Tilde{\mathcal{F}}}),
\end{align}
\end{subequations}
where we have used \eqref{ineq}, \eqref{Enonsusy}, \eqref{Enonsusycalib}, and Stokes' theorem.
This entails
\begin{align} E(\Sigma',\mathcal{F}')-E(\Sigma_{\Tilde{\text{c}}},\mathcal{F}_{\Tilde{\text{c}}})\geq T_p\int_{\Tilde\Sigma}\dd(\Xi\wedge \e^{-\Tilde{\mathcal{F}}}).\label{boundE}
\end{align}
Similarly, for anti D$p$-branes in the generalized homology class of an anti D$p$-brane wrapping an almost anti-calibrated cycle $\Sigma_{\Tilde{\text{ac}}}$, we get
\begin{align} E(\Sigma',\mathcal{F}')-E(\Sigma_{\Tilde{\text{ac}}},\mathcal{F}_{\Tilde{\text{ac}}})\geq -T_p\int_{\Tilde\Sigma}\dd(\Xi\wedge \e^{-\Tilde{\mathcal{F}}}).\label{boundEanti}
\end{align}
We could therefore conclude that an (anti) D$p$-brane wrapping an almost (anti-)calibrated generalized cycle is energy minimizing within its generalized homology class if
\begin{align}
     \hspace{-1.1cm}\text{D$p$-branes:}\, \int_{\Tilde\Sigma}\dd(\Xi\wedge \e^{-\Tilde{\mathcal{F}}})&\geq 0\qquad \forall\, (\Sigma',\mathcal{F}'),\, (\Tilde{\Sigma},\Tilde{\mathcal{F}})\, \, \text{such that}\begin{cases} & \partial\Tilde{\Sigma}=\Sigma'-\Sigma_{\Tilde{\text{c}}}\\ & \Tilde{\mathcal{F}}|_{\Sigma_{\Tilde{\text{c}}}}=\mathcal{F}_{\Tilde{\text{c}}},\, \Tilde{\mathcal{F}}|_{\Sigma'}=\mathcal{F}'\end{cases}\label{lastboundmink}\\
     \hspace{-1.1cm}\text{anti D}p\text{-brane:}\,  \int_{\Tilde\Sigma}\dd(\Xi\wedge \e^{-\Tilde{\mathcal{F}}})&\leq 0\qquad \forall\, (\Sigma',\mathcal{F}'),\, (\Tilde{\Sigma},\Tilde{\mathcal{F}})\, \, \text{such that}\begin{cases} & \partial\Tilde{\Sigma}=\Sigma'-\Sigma_{\Tilde{\text{ac}}}\\ & \Tilde{\mathcal{F}}|_{\Sigma_{\Tilde{\text{ac}}}}=\mathcal{F}_{\Tilde{\text{ac}}},\, \Tilde{\mathcal{F}}|_{\Sigma'}=\mathcal{F}'\end{cases}\label{lastboundmink2}
\end{align}
From the point of view of calibration theory, the supersymmetry breaking term $\dd(\e^{-B}\Xi)$ entering \eqref{nonsusy3} prevents the pure spinor $\e^{4A-\phi}\e^{-B}\text{Im}\Phi_1$ to be a generalized calibration for space-filling D$p$-branes. In general, it therefore spoils the stability of D$p$-branes wrapping almost calibrated generalized cycles. The above bound now gives a criterion for the supersymmetry breaking term and the almost (anti-)calibrated generalized cycle to preserve the classical stability of a (anti) D$p$-brane wrapping said cycle.

However, failure to meet this criterion doesn't necessarily mean that the said (anti) D$p$-brane is unstable.

Regarding this criterion, a particularly simple class of supersymmetry breaking-vacua is the one where $\Psi\neq0$, $\Upsilon\neq0$, and $\Xi=0$. For such backgrounds, almost calibrated space-filling D$p$-branes are stable, as they automatically satisfy \eqref{lastboundmink}. There are several examples of such backgrounds in the literature \cite{Lust:2008zd,Legramandi:2019ulq,Menet:2023rnt,Menet:2023rml}.

One of the strengths of this formalism is the ability to tune $\Xi$ to meet this criterion via the gauge freedom of the RR potentials, as we will illustrate in the next section. Indeed, we will do so and use this criterion to demonstrate the stability of the space-filling D6-branes present in some non-supersymmetric type IIA solutions of \cite{Macpherson:2024frt}.
\subsection{A \texorpdfstring{Minkowski$_4$ application}{A Minkowski4 application}}
In this section, we evaluate the stability criterion constructed in \ref{mink} for the space-filling D-branes present in some solutions of \cite{Macpherson:2024frt}. 
The authors of this paper construct AdS$_5$ solutions of type IIA supergravity by dimensionally reducing in various ways the Gaiotto--Maldacena (GM) class of $\mathcal{N}=2$ AdS$_5$ solutions of eleven-dimensional supergravity \cite{Gaiotto:2009gz}, itself obtained by imposing a ${\rm U}(1)$ isometry on the general classification.

\subparagraph{The supersymmetric solutions.} The 
$\mathcal{N}=2$ supersymmetric reduction to IIA of the GM class is:
\be
\dd s^2 = f_1^{\frac{3}{2}} f_5^{\frac{1}{2}}\left[
4\dd s^2_{\text{AdS}_5} +f_2 \dd s^2_{S^2} + f_4(\dd\sigma^2 + \dd\eta^2) + f_3 \dd \chi^2
\right]\label{susysolmmn}\ee
\be \e^{\frac{4}{3} \phi} = f_1 f_5 \qquad
H = \dd f_8  \wedge \text{vol}(S^2)\qquad
C_1=f_6 \dd\chi \qquad
C_3 = f_7\, \dd\chi \wedge \text{vol}_{S^2}\nonumber,
\ee
with vol$_{S^2}=\sin{\theta}\dd\theta\wedge\dd\phi$, and the functions $f_i$ all depending on a single function $V=V(\sigma,\eta)$:
\be
f_1 = \kappa^{\frac{2}{3}} \left(\frac{\dot{V}\Tilde{\Delta}}{2V''}\right)^{\frac{1}{3}}\qquad
f_2 = \frac{2V'' \dot{V}}{\tilde{\Delta}}\qquad
f_3 = \frac{4\sigma^2}{\Lambda} \qquad
f_4 = \frac{2V''}{\dot{V}} \qquad
f_5 = \frac{2\Lambda V''}{\dot{V}\tilde{\Delta}}\nonumber \ee
\be f_6 = \frac{2\dot{V} \dot{V}'}{V'' \Lambda} \qquad
f_7 = -\frac{4\kappa \dot{V}^2 V''}{\tilde{\Delta}} \qquad
f_8 = 2\kappa \left( \frac{\dot{V} \dot{V}'}{\tilde{\Delta}} - \eta \right)\ee
\be\tilde{\Delta} = \Lambda (V'')^2 + (\dot{V}')^2, \quad \Lambda = \frac{2\dot{V} - \ddot{V}}{V''},
\nonumber\ee
where $V$ should satisfy the Laplace equation in cylindrical coordinates:
\be \label{laplace}
\ddot{V} + \sigma^2 V'' = 0\,,
\qquad
\dot{V} \equiv \sigma \, \partial_\sigma V, \quad V' \equiv \partial_\eta V\,.\ee
The boundary conditions 
\be \dot{V}|_{\eta=0,P}=0,\qquad\dot{V}|_{\sigma=0}=\mathcal{R}(\eta),\label{bc}\ee
ensure that the metric remains regular with the shrinking of the two-sphere \cite{Gaiotto:2009gz}, with $\eta$ belonging to a finite interval $[0,P]$. $\mathcal{R}(\eta)$ is a function related to the M-theory orbifold singularities, and is highly constrained by flux quantization. See \cite{Macpherson:2024frt,Gaiotto:2009gz,Aharony:2012tz} for detailed discussions.

\cite[Sec.~2.3]{Macpherson:2024frt} parameterizes the piecewise-linear function as
\begin{equation}
\mathcal{R}(\eta) = 
\begin{cases}
N_1 \eta & \eta \in [0,1] \\
N_k + (N_{k+1} - N_k)(\eta - k) & \eta \in [k, k+1] \\
N_{P-1}(P - \eta) & \eta \in [P-1, P]\,.
\end{cases}
\end{equation}
The solution to the Laplace equation \eqref{laplace} respecting the boundary conditions \eqref{bc} is
\begin{equation}
V(\sigma, \eta) = - \sum_{n=1}^{\infty} \mathcal{R}_n \sin\left(\frac{n \pi}{P} \eta \right) K_0\left( \frac{n \pi}{P} \sigma \right)\,,\label{psol}
\end{equation}
where $K_0$ is a modified Bessel function of the second kind, and
\begin{equation}
\mathcal{R}_n = \frac{2}{P} \int_0^P \mathcal{R}(\eta) \sin\left( \frac{n \pi}{P} \eta \right) d\eta 
= \frac{2P}{(n \pi)^2} \sum_{k=1}^P b_k \sin\left( \frac{n \pi k}{P} \right)\,,\quad 
b_k = 2N_k - N_{k+1} - N_{k-1}.
\end{equation}
This ansatz divides the $\eta$ direction into $P$ unit cells with $k=0,...,P-1$, and we define the $B$ field in the cell with $\eta\in[k,k+1]$ as \be B = (2\kappa k + f_8) \text{vol}_{S^2}\,,\ee
which means that a large gauge transformation $B_2\rightarrow B_2+2\pi\kappa\, \text{vol}_{S^2}$ is performed when traversing between cells in the direction increasing $\eta$.

We will be mainly interested in the behaviour of this solution in the neighbourhood of the region $\{\sigma=0,\eta=k\}$, for reasons that will soon become clear. We expand around this region as $ \sigma = r \sin \alpha,\eta = k - r \cos \alpha$ for $r$ small, yielding \cite{Macpherson:2024frt}
\be
\dot{V} = N_k, \quad V'' = \frac{b_k}{2r}, \quad \dot{V}' = \frac{b_k}{2} \left( 1 + \cos \alpha \right) + N_{k+1} - N_k.\label{nearD}
\ee
To leading order, the solution tends to
\begin{align} 
\frac{\dd s^2}{2\kappa \sqrt{N_k}} = \frac{1}{\sqrt{\frac{b_k}{r}}} \left( 4 \dd s^2 _{\text{AdS}_5} + \dd s^2_{S^2} \right) + \frac{\sqrt{\frac{b_k}{r}}}{N_k}\left( \dd r^2 + r^2 \dd s^2_{\tilde{S}^2} \right)\,,\qquad
\e^{-\phi} = \left( \frac{N_k b_k^3}{k 2^6 \kappa^2 r^3} \right)^{\frac{1}{4}},
\end{align}
from which we can see that there are D6-branes wrapping AdS$_5\times S^2$ at the locus $\{\sigma=0,\eta=k\}$. In this context, $b_k$ is the charge of the D6-branes \cite{Macpherson:2024frt}. $\tilde{S}^2$ is spanned by $(\alpha,\chi)$.

\subparagraph{The non-supersymmetric solutions.} We now focus on a non-supersymmetric class of solutions presented in \cite{Macpherson:2024frt} and preserving an SU$(2)\times$U$(1)$ isometry. It corresponds to a different dimensional reduction of the GM class than the one used for the supersymmetric solution, and it is parametrised by a supersymmetry breaking parameter $\xi$:
\be
\begin{split} 
&\dd s^2 = f_1^{\frac{3}{2}} f_5^{\frac{1}{2}}\sqrt{\Delta}\left[
4\dd s^2_{\text{AdS}_5} +f_2 \dd s^2_{S^2} + f_4(\dd \sigma^2 + \dd \eta^2) + \frac{f_3}{\Delta} \dd \chi^2\,,
\right]\\
\label{nonsusysolmmn}
\e^{\frac{4}{3} \phi} &= f_1 f_5 \Delta\qquad 
\Delta = \left(1 + \xi f_6 \right)^2 + \frac{\xi^2 f_3}{f_5}\,, \qquad
H = \dd(f_8 + \xi f_7) \wedge \text{vol}_{S^2}\,, \\
&C_1= \left( f_6 + \xi \left( \frac{f_3}{f_5} + f_6^2 \right) \right) \dd\chi\,, \qquad
C_3 = f_7\, \dd\chi \wedge \text{vol}_{S^2}\,.
\end{split}
\ee
(See \cite{Macpherson:2024frt} for details and for the discussion of the precise dimensional reduction.)
This class of vacua is a parametric deformation of the supersymmetric class of solutions \eqref{susysolmmn}.

    We will focus here on the solutions satisfying \eqref{psol}, and thus \eqref{nearD} in the vicinity of $\{\sigma=0,\eta=k\}$. We now take the $B$ field in the $k$-th cell to be \be B=(2\kappa k+f_8 + \xi f_7) \wedge \text{vol}_{S^2}.\label{B0mmn}\ee
We will be specifically interested in the properties of the stack of space-filling D6-branes, once supersymmetry is broken.
\vspace{0.2cm}

For starters, in the vicinity of $\{\sigma=0,\eta=k\}$, the solution now tends to \cite{Macpherson:2024frt}
\begin{equation}
\begin{split} 
\hspace{-0.4cm}\frac{\dd s^2}{2\kappa \sqrt{N_k}} &= \sqrt{\Delta_k}\left[\frac{1}{\sqrt{\frac{b_k}{r}}} \left( 4 \dd s^2_{\text{AdS}_5} + \dd s^2_{S^2} \right) + \frac{\sqrt{\frac{b_k}{r}}}{N_k}\left( \dd r^2 + r^2 \left(\dd\alpha^2+\frac{\sin{\alpha}^2}{\Delta_k}\dd\chi^2\right) \right)\right]\\
\hspace{-0.4cm}\e^{-\phi} &= \left(\frac{N_k b_k^3}{k 2^6 \kappa^2 r^3} \right)^{\frac{1}{4}}\Delta_k^{-\frac{3}{4}},
\end{split}
\end{equation}
at leading order in $r$, with
\begin{equation}
\hspace{-.2cm}\Delta_k = \frac{1}{4} \xi^2 b_k^2 \sin^2 \alpha + \left(1 + \xi g(\alpha)\right)^2, \quad 
g(\alpha) = \cos^2\left(\frac{\alpha}{2}\right)(N_k - N_{k-1}) + \sin^2\left(\frac{\alpha}{2}\right)(N_{k+1} - N_k).
\end{equation}
This metric suggests that the stack of D6-branes extended along AdS$_5\times S^2$ at the $\{\sigma=0,\eta=k\}$ locus now backreacts on a cone with a $\mathbb{WCP}^1_{[l_{k-1},l_{k}]}$ spindle base, spanned by $(\alpha,\chi)$, with $l_{k-1}=\Delta_k(\alpha=0)$, $l_{k}=\Delta_k(\alpha=\pi)$. See \cite{Macpherson:2024frt} for a detailed discussion on the spindle, including how it modifies the D6-branes charge into $\frac{b_k}{l_kl_{k-1}}$.

Regarding the potential stability of this stack of D6-branes, in \cite{Macpherson:2024frt} a particular gauge is discussed for the RR potential such that the action of a probe D6-brane placed at the locus of the stack vanishes: $S_{\text{DBI}}+S_\text{WZ}=0$, suggesting that the contracting and expanding forces felt by the probe brane cancel each other out.

\subparagraph{Stability from pure spinors.} We now turn to the pure spinor formalism to investigate this further, in particular considering these $\text{AdS}_5$ solutions as Mink$_4$ solutions as in Sec.~\ref{sub:ads5}; in particular,
\be \dd s^2_{\text{AdS}_5} = \e^{2\rho}\dd s^2_{\text{Mink}_4} + \dd\rho^2,\ee
with the ten-dimensional metric and fluxes decompositions \eqref{10dmet} and \eqref{RR}.
The radial AdS direction is therefore now understood as being part of the six-dimensional internal space.

We introduce the following pure spinors, defined in terms of an internal SU$(2)$ structure $(z,j,\omega)$
\be \Phi_+=\omega\wedge \e^{\frac{1}{2}z\wedge\bar{z}}\qquad\Phi_-=\ii z\wedge \e^{-\ii j},\label{spinorsmmn}\ee
while we write the SU$(2)$ structure via a complex vielbein $(E_1,E_2,E_3)$ as 
\be \omega=E_1\wedge E_2\qquad j=\frac{\ii}{2}(E_1\wedge\bar{E}_1+E_2\wedge\bar{E}_2)\qquad z=E_3.\ee

Starting with the class of supersymmetric solutions \eqref{susysolmmn}, we introduce the following complex vielbein
\begin{subequations}
\begin{align}
E_1 &= 
\frac{2\, \e^{\ii \phi}\, \kappa^2 \sigma \cos\theta\, \dot{V}}{f_1^{9/4} \sqrt{f_3} f_5^{1/4}}\, \mathrm{d}\theta
+ \frac{4\, \e^{\ii \phi}\, \kappa^2 \sigma \sin\theta\, \dot{V}}{f_1^{9/4} \sqrt{f_3} f_5^{1/4}}\, \mathrm{d}\rho + \frac{2\ii\, \e^{\ii \phi}\, \kappa^2 \sigma \sin\theta\, \dot{V}}{f_1^{9/4} \sqrt{f_3} f_5^{1/4}}\, \mathrm{d}\phi\nonumber\\
&\quad + \frac{2\, \e^{\ii \phi}\, \kappa^2 \sigma \sin\theta\, \dot{V}'}{f_1^{9/4} \sqrt{f_3} f_5^{1/4}}\, \mathrm{d}\eta - \frac{2\, \e^{\ii \phi}\, \kappa^2 \sigma^2 \sin\theta\, V''}{f_1^{9/4} \sqrt{f_3} f_5^{1/4}}\, \mathrm{d}\sigma \label{E1}\\
E_2 &= 
- \e^{\ii \chi} f_1^{3/4} \sqrt{f_3} f_5^{1/4}\, \mathrm{d}\rho
- \ii\, \e^{\ii \chi} f_1^{3/4} \sqrt{f_3} f_5^{1/4}\, \mathrm{d}\chi
- \frac{\e^{\ii \chi} f_1^{3/4} \sqrt{f_3} f_5^{1/4}}{\sigma}\, \mathrm{d}\sigma \label{E2}\\
E_3 &= 
- \frac{\ii\, f_5^{1/4} \kappa \sin\theta\, \dot{V}}{f_1^{3/4}}\, \mathrm{d}\theta 
+ \frac{\kappa  \left(\ii f_1^{3/2} \sqrt{f_5} f_6 \cos\theta + 2 \kappa \, \dot{V}\right) \dot{V}'}{f_1^{9/4} f_5^{1/4} f_6}\, \mathrm{d}\eta\nonumber\\
&\quad- \frac{2\, \kappa \left(-\ii f_1^{3/2} f_4 \sqrt{f_5} \cos\theta\, \dot{V} + 2 \kappa\, \dot{V}'\right)}{f_1^{9/4} f_4 f_5^{1/4}}\, \mathrm{d}\rho + \frac{\kappa \sigma \left(\kappa \, \dot{V}' - \ii f_1^{3/2} \sqrt{f_5} \cos\theta\, V''\right)}{f_1^{9/4} f_5^{1/4}}\, \mathrm{d}\sigma.\label{E3}
\end{align}
\end{subequations}
The pure spinors \eqref{spinorsmmn} constructed out of the above vielbein preserve supersymmetry: they solve the pure spinor equations \eqref{susymink}.
The cycle $\Sigma=(\rho,S^2)$, with vol$_\Sigma=\sin{\theta}\dd\rho\wedge\dd\theta\wedge\dd\phi$ is calibrated \eqref{kappa1}
precisely at the locus $\{\sigma=0,\eta=k\}$ \cite{Macpherson:2024frt}. The stack of space-filling D6-branes wrapping $\Sigma$ at $\{\sigma=0,\eta=k\}$ is therefore BPS.

Let us now move to the case of the supersymmetry-breaking class of solutions \eqref{nonsusysolmmn}. One can then define the corresponding following non-supersymmetric vielbein 
\begin{subequations}\label{eq:Ei-nonsusy}
\begin{align}
E_1 &= 
\frac{2 \e^{\ii \phi} \kappa^2 \sigma \cos\theta\, \Delta^{1/4} \dot{V}}{f_1^{9/4} \sqrt{f_3} f_5^{1/4}}\, \mathrm{d}\theta  + \frac{4 \e^{\ii \phi} \kappa^2 \sigma \sin\theta\, \Delta^{1/4} \dot{V}}{f_1^{9/4} \sqrt{f_3} f_5^{1/4}}\, \mathrm{d}\rho+ \frac{2i \e^{\ii \phi} \kappa^2 \sigma \sin\theta\, \Delta^{1/4} \dot{V}}{f_1^{9/4} \sqrt{f_3} f_5^{1/4}}\, \mathrm{d}\phi \notag \\
&\quad + \frac{2 \e^{\ii \phi} \kappa^2 \sigma \sin\theta\, \Delta^{1/4} \dot{V}'}{f_1^{9/4} \sqrt{f_3} f_5^{1/4}}\, \mathrm{d}\eta - \frac{2 \e^{\ii \phi} \kappa^2 \sigma^2 \sin\theta\, \Delta^{1/4} V''}{f_1^{9/4} \sqrt{f_3} f_5^{1/4}}\, \mathrm{d}\sigma\,, \label{E10} \\[1em]
E_2 &=- \e^{\ii \chi} f_1^{3/4} \sqrt{f_3} f_5^{1/4} \Delta^{1/4}\, \mathrm{d}\rho- \frac{i \e^{\ii \chi} f_1^{3/4} \sqrt{f_3} f_5^{1/4}}{\Delta^{1/4}}\, \mathrm{d}\chi- \frac{\e^{\ii \chi} f_1^{3/4} \sqrt{f_3} f_5^{1/4} \Delta^{1/4}}{\sigma}\, \mathrm{d}\sigma \,,\label{E20} \\[1em]
\label{E30}E_3 &= 
- \frac{\ii f_5^{1/4} \kappa\sin\theta\, \Delta^{1/4} \dot{V}}{f_1^{3/4}}\, \mathrm{d}\theta + \frac{\kappa \left( \ii f_1^{3/2} \sqrt{f_5} f_6 \cos\theta + 2 \kappa \dot{V} \right) \dot{V}' \Delta^{1/4}}{f_1^{9/4} f_5^{1/4} f_6}\, \mathrm{d}\eta \\
&\quad + \frac{2\ii \kappa \left( f_1^{3/2} f_4 \sqrt{f_5} \cos\theta \dot{V} + 2i \kappa \dot{V}' \right) \Delta^{1/4}}{f_1^{9/4} f_4 f_5^{1/4}}\, \mathrm{d}\rho \notag
+ \frac{\kappa \sigma \left( \kappa  \dot{V}' - \ii f_1^{3/2} \sqrt{f_5} \cos\theta V'' \right) \Delta^{1/4}}{f_1^{9/4} f_5^{1/4}}\mathrm{d}\sigma \,.
\end{align}
\end{subequations}
This is a parametric deformation in $\xi$ of the supersymmetric vielbein \eqref{E1}, \eqref{E2} and \eqref{E3}. It is actually merely a global $\Delta^{1/4}$ rescaling of the supersymmetric vielbein, with the exception of the $\chi$ direction in $E_2$. The pure spinors defined with this vielbein now breaks all the supersymmetry  conditions \eqref{susymink}. 

At this stage, these pure spinors characterise the general class of solutions \eqref{nonsusysolmmn}. However, the supersymmetry breaking terms in this setting are lengthy and not particularly insightful. 

We now focus on the solutions \eqref{psol}, and in particular on their behaviour in the region of the internal space close to  $\{\sigma=0,\eta=k\}$. We will assess the potential stability of a probe D6-brane placed at that locus by comparing its energy to that of nearby probe D6-branes, using the bound \eqref{boundE} derived in the previous section.

First of all, given that in this region the geometry of such solutions tends to the near horizon geometry of a stack of D6-branes sitting at $\{\sigma=0,\eta=k\}$, the DBI action for a probe D6-brane vanishes at this locus.\footnote{The world-sheet flux $\mathcal{F}$ contribution to the DBI action also vanishes, given the $B$ field \eqref{B0mmn}.} Furthermore, using the non-supersymmetric pure spinors defined through the complex vielbein \eqref{eq:Ei-nonsusy}, one can show that the would-be space-filling calibration form $\e^{4A-\phi}\e^{-B}\text{Im}\Phi_-$ also vanishes at the locus $\{\sigma=0,\eta=k\}$. 

The cycle $\Sigma=(\rho,S^2)$ is therefore almost calibrated:
\be  \sqrt{\text{det}(g|_{(\rho,S^2)}+\mathcal{F})}\dd^{p-3}\xi=\left.\text{Im}\Phi_-|_{(\rho,S^2)}\wedge \e^{-\mathcal{F}}\right|_{\text{top}}\ee
precisely at the locus $\{\sigma=0,\eta=k\}$.

\begin{figure}[!ht]
   \centering
\includegraphics[width=0.55\linewidth]{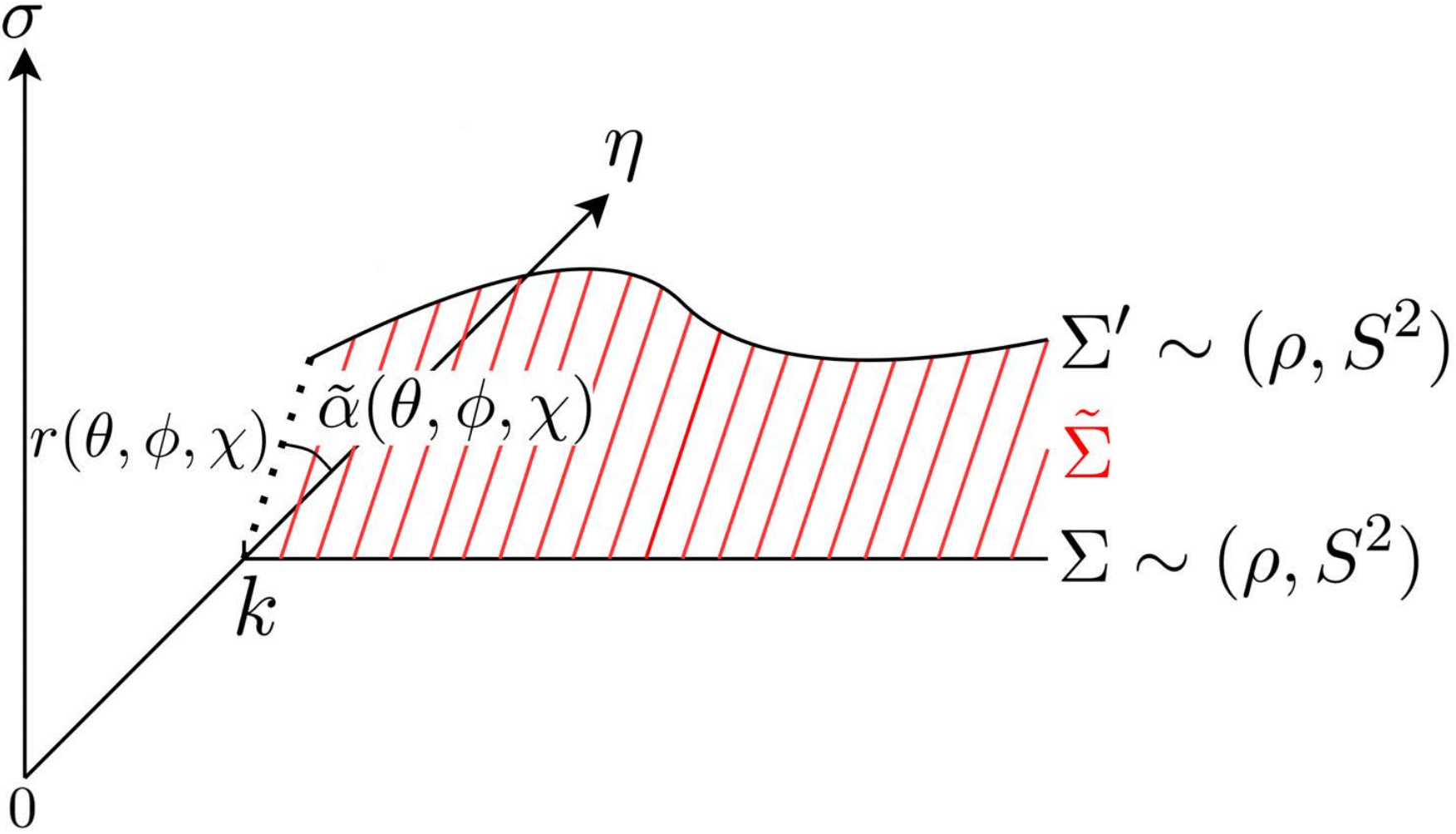} 
   \caption{\small The probe D6-brane wrapping $\Sigma=(\rho,S^2)$ at the locus $\{\sigma=0,\eta=k\}$, and a neighbouring probe D6-brane wrapping $\Sigma'=(\rho,S^2)$ at $\{\sigma = r \sin \alpha,\eta = k - r \cos \alpha\}$, where $r$ is small and $r$ and $\alpha$ are arbitrary functions of the internal coordinates. (We don't take $r$ and $\alpha$ to be arbitrary functions of $\rho$, the radial direction of AdS$_5$.) We introduced $\Tilde{\alpha}=\pi-\alpha$ for visual clarity.}
    \label{fig:Wiggly brane}
\end{figure}

We now introduce a secondary probe D6-brane, wrapping $\Sigma'$, in the generalized homology class of the one wrapping $\Sigma$: there is therefore a generalized cycle $(\Tilde{\Sigma},\Tilde{\mathcal{F}})$ such that $\partial \Tilde{\Sigma}=\Sigma'-\Sigma$ and $\Tilde{\mathcal{F}}|_{\Sigma}=\mathcal{F}$, $\Tilde{\mathcal{F}}|_{\Sigma'}=\mathcal{F}'$. We take this probe brane to be in the vicinity of the locus, as shown in Fig.~\ref{fig:Wiggly brane}.
In order to assess the potential stability of our probe D6-brane sitting at $\{\sigma=0,\eta=k\}$, we compare its energy to the one of the probes in its generalized homology class. To do so, we make use of the bound \eqref{boundE}.
We need to write down the supersymmetry breaking term $\Xi$ from \eqref{susybr3}, responsible for the violation of the third pure spinor equation \eqref{susy3},
which calibrates space-filling D$p$-branes in the supersymmetric case. The RR potentials are given in \cite{Macpherson:2024frt}, up to their closed components. We introduce
\be 
\Tilde{C}_\text{cl}=\e^{4\rho}p\, \dd\rho\wedge\text{vol}_{S^2}+\frac{1}{4}\e^{4\rho}\wedge\text{vol}_{S^2}\wedge\dd p,
\ee
with $p=p(\sigma,\eta)$ an arbitrary function. We choose it to be
\be p=-\frac{\kappa^3\xi^2}{\sqrt{2V''}},\ee
so that, at leading order in $r$:
\be \e^{-B}\Xi|_{(\rho,S^2)}=\e^{4\rho}\kappa^3\xi^2\sqrt\frac{r(\theta,\phi,\chi)}{b_k}\dd\rho\wedge\text{vol}_{S^2}.\ee
Given that this supersymmetry breaking term vanishes at the locus $\{\sigma=0,\eta=k\}$, the bound \eqref{boundE} simplifies into
\begin{align} E(\Sigma',\mathcal{F}')-E(\Sigma,\mathcal{F})&\geq T_6\int_{\Sigma'}\Xi|_{\Sigma'}\wedge \e^{-\mathcal{F}'}\\
&\geq T_6 \kappa^3\xi^2 \int \e^{4\rho}\sqrt\frac{r(\theta,\phi,\chi)}{b_k}\dd\rho\wedge\text{vol}_{S^2}\,.
\end{align}
The right-hand side of this bound being always positive,\footnote{We have considered throughout the brane energies per unit of external volume. The $\rho$ direction should here be thought of as an external dimension, discarding its integral.} we conclude that the probe D6-brane at the locus $\{\sigma=0,\eta=k\}$ is classically stable, as it minimizes its energy within its homology class.

\section*{Acknowledgments}
We thank N.~Macpherson, P.~Merrikin, and V.~Van Hemelryck for correspondence and discussions.
We are supported in part by the INFN, and by the MUR-PRIN contract 2022YZ5BA2.

\bibliographystyle{JHEP}
\bibliography{biblio}

\end{document}